\documentclass[11pt]{article}
\usepackage{eurosym}
\usepackage{amsfonts}
\usepackage{amssymb}
\usepackage{graphicx}
\usepackage{amsmath}
\usepackage{makeidx}
\usepackage{indentfirst}
\usepackage[T1]{fontenc}
\usepackage[utf8]{inputenc}

\newcounter{resultnum}[section]
\newcounter{conclusionnum}[section]
\newcounter{conditionnum}[section]
\newcounter{conjecturenum}[section]
\newcounter{examplenum}[section]
\newcounter{exercisenum}[section]
\newcounter{lemmanum}[section]
\newcounter{notationnum}[section]
\newcounter{theoremnum}[section]
\newcounter{definitionnum}[section]
\newcounter{corollarynum}[section]
\newcounter{remarknum}[section]
\newcounter{propositionnum}[section]
\newcounter{acknowledgementnum}[section]
\newcounter{algorithmnum}[section]
\newcounter{axiomnum}[section]
\newcounter{casenum}[section]
\newcounter{claimnum}[section]
\newcounter{summarynum}[section]
\newcounter{problemnum}[section]
\begin{document}

\title{The fundamental physical importance of generic off-diagonal solutions%
\\
and Grigori Perelman entropy in the Einstein gravity theory}
\date{v1 closed to the version published online in GRG:\ Aug 1, 2025}
\author{ \textbf{Sergiu I. Vacaru} \thanks{%
emails: sergiu.vacaru@fulbrightmail.org ; sergiu.vacaru@gmail.com }  \\ %
{\small \textit{\ Department of Physics, Kocaeli University, Kocaeli, 41001,
T\"{u}rkiye; }}  \\  {\small \textit{Department of Physics, California
State University at Fresno, Fresno, CA 93740, USA}} 
\\  {\small \textit{Astronomical Observatory, Taras Shevcenko National University of Kyiv, Kyiv 01601, Ukraine}}
\and {\textbf{El\c{s}en
Veli Veliev }} \thanks{%
email: elsen@kocaeli.edu.tr and elsenveli@hotmail.com} \\
{\small \textit{\ Department of Physics,\ Kocaeli University, Kocaeli,
41001, T\"{u}rkiye }} \vspace{.1 in} }
\maketitle

\begin{abstract}
The gravitational field equations in general relativity (GR) consist of a
sophisticated system of nonlinear partial differential equations. Solving
such equations in some generic off-diagonal forms is usually a hard analytic
or numeric task. Physically important solutions in GR were constructed using
diagonal ansatz for metrics with maximum 4 independent coefficients. The
Einstein equations can be solved in exact or parametric forms determined by
some integration constants for corresponding assumptions on spherical or
cylindric spacetime symmetries. The anholonomic frame and connection
deformation method allows us to construct generic off-diagonal solutions
described by 6 independent coefficients of metrics depending, in general, on
all spacetime coordinates. New types of exact and parametric solutions are
determined by generating and integration functions and (effective)
generating sources. They may describe vacuum gravitational and matter fields
solitonic hierarchies; locally anisotropic polarizations of physical
constants for black holes, wormholes, black toruses, or cosmological
solutions; various types of off-diagonal deformations of horizons etc. The
additional degrees of freedom (related to off-diagonal coefficients) can be
used to describe dark energy and dark matter configurations and elaborate
locally anisotropic cosmological scenarios. In general, the generic
off-diagonal solutions do not involve certain hypersurface or holographic
configurations and can't be described in the framework of the
Bekenstein-Hawking thermodynamic paradigm. We argue that generalizing the
concept of G. Perelman's entropy for relativistic Ricci flows allows us to
define and compute geometric thermodynamic variables for all possible
classes of solutions in GR. 

\end{abstract}

\tableofcontents



\section{Introduction}

\label{sec1}At the end of 1915, A. Einstein and D. Hilbert completed the
formulation of the general relativity (GR) theory. The monographs \cite%
{hawrking73,misner73,wald82,kramer03} contain main historical remarks, main
results and various cosmological and astrophysical applications of GR. To
elaborate on methods of constructing exact and parametric solutions of
gravitational and matter field equations is one of the most important tasks
in mathematical relativity. Various geometric, analytic and numeric methods
for finding solutions for Einstein equations have been elaborated for
diagonal ansatz for metrics and using the Levi-Civita, LC, connection. By
definition, such a linear connection is metric compatible and with zero
torsion. Different black hole (BH), wormhole (WH), black torus (BT), and
cosmological solutions were generated for prescribed spherical, cylindrical,
toroid, or other types of global and local symmetries. For more than 80
years, GR was considered as the standard theory of gravity even various
modified gravity theories, MGTs, were elaborated using non-Riemannian
geometries. We mention the string and gauge gravity theories, metric-affine
gravity models with torsions and nonmetricity, generalized Finsler gravity
theories etc. \cite{hehl95,blum12,vacaru25,vacaru18}, see also references
therein. 

\vskip5pt The discovery of late-time cosmic acceleration \cite%
{riess98,perlmutter99} resulted in extensive research on MGTs \cite%
{sotiriou10,nojiri11,capo11,clifton12,harko14} and dark energy (DE) and dark
matter (DM) physics \cite{copeland06} (we cite some early works); and on
geometric and quantum information flows \cite{partner06}, see references
therein. The anholonomic frame and connection deformation method (AFCDM) was
elaborated as a geometric and analytic method for constructing generic
off-diagonal solutions in GR and MGTs, see recent reviews of results in \cite%
{vacaruplb16,vbubuianu17,vacaru18,partner02}. We note that a generic
off-diagonal metric can't be diagonalized by coordinate transforms in a
finite spacetime region and the coefficients of metrics and (non) linear
connections may depend, in general, on all spacetime coordinates. The main
reason to study off-diagonal configurations in GR is that (in general) the
corresponding exact or parametric solutions are described by 6 independent
coefficients (from 10 ones for a symmetric metric tensor on a Lorentz
manifold). These are different from the cases of quasi-stationary and
cosmological solutions described by diagonal ansatz with a maximum of 4
independent coefficients of metrics. We argue that two additional degrees of
freedom allow us to describe new models of relativistic physics and
cosmology involving nonlinear off-diagonal gravitational and (effective)
matter field interactions. This can be used, for instance, for elaborating
various models of nonlinear classical and quantum theories, locally
anisotropic thermodynamics and effective DE and DM physics. We mention here
various classes of generic off-diagonal solutions involving nontrivial
gravitational vacuum and pattern--creating structures (for instance, time
quasi-crystal-like), with locally anisotropic polarizations of physical
constants; or describing moving BHs, black ellipsoid/torus configurations
and nonholonomic WHs. For details, we cite \cite%
{partner06,vacaruplb16,vbubuianu17,vacaru18,partner02,sv11,nonassocFinslrev25,vacaru25}%
. 

\vskip5pt This paper contains a brief overview and new results on
applications of the AFCDM for generating new classes of generic off-diagonal
quasi-stationary and locally anisotropic cosmological spacetimes in GR. In a
more general and different form, for nonassociative Finsler-like MGTs, such
methods and rigorous proofs are reviewed in \cite{nonassocFinslrev25}. We
note that to decouple in off-diagonal form the Einstein equations is not
possible if we work only with the LC-connection $\nabla \lbrack \mathbf{g}].$
The main idea in our geometric approach is to use some canonical distortions
of linear connections, $\nabla \lbrack \mathbf{g}]\rightarrow $ $\widehat{%
\mathbf{D}}[\mathbf{g}]=\nabla \lbrack \mathbf{g}]+$ $\widehat{\mathbf{Z}}[%
\mathbf{g}],$ which allows a general off-diagonal decoupling of the Einstein
equations. The corresponding systems of nonlinear partial differential
equations, PDEs, are written in terms of $\widehat{\mathbf{D}}[\mathbf{g}]$
and using effective sources encoding terms defined by the distortion tensor $%
\widehat{Z}[\mathbf{g}].$ In this approach, $\widehat{\mathbf{D}}[\mathbf{g}%
] $ and $\widehat{\mathbf{Z}}[\mathbf{g}]$ are defined by the same metric
structure as $\nabla \lbrack \mathbf{g}]$ and also in a metric compatible
form, $\widehat{\mathbf{D}}\mathbf{g}=0.$ For "hat" geometric variables, the
nontrivial torsion structure $\widehat{\mathbf{T}}[\widehat{\mathbf{D}}]$ is
induced in a nonholonomic (i.e. non-integrable, equivalently, anholonomic)
form by generic off-diagonal coefficients of $\mathbf{g.}$ This is different
from the Einstein-Cartan or string gravity when additional (effective)
sources are considered for torsion fields. In this work, we use "boldface"
symbols for spaces and geometric/physical objects if the geometric
constructions can be adapted to a conventional 2+2 splitting stated by
certain type nonholonomic frames (equivalently, tetrads). We may consider $%
\widehat{\mathbf{D}}$ as an auxiliary linear connection which allows us to
decouple and integrate in certain general forms the Einstein equations in
"hat"-variables. Such generic off-diagonal solutions are determined by
generating and integration functions and generating (effective) sources
depending, in principle, on all spacetime coordinates. After certain classes
of solutions have been constructed in a general off-diagonal form, we can
consider sub-classes of generating/ integrating data when the torsion $%
\widehat{\mathbf{T}}[\widehat{\mathbf{D}}]=0,$ i.e. to extract
LC-configurations. We emphasize that if we work from the very beginning with
$\nabla \lbrack \mathbf{g}]$ we are not able to decouple for generic
off-diagonal ansatz the corresponding systems of nonlinear PDEs. Certain
general classes of exact or parametric solutions have to be found for  $%
\widehat{\mathbf{D}}$ and then restricted to LC configurations if we want to
generate solutions in the GR theory. 

The AFCDM is very different from other methods of constructing exact solutions defined by diagonal ansatz with
certain global or local symmetries \cite{kramer03}. Those methods do not
involve hat-variables and distorted linear connections as we defined in \cite%
{vacaru18,partner06,vbubuianu17,partner02,nonassocFinslrev25,v25c}. We
emphasize that if we apply the AFCDM on Lorentz manifolds by considering
nonholonomic frames and distortions of the LC connection to auxiliary
connections with nonholonomically induced torsion, we stay almost "closed"
to GR. Such nonholonomic torsion structures are defined by off-diagonal
terms of the metrics and all classes of generated solutions can be
restricted to possess zero torsion (the metrics remaining, in general,
generic off-diagonal). The "trick" with "hat" connections is necessary for
general decoupling the Einstein equations with respect to certain
nonholonomic frames (if we work only with the LC-connection this is not
possible). Haven decoupled the system of nonlinear PDEs, we can integrate it
in off-diagonal form and always impose additional nonholonomic constraints
on the generating functions and generating sources to construct
exact/parametric solutions in GR. Such an approach became trivial if we fix
from the very beginning only diagonal ansatz for metrics, which results in
systems of nonlinear ODEs (they are defined by integration constants). 

\vskip5pt The main goal of this paper is to analyze the most general
geometric properties and physical implications of generic off-diagonal
solutions in GR. We shall omit tedious proofs and cumbersome formulas
presented (in more general forms, for various MGTs) in \cite%
{vbubuianu17,partner02}. A generalized abstract geometric and abstract index
formalism (similar to that in \cite{misner73}) will be used to simplify the
formulas and proofs. We shall provide explicit formulas describing how
physically important BH, WH, BT, and cosmological solutions are
nonholonomically deformed into new classes of off-diagonal solutions in GR.
Certain models with nontrivial gravitational vacuum, polarization of
physical constants, and deformation of horizons will be elaborated. Because
the general off-diagonal solutions in GR do not involve, in general, certain
horizon, duality, or holographic configurations, the Bekenstein-Hawking
thermodynamic paradigm \cite{bek2,haw2} is not applicable to characterise
the bulk of such soluitons. We argue that other types of thermodynamic
variables defined by G. Perelman \cite{perelman1} can be used for
characterizing the fundamental properties of off-diagonal solutions \cite%
{gheorghiuap16,partner06}. In our works, we do not attempt to formulate and
prove relativistic or non-Riemannian generalizations of the Poincar\'{e}%
-Thurston conjecture \cite{perelman1,hamilton82}. Our models with
nonholonomic generalizations of the Ricci flow theory are formulated to
apply the AFCDM for constructing new classes of off-diagonal solutions for
nonholonomic Ricci solitons. Such geometric configurations are equivalent
to the Einstein equations with nontrivial cosmological constants if certain
additional conditions are imposed. We show how to define and compute the G.
Perelman thermodynamic variables for respective classes of physical
important solutions. 

\vskip5pt The paper is organized as follows. In section \ref{sec02}, we
outline the main concepts and necessary formulas for formulating and
applying the AFCDM in GR. The formalism of nonholonomic 2+2 variables with
associated nonlinear connection structures and distortion of linear
connection is outlined. We explain how the nonholonomic Einstein equations
can be decoupled in hat variables and provide the formulas for generating
off-diagonal quasi-stationary solutions. The importance of new types of
nonlinear symmetries is emphasized. We show how the quasi-stationary metrics
can be dualized for a time-like coordinate to generate locally anisotropic
cosmological solutions. In section \ref{sec03}, we provide explicit formulas
for constructing off-diagonal deformations of BH, WH, BT and cosmological
solutions in GR. Section \ref{sec04} is devoted to a relativistic
generalization of the G. Perelman thermodynamics for nonholonomic geometric
flows and nonholonomic Ricci solitons. We show how such thermodynamic
variables can be computed for respective classes of off-diagonal solutions
(constructed in this work). The main results are discussed and concluded in
section \ref{sec05}. Appendix \ref{appendixa} contains technical results and
necessary formulas for proving a general decoupling and integration property
of the Einstein equations. Tables 1-3 from Appendix \ref{appendixb}
summarize the AFCDM for generating off-diagonal quasi-stationary and locally
anisotropic solutions in GR. 

\section{The AFCDM as a geometric method for constructing off-diagonal
solutions}

\label{sec02}We review the main geometric constructions defining the AFCDM
for constructing exact and parametric solutions in GR using nonholonomic 2+2
variables with canonical distortions of linear connections. Such a geometric
and analytic method allows us to prove certain general decoupling and
integration properties of the gravitational field equations (see subsections
below) and of the relativistic geometric flow equations (see Section \ref%
{sec04}). Detailed proofs for extra dimensions and MGTs can be found in \cite%
{vbubuianu17,vacaru18,partner02,nonassocFinslrev25,partner06}. Appendices %
\ref{appendixa} and \ref{appendixb} summarize the main formulas and proofs
for four-dimensional (4-d) Lorentz manifolds. 

\subsection{Nonholonomic 2+2 variables on Lorentz manifolds}

The GR theory is formulated in standard geometric form for Lorentz manifolds
(i.e. spacetimes) determined by geometric data $(V,\mathbf{g})$ \cite%
{hawrking73,misner73,wald82}. In such an approach, $V$ is a 4-d
pseudo-Riemannian manifold of necessary smooth (differentiability) class
enabled with a symmetric metric tensor of signature $(+++-),$ 
\begin{equation}
\mathbf{g}=g_{\alpha ^{\prime }\beta ^{\prime }}(u)e^{\alpha ^{\prime
}}\otimes e^{\beta ^{\prime }}.  \label{mst}
\end{equation}%
In (\ref{mst}), some general co-frames $e^{\alpha ^{\prime }}$ are dual to
frame bases $e_{\alpha ^{\prime }}$ and the Einstein convention of
summarizing "up-low" repeating indices is used. On a coordinate neighborhood 
$U\subset V,$ we can use local coordinates labeled as $u=\{u^{\alpha
}=(x^{i},y^{a})\}.$ Such coordinates are used for a conventional $2+2$
splitting into h-coordinates, $x=(x^{i}),$ and v-coordinates, $y=(y^{a}),$
for indices $j,k,...=1,2$ and $a,b,c,...=3,4,$ when $\alpha ,\beta
,...=1,2,3,4.$ To compute the coefficients of geometric objects on $V,$ we
can use local coordinate bases and, respective, co-bases, $%
e_{\alpha}=\partial _{\alpha }=\partial /\partial u^{\beta }$ and $%
e^{\beta}=du^{\beta }.$ We can also consider transforms to arbitrary frames
(tetrads) defined as $e_{\alpha ^{\prime }}=e_{\ \alpha ^{\prime}}^{\alpha
}(u)e_{\alpha }$ and $e^{\alpha ^{\prime }}=e_{\alpha \ }^{\ \alpha ^{\prime
}}(u)e^{\alpha }.$ By definition, such (co) bases are orthonormal if $%
e_{\alpha \ }^{\ \alpha ^{\prime }}e_{\ \alpha ^{\prime}}^{\beta }=\delta
_{\alpha }^{\beta },$ where $\delta _{\alpha }^{\beta }$ is called the
Kronecker symbol. 

A 2+2 spacetime decomposition can be defined in coordinate-free form using
nonholonomic 2+2 distributions on $V$. Let us explain how this geometric
formalism is elaborated for 4-d Lorentz manifolds. We can always introduce
in a global form on $V$ a conventional nonlinear connection structure
(N-connection). Such a geometric object was used in coordinate form by E.
Cartan \cite{cartan35} (on tangent bundles) and defined in rigorous
mathematical form in \cite{ehresmann55}. Using point-by-point decompositions
of the tangent bundles, $TV:=\bigcup\nolimits_{u}T_{u}V,$ an N-connection is
defined as a non--integrable (equivalently, nonholonomic, or anholonomic)
conventional horizontal and vertical splitting. We write this, in brief, as
a h- and v--decomposition into respective 2-d and 2-d subspaces, $hV$ and $%
vV.$ This is equivalent to the condition that a Whitney sum 
\begin{equation}
\mathbf{N}:\ TV=hV\oplus vV  \label{ncon}
\end{equation}%
is globally defined for $V$ and $TV.$ A N-connection (\ref{ncon}) consists
an example of nonholonomic distribution defining a fibered 2+2 structure on
a Lorentz spacetime manifold \cite{vacaruplb16,vbubuianu17}. This is
different from the Finsler geometry, see details in \cite{vacaru18} when the
N-connections are defined by splitting of type $TTV=hTV\oplus vTV.$ In those
geometries with additional velocity/momentum-like variables, the second
tangent bundle $TTV$ and (in some equivalent forms) the dual constructions
for $TT^*V$) are enabled by nonholonomic distributions defined by a certain
splitting with exact sequences. A N-connection (\ref{ncon}) is defined by a
nonholonomic distribution stated by a set of coefficients $N_{i}^{a}(u)$
when in local coordinate form 
\begin{equation}
\mathbf{N}=N_{i}^{a}(x,y)dx^{i}\otimes \partial /\partial y^{a}.
\label{nconcoef}
\end{equation}%
%

The term \textit{nonholonomic Lorentz manifold} (denoted $\mathbf{V}$) can
be used and if such a spacetime is enabled with a h- and v-splitting (\ref%
{ncon}) at least on a neighbourhood $U\subset V.$ For general covariant
frame and coordinate transforms, we can consider general frame structures, $%
e_{\alpha ^{\prime }}=(e_{i^{\prime }},e_{a^{\prime }})$ and $e^{\beta
^{\prime }}= (e^{i^{\prime }},e^{a^{\prime }}).$ A decoupling of fundamental
physical equations and generating off-diagonal solutions are possible only
for special nonholonomic frames as we define below. Here we note that we
shall omit priming, underlying, overlying etc. of indices if that do not
result in ambiguities. "Boldface" symbols will be used to emphasize that
certain spaces or geometric objects are enabled (or adapted) with (to) an
N-connection structure. 

A set of N-connection coefficients $\mathbf{N}=\{N_{i}^{a}\}$ (\ref{nconcoef}%
) allows us to define N--elongated (equivalently, N-adapted) local bases as
certain partial derivative operators, $\mathbf{e}_{\nu },$ and dual (co-)
bases and instead of standard differentials, $\mathbf{e}^{\mu }.$ We can
work with linear on $N_{i}^{a}$ local frames: 
\begin{eqnarray}
\mathbf{e}_{\nu } &=&(\mathbf{e}_{i},e_{a})=(\mathbf{e}_{i}=\partial
/\partial x^{i}-\ N_{i}^{a}(u)\partial /\partial y^{a},\ e_{a}=\partial
_{a}=\partial /\partial y^{a}),\mbox{ and  }  \label{nader} \\
\mathbf{e}^{\mu } &=&(e^{i},\mathbf{e}^{a})=(e^{i}=dx^{i},\ \mathbf{e}%
^{a}=dy^{a}+\ N_{i}^{a}(u)dx^{i}).  \label{nadif}
\end{eqnarray}%
A N-elongated basis (\ref{nader}) satisfies nonholonomic relations 
\begin{equation}
\lbrack \mathbf{e}_{\alpha },\mathbf{e}_{\beta }]=\mathbf{e}_{\alpha }%
\mathbf{e}_{\beta }-\mathbf{e}_{\beta }\mathbf{e}_{\alpha }=W_{\alpha \beta
}^{\gamma }\mathbf{e}_{\gamma },  \label{nonholr}
\end{equation}%
when the (antisymmetric) nontrivial anholonomy coefficients are computed 
\begin{equation}
W_{ia}^{b}=\partial _{a}N_{i}^{b},W_{ji}^{a}=\Omega _{ij}^{a}=\mathbf{e}%
_{j}\left( N_{i}^{a}\right) -\mathbf{e}_{i}(N_{j}^{a}).  \label{anhcoef}
\end{equation}%
Here $\Omega _{ij}^{a}$ define the coefficients of an N-connection curvature 
$\Omega $. A N-adapted base $\mathbf{e}_{\alpha }\simeq \partial
_{\alpha}=\partial /\partial u^{\alpha }$ is holonomic if and only if all
anholonomic coefficients (\ref{anhcoef}) vanish. In such cases, usual
partial derivatives $\partial _{\alpha }$ can be obtained for certain
coordinate transforms. In curved spacetime coordinates, for holonomic bases,
the coefficients $N_{j}^{a}$ may be non-zero even if all $W_{\alpha \beta
}^{\gamma }=0.$ Such an example is given by rotating coordinates which
result in effective off-diagonal terms to certain diagonal metrics. 

To emphasize that certain geometric objects are defined on a nonholonomic
Lorentz manifold $\mathbf{V}$ enabled with a N-connection structure we shall
call them d-objects. Similarly, terms like d-vectors, d-tensors, d-spinors
etc. can be used. If necessary, N-adapted geometric constructions can be
performed on  tangent, $T\mathbf{V,}$ and cotangent, $T^{\ast }\mathbf{V}$,
bundles; and their tensor products, for instance, $T\mathbf{V\otimes }%
T^{\ast }\mathbf{V.}$ For d-objects, the N-adapted coefficients are computed
for frames (\ref{nader}) and (\ref{nadif}), or their tensor products etc. In
abstract form, for instance, we can write a d--vector as $\mathbf{X}=(hX,vX)$%
. 

Using above convention on d-objects, a spacetime metric $\mathbf{g}$ (\ref%
{mst}) can be written equivalently as a d--metric, 
\begin{eqnarray}
\ \mathbf{g} &=&(hg,vg)=\ g_{ij}(x,y)\ e^{i}\otimes e^{j}+\ g_{ab}(x,y)\ 
\mathbf{e}^{a}\otimes \mathbf{e}^{b}  \label{dm} \\
&=&\underline{g}_{\alpha \beta }(u)du^{\alpha }\otimes du^{\beta }.
\label{cm}
\end{eqnarray}%
We shall use also notations of type $hg=\{\ g_{ij}\}$ and $\ vg=\{g_{ab}\}.$
The off-diagonal coefficients in (\ref{cm}) are computed if we introduce the
coefficients of (\ref{nadif}) into (\ref{dm}) with a corresponding
regrouping for a coordinate dual basis. This way, we obtain 
\begin{equation}
\underline{g}_{\alpha \beta }=\left[ 
\begin{array}{cc}
g_{ij}+N_{i}^{a}N_{j}^{b}g_{ab} & N_{j}^{e}g_{ae} \\ 
N_{i}^{e}g_{be} & g_{ab}%
\end{array}%
\right] .  \label{ansatz}
\end{equation}
A metric $\mathbf{g}=\{\underline{g}_{\alpha \beta }\}$ (\ref{ansatz}) is
generic off--diagonal if the anholonomy coefficients $W_{\alpha
\beta}^{\gamma }$ (\ref{anhcoef}) are not identical to zero. For 4-d
(pseudo) Riemannian spaces, such a matrix can not be diagonalized via
coordinate transforms. The above formulas for d-metrics and off-diagonal
metrics can be defined for a prescribed set of coefficients $N_{i}^{a}.$
Parameterizations of type (\ref{ansatz}) are used in MGTs, for instance, in
1) in Kaluza-Klein gravity when $N_{j}^{e}=A_{j}^{e}$ are identified as
certain gauge fields after compactification on $y$-coordinates (usually,
there are considered higher dimension spacetimes). \ 2) Typically, in
Finsler like gravity theories, $N_{j}^{e}$ are constructed in certain forms
which define respective Finsler-Lagrange-Hamilton theories, see details in 
\cite{vacaru18,vacaruplb16,vbubuianu17}. In GR, the N-coefficients can be
related to generic off-diagonal terms and used for N-adapted geometric
constructions which allow to prove decoupling properties of the Einstein
equations written in hat-variables (see next subsections). The constructions
are very cumbersome if we use coordinate bases and off-diagonal metrics $%
\mathbf{g}=\{\underline{g}_{\alpha \beta }\}$ (\ref{ansatz}). To prove
general decoupling and integrability properties is not possible if we use
only the LC-connection $\nabla $ (in a certain sense, the zero torsion
conditions do not allow it). Here we also note that the components of the
inverse metric $\underline{g}^{\alpha \beta }$ (in general, with
off-diagonal terms) are computed for nondegenerated metric structures
following standard formulas $\underline{g}^{\alpha \beta }\underline{g}%
_{\gamma \beta }=\delta _{\beta }^{\alpha }.$ The inverse d-metrics and
their h- and v-coefficients, $\mathbf{g}^{\alpha \beta }=(g^{ij},g^{ab}),$
are defined in similar forms. 

\subsubsection{Fundamental geometric and physical d-objects}

Prescribing an N-connection structure on a nonholonomic $\mathbf{V,}$ we can
define certain classes of linear connections which are, or not, N-adapted.
For instance, a LC connection $\nabla $ does not preserve an h- and
v-splitting and is not adapted. To perform a covariant differential and
integral calculus in N-adapted form we need a more special class of
distinguished connections, d-connections.

A \textbf{\ d--connection} $\mathbf{D}=(hD,vD)$ is a linear connection
preserving under parallelism the N--connection splitting (\ref{ncon}). It
defines a covariant N--adapted derivative $\mathbf{D}_{\mathbf{X}}\mathbf{Y}$
of a d--vector field $\mathbf{Y}=hY+vY$ in the direction of a d--vector $%
\mathbf{X}=hX+vC.$ For N--adapted frames (\ref{nader}) and (\ref{nadif}),
any covariant d-derivative $\mathbf{D}_{\mathbf{X}}\mathbf{Y}$ can be
computed as in GR \cite{misner73} and, in a more general sense as in metric
affine gravity \cite{hehl95,vacaru18}. The N-adapted coefficients involve
respective h- and v-indices, 
\begin{equation}
\mathbf{D}=\{\mathbf{\Gamma }_{\ \alpha \beta }^{\gamma }=(L_{jk}^{i},\acute{%
L}_{bk}^{a};\acute{C}_{jc}^{i},C_{bc}^{a})\},\mbox{ where }hD=(L_{jk}^{i},%
\acute{L}_{bk}^{a})\mbox{ and }vD=(\acute{C}_{jc}^{i},C_{bc}^{a}),
\label{hvdcon}
\end{equation}%
see details in \cite%
{vacaruplb16,vbubuianu17,nonassocFinslrev25,partner02,partner06}. The
convention for abstract or coordinate indices is $i,j,...=1,2$ and $%
a,b,...=3,4.$

As a N-adapted linear connection, any d--connection $\mathbf{D}$ is
characterized by three fundamental geometric d-objects. In abstract form, we
define: 
\begin{eqnarray}
\mathcal{T}(\mathbf{X,Y})&:= &\mathbf{D}_{\mathbf{X}}\mathbf{Y}-\mathbf{D}_{%
\mathbf{Y}}\mathbf{X}-[\mathbf{X,Y}],\mbox{ torsion d-tensor,  d-torsion};
\label{fundgeom} \\
\mathcal{R}(\mathbf{X,Y})&:=&\mathbf{D}_{\mathbf{X}}\mathbf{D}_{\mathbf{Y}}-%
\mathbf{D}_{\mathbf{Y}}\mathbf{D}_{\mathbf{X}}-\mathbf{D}_{\mathbf{[X,Y]}},%
\mbox{ curvature d-tensor, d-curvature};  \notag \\
\mathcal{Q}(\mathbf{X})&:= &\mathbf{D}_{\mathbf{X}}\mathbf{g,}%
\mbox{nonmetricity d-fiels, d-nonmetricity}.  \notag
\end{eqnarray}%
The N--adapted coefficients of such geometric d-objects can be computed in
explicit form by introducing $\mathbf{X}=\mathbf{e}_{\alpha }$ and $\mathbf{Y%
}=\mathbf{e}_{\beta },$ defined by (\ref{nader}), and considering
h-v-splitting $\mathbf{D}=\{\mathbf{\Gamma }_{\ \alpha \beta }^{\gamma }\}$ (%
\ref{hvdcon}) into above formulas, 
\begin{eqnarray}
\mathcal{T} &=&\{\mathbf{T}_{\ \alpha \beta }^{\gamma }=\left( T_{\
jk}^{i},T_{\ ja}^{i},T_{\ ji}^{a},T_{\ bi}^{a},T_{\ bc}^{a}\right) \};
\label{fundgeomc} \\
\mathcal{R} &\mathbf{=}&\mathbf{\{R}_{\ \beta \gamma \delta }^{\alpha }%
\mathbf{=}\left( R_{\ hjk}^{i}\mathbf{,}R_{\ bjk}^{a}\mathbf{,}R_{\ hja}^{i}%
\mathbf{,}R_{\ bja}^{c}\mathbf{,}R_{\ hba}^{i},R_{\ bea}^{c}\right) \mathbf{%
\};}  \notag \\
\ \mathcal{Q} &=&\mathbf{\{Q}_{\ \alpha \beta }^{\gamma }=\mathbf{D}^{\gamma
}\mathbf{g}_{\alpha \beta }=(Q_{\ ij}^{k},Q_{\ ij}^{c},Q_{\ ab}^{k},Q_{\
ab}^{c})\}.  \notag
\end{eqnarray}%
We say that any geometric data $\left( \mathbf{V},\mathbf{N},\mathbf{g,D}%
\right) $ define a nonholonomic, or N-adapted, metric-affine structure
(equivalently, metric-affine d-structure) determined by a d-metric and a
d-connection stated independently. If we work with a LC connection $\nabla ,$
defined by the conditions $\mathcal{Q}[\nabla ]=\nabla \mathbf{g=0}$ and $%
\mathcal{T}[\nabla ]=0,$ we obtain abstract and index formulas as in GR. In
our works, for tensors and not d-tensors, we use not boldface symbols and
consider functional dependencies $[\nabla ]$ or abstract left labels, for
instance, $\mathcal{R}[\nabla ] =\ _{\nabla}\mathcal{R}=\{ R_{\ \beta \gamma
\delta }^{\alpha }=\mathbf{\ }_{\nabla }R_{\ \beta \gamma \delta }^{\alpha
}\}.$ Explicit formulas for the coefficients (\ref{fundgeomc}) of different
types of d-connections and linear connections are provided in \cite%
{vacaru18,vbubuianu17,partner02,nonassocFinslrev25}, see also Appendix \ref%
{appendixa}. 

\subsubsection{The Einstein equations in nonholonomic canonical variables}

For any d-metric structure $\mathbf{g}$ (\ref{dm}), we can define two
important linear connection structures: 
\begin{equation}
(\mathbf{g,N})\rightarrow \left\{ 
\begin{array}{cc}
\mathbf{\nabla :} & \mathbf{\nabla g}=0;\ _{\nabla }\mathcal{T}=0,\ \mbox{\
the LC--connection }; \\ 
\widehat{\mathbf{D}}: & \widehat{\mathbf{Q}}=0;\ h\widehat{\mathcal{T}}=0,v%
\widehat{\mathcal{T}}=0,\ hv\widehat{\mathcal{T}}\neq 0,\mbox{ the canonical
d-connection}.%
\end{array}%
\right.  \label{twocon}
\end{equation}
The LC connection is the standard one in (pseudo) Riemannian geometry but it
does not allow decoupling of (modified) gravitational equations for generic
off-diagonal metrics. In our works, we prefer to work with the auxiliary hat
connection, i.e. a d-connection, $\widehat{\mathbf{D}}.$ It is related to
the LC connection by a canonical distortion relation, 
\begin{equation}
\widehat{\mathbf{D}}[\mathbf{g}]=\nabla \lbrack \mathbf{g}]+\widehat{%
\mathcal{Z}}[\mathbf{g}],  \label{canondistrel}
\end{equation}%
where $\widehat{\mathcal{Z}}=\{\widehat{\mathbf{Z}}_{\ \alpha \beta}^{\gamma
}\}$ is the canonical distortion d-tensor. We write $[\mathbf{g}]$ to
emphasize that all three geometric objects are determined by the same metric
structure $\mathbf{g}$. For geometric objects adapted to an N-connection, we
use boldface symbols. In the next section, we provide all necessary
definitions and abstract/index formulas. A less known fact is that GR can be
defined equivalently using both types of geometric data $[\mathbf{g},\nabla ]
$ and (or) $[\mathbf{g},\mathbf{N},\widehat{\mathbf{D}}].$ The hat variables
are used in our and co-authors' works for elaborating the AFCDM. They allow
to decouple and integrate the Einstein equations with nontrivial
N-connection structure $\mathbf{N}$ "absorbing" in a sense the off-diagonal
terms in $\mathbf{g}=\{\underline{g}_{\alpha \beta }\}$ (\ref{ansatz}). It
should be noted that the distortions (\ref{canondistrel}) involve a
canonical d-torsion structure, $\widehat{\mathcal{T}}=\{\widehat{\mathbf{T}}%
_{\ \alpha \beta }^{\gamma }\}$, as we stated in (\ref{twocon}). Such
geometric d-objects are different from those in the Einstein-Cartan theory
because $\widehat{\mathcal{T}}$ and $\widehat{\mathcal{Z}}$ are
nonholonomically induced by N-connection and d-metric coefficients. We do
not need additional (spin-like) sources for the canonical d-torsion. In a
general form, we can include the distortions of the Ricci tensor, (to be
defined below) determined by (\ref{canondistrel}) as certain effective
matter sources in the Einstein equations for $[\mathbf{g},\nabla ]$. An
alternative variant for extracting LC configurations is to impose additional
constraints on generating and integration functions for respective solutions
which result in zero distortion d-tensors, 
\begin{equation}
\widehat{\mathbf{Z}}=0,\mbox{ which is equivalent to }\ \widehat{\mathbf{D}}%
_{\mid \widehat{\mathcal{T}}=0}=\nabla .  \label{lccond}
\end{equation}

A canonical d-connection (\ref{twocon}) is defined by N-adapted coefficients 
$\widehat{\mathbf{D}}=\{\widehat{\mathbf{\Gamma }}_{\ \alpha \beta }^{\gamma
}=(\widehat{L}_{jk}^{i},\widehat{L}_{bk}^{a},\widehat{C}_{jc}^{i},\widehat{C}%
_{bc}^{a})\}$ computed for a d--metric $\mathbf{g}=[g_{ij},g_{ab}]$ (\ref{dm}%
) using N--elongated partial derivatives (\ref{nader}). In explicit form, 
\begin{eqnarray}
\widehat{L}_{jk}^{i} &=&\frac{1}{2}g^{ir}(\mathbf{e}_{k}g_{jr}+\mathbf{e}%
_{j}g_{kr}-\mathbf{e}_{r}g_{jk}),  \label{cdc} \\
\widehat{L}_{bk}^{a} &=&e_{b}(N_{k}^{a})+\frac{1}{2}g^{ac}(\mathbf{e}%
_{k}g_{bc}-g_{dc}\ e_{b}N_{k}^{d}-g_{db}\ e_{c}N_{k}^{d}),  \notag \\
\widehat{C}_{jc}^{i} &=&\frac{1}{2}g^{ik}e_{c}g_{jk},\ \widehat{C}_{bc}^{a}=%
\frac{1}{2}g^{ad}(e_{c}g_{bd}+e_{b}g_{cd}-e_{d}g_{bc}).  \notag
\end{eqnarray}
In a similar form, we can compute the coefficients of an LC connection $%
\nabla =\{\Gamma _{\ \alpha \beta }^{\gamma }\},$ see general coefficient or
N-adapted formulas in \cite{misner73,vacaru18}. The N-adapted coefficients
of the canonical distortion d-tensor in (\ref{canondistrel}) can be found as 
$\widehat{\mathbf{Z}}=\{\widehat{\mathbf{Z}}_{\ \alpha \beta }^{\gamma }=%
\widehat{\mathbf{\Gamma }}_{\ \alpha \beta }^{\gamma }-\Gamma _{\ \alpha
\beta }^{\gamma }\}.$

Using $\widehat{\mathbf{\Gamma }}_{\ \alpha \beta }^{\gamma }$\ (\ref{cdc})
instead of the coefficients of a general d-connection $\mathbf{D=\{\Gamma }%
_{\ \alpha \beta }^{\gamma }\}$, we can compute the N-adapted coefficients
of canonical fundamental d--objects (\ref{fundgeom}). Such hat symbols are
used, for instance, for $\widehat{\mathcal{R}}=\{\widehat{\mathbf{R}}_{\
\beta \gamma \delta }^{\alpha }=(\widehat{R}_{\ hjk}^{i},\widehat{R}_{\
bjk}^{a},...)\},\ \widehat{\mathcal{T}}=\{\widehat{\mathbf{T}}_{\ \alpha
\beta }^{\gamma }=(\widehat{T}_{\ jk}^{i},\widehat{T}_{\ ja}^{i},...)\},$
for $\widehat{\mathcal{Q}}=\{\widehat{\mathbf{Q}}_{\gamma \alpha \beta }=(%
\widehat{Q}_{kij}=0,\widehat{Q}_{kab}=0)=0.$ Considering the canonical
distortion relation for linear connections (\ref{canondistrel}), we can
compute respective canonical distortions of the fundamental geometric
d-objects for $\nabla $. Such formulas relate, for instance, two different
curvature tensors, $\ _{\nabla }\mathcal{R}=\{\ _{\nabla }R_{\ \beta \gamma
\delta }^{\alpha }\}$ and $\ \widehat{\mathcal{R}}=\{\widehat{\mathbf{R}}_{\
\beta \gamma \delta }^{\alpha }\}$ etc. 

We can define the canonical Ricci d-tensor using a respective contraction on
the 1st and 4th indices of the canonical curvature d-tensor,%
\begin{equation}
\widehat{\mathbf{R}}ic=\{\widehat{\mathbf{R}}_{\ \beta \gamma }:=\widehat{%
\mathbf{R}}_{\ \beta \gamma \alpha }^{\alpha }\}.  \label{criccidt}
\end{equation}%
In general, this d-tensor is not symmetric, $\widehat{\mathbf{R}}_{\ \beta
\gamma }\neq \widehat{\mathbf{R}}_{\ \gamma \beta }.$ This is typical for
nonholonomic geometric objects. The canonical scalar curvature is defined as%
\begin{equation}
\widehat{R}sc:=\mathbf{g}^{\alpha \beta }\widehat{\mathbf{R}}_{\ \alpha
\beta }.  \label{criccidsc}
\end{equation}%
The (nonholonomic) canonical Einstein d-tensor can be defined in abstract
geometric form as in \cite{misner73} but using (\ref{criccidt}) and (\ref%
{criccidsc}), 
\begin{equation}
\widehat{\mathbf{E}}n:=\widehat{\mathbf{R}}ic-\frac{1}{2}\mathbf{g}\widehat{R%
}sc=\{\widehat{\mathbf{R}}_{\ \beta \gamma }-\frac{1}{2}\mathbf{g}_{\ \beta
\gamma }\widehat{R}sc\}.  \label{criccdsc}
\end{equation}%
We can also introduce in N-adapted form necessary types of energy-momentum
sources $\widehat{\mathbf{T}}_{\alpha \beta }$ and postulate the Einstein
equations in hat variables, 
\begin{equation}
\widehat{\mathbf{E}}n_{\alpha \beta }=\varkappa \widehat{\mathbf{T}}_{\alpha
\beta }.  \label{einstceq1}
\end{equation}%
In these formulas, the constant $\varkappa $ can be related to the Newton
gravitational constant. Such nonholonomic equations can be equivalent to the
standard Einstein ones in GR if $\widehat{\mathbf{T}}_{\alpha \beta }$ is
constructed in a way to include as sources the energy-momentum tensors for
matter (in N-adapted bases) but also the distortion terms coming from $%
En_{\alpha \beta }[\nabla ]$ and $T_{\alpha \beta }[\nabla ].$ Such
distortions have to be computed using (\ref{canondistrel}). Alternatively,
we obtain an equivalence with the gravitational field equations in GR if we
add to the nonholonomic constraints (\ref{lccond}) for extracting LC
configurations. 

Let us explain how the nonholonomic canonical gravitational equations (\ref%
{einstceq1}) can be proven in N-adapted variational form. We can introduce
conventional gravitational and matter fields Lagrange densities, $\ ^{g}L(%
\widehat{\mathbf{R}}ic)$ (as in GR with $\ ^{g}L(R)$) and postulating a $\
^{m}L(\varphi ^{A}, \mathbf{g}_{\beta \gamma }).$ The stress-energy d-tensor
of matter fields $\varphi ^{A}$ (labelled by a general index $A$) is defined
and computed as in GR but with respective dyadic decompositions, 
\begin{equation}
\mathbf{T}_{\alpha \beta }=-\frac{2}{\sqrt{|\mathbf{g}_{\mu \nu }|}}\frac{%
\delta (\ ^{m}L\sqrt{|\mathbf{g}_{\mu \nu }|})}{\delta \mathbf{g}^{\alpha
\beta }}.  \label{emdt}
\end{equation}%
We define the trace $T:=\mathbf{g}^{\alpha \beta }\mathbf{T}_{\alpha \beta }$
and construct certain effective sources of Ricci d-tensors, $\widehat{%
\mathbf{Y}}[\mathbf{g,}\widehat{\mathbf{D}}]\simeq \{\mathbf{T}_{\alpha
\beta }-\frac{1}{2}\mathbf{g}_{\alpha \beta }T\}.$ In various physical
theories, one considers more general $\ ^{m}L,$ for instance, depending on
some covariant/spinor derivatives. For simplicity, we do not consider such
generalizations in this work such cases. 

For our purposes, we consider (effective) sources $\widehat{\mathbf{Y}}[%
\mathbf{g,}\widehat{\mathbf{D}}]= \{\Upsilon _{~\delta }^{\beta }(x,y)\}$
parameterized with respect to N-adapted frames (\ref{nader}) and (\ref{nadif}%
) in such forms: 
\begin{equation}
\widehat{\Upsilon }_{~\delta }^{\beta }=diag[\Upsilon _{\alpha }:\Upsilon
_{~1}^{1}=\Upsilon _{~2}^{2}=~^{h}\Upsilon (x^{k});\Upsilon
_{~3}^{3}=\Upsilon _{~4}^{4}=~^{v}\Upsilon (x^{k},y^{a})].  \label{esourc}
\end{equation}%
Such formulas can be obtained for some general classes of energy-momentum
tensors by using respective frame/coordinate transforms if such conditions
are not satisfied for a $\Upsilon _{\beta \delta }.$ Such an assumption
means that we generate off-diagonal solutions for certain classes of
nonholonomic transforms and constraints when the effective sources are
determined by \textbf{two generating sources} $\ ^{h}\Upsilon (x^{k})$ and $%
\ ^{v}\Upsilon (x^{k},y^{a})$. It imposes certain nonholonomic constraints
on $\mathbf{T}_{\alpha \beta }$, possible cosmological constant $\Lambda $
and more special splitting of constants into h- and v-components. Such
constraints may involve distortion d-tensors $\widehat{\mathbf{Z}}[\mathbf{g}%
] $ and other values included in $\widehat{\mathbf{Y}}.$ Nevertheless, the
conditions (\ref{esourc}) allow us to decouple and integrate in general
explicit forms the geometric flows and gravitational and matter field
equations in many cases. For instance, this is possible if we consider that $%
\widehat{\mathbf{Y}}[\mathbf{g,}\widehat{\mathbf{D}},\kappa ]$ contains a
small parameter $\kappa $, or if the gravitational and matter field dynamics
is subjected to certain convenient classes of constraints, trapping
hypersurface conditions, ellipsoid symmetries etc. as we reviewed in \cite%
{vacaruplb16,vbubuianu17,partner02}. In such cases, the solutions can be
constructed exactly or recurrently using power decompositions $\kappa
^{0},\kappa ^{1},\kappa ^{2},...$ For various purposes, $\kappa $ can be
related to a gauge field source or an other parameter for constructing
ellipsoid deformations etc. We say that the corresponding classes of
solutions are exact or parametric, for instance, for linear dependencies on $%
\kappa ^{0}$ and $\kappa ^{1}.$ 

For N-adapted frames (\ref{nader}) and (\ref{nadif}), the Einstein equations
(\ref{einstceq1}) can be written in a form which is more convenient for
decoupling, 
\begin{eqnarray}
\widehat{\mathbf{R}}_{\ \ \beta }^{\alpha } &=&\widehat{\mathbf{\Upsilon }}%
_{\ \ \beta }^{\alpha },  \label{cdeq1} \\
\widehat{\mathbf{T}}_{\ \alpha \beta }^{\gamma } &=&0,%
\mbox{ if we extract
LC configuations with }\nabla .  \label{lccond1}
\end{eqnarray}%
In this system of nonlinear PDEs, the generating sources $\widehat{\mathbf{%
\Upsilon }}_{\ \ \beta }^{\alpha }=[\ ^{h}\Upsilon \delta _{\ \ j}^{i},\
^{v}\Upsilon \delta _{\ \ b}^{a}]$ (\ref{esourc}) and the equations (\ref%
{lccond1}) are equivalent to (\ref{lccond}); the induced nonholonomic
d-torsion $\widehat{\mathcal{T}}=\{\widehat{\mathbf{T}}_{\ \alpha
\beta}^{\gamma }[\mathbf{g,N,}\widehat{\mathbf{D}}]\}$ is defined as in (\ref%
{fundgeom}). 

Let us discuss the conservation laws for (\ref{cdeq1}). Here we note that,
in general, $\widehat{\mathbf{D}}^{\beta }\widehat{\mathbf{E}}_{\ \
\beta}^{\alpha }\neq 0$ and $\widehat{\mathbf{D}}^{\beta }\widehat{\mathbf{%
\Upsilon }}_{\ \ \beta }^{\alpha }\neq 0,$ which is different from the
Einstein and energy-momentum tensors written in standard form in GR.
Non-zero covariant divergences are typical for nonholonomic systems. For
instance, in nonholonomic mechanics, the conservation laws are not standard
ones. In continuous mechanics, we can introduce the so-called Lagrange
multiples associated with certain classes of nonholonomic constraints.
Solving the constraint equations, we re-define the variables, then introduce
new effective Lagrangians and, finally, we can define standard conservation
laws. This can be performed in explicit general forms only for some "toy"
models with effective and real matter fields in GR. In nonholonomic
canonical variables, using distortions of connections (\ref{canondistrel}),
we can rewrite (\ref{cdeq1}) in terms of $\nabla ,$ when $\nabla ^{\beta
}E_{\ \ \beta }^{\alpha }=\nabla ^{\beta }T_{\ \ \beta }^{\alpha }=0.$ We
conclude that there are no conceptual problems with the definition of
conservation laws for matter fields using two different linear connections (%
\ref{twocon}) which are defined by the same metric structure $\mathbf{g.}$ 

\subsection{Off-diagonal decoupling and integration of Einstein equations}

\label{ssdecintei} In this subsection, we show how the canonical distorted
Einstein equations (\ref{cdeq1}) can be decoupled and integrated in general
off-diagonal forms. We provide N-adapted coefficient formulas, analyze
respective nonlinear and dual symmetries and classify certain important
variants of parametrization of such solutions. Some general conditions of
extracting LC configurations and respective off-diagonal metrics in GR are
stated. In Appendix \ref{appendixa}, we provide technical computations of
the coefficients of Ricci d-tensors for quasi-stationary configurations. 

\subsubsection{Motivations for constructing off-diagonal solutions}

The Schwarzschild BH metric consists of the most important example of
solutions constructed by using a diagonal ansatz which transforms the
Einstein equations into a system of nonlinear ODEs. It is an example of
vacuum Einstein spaces with zero cosmological constant defined by the
conditions $Ric[\nabla ]=0$. We can use the formulas for the Ricci d-tensor (%
\ref{criccidt}) but for the LC connection in a coordinate base. In spherical
coordinates $u^{\alpha }=(r,\theta ,\varphi ,t)$, such coefficients for the
diagonal static solution can be written as%
\begin{equation}
g_{\alpha \beta }=diag[g_{1}(r)=(1-\frac{r_{s}}{r})^{-1},g_{2}(r,\theta
)=r^{2},g_{3}(u^{\gamma })=r^{2}\sin ^{2}\theta ,g_{4}(u^{\gamma
})=-f(r)=-(1-\frac{r_{s}}{r})].  \label{sch}
\end{equation}%
The corresponding quadratic line element (with zero N-coefficients for $%
\mathbf{g}=\{\underline{g}_{\alpha \beta }\}$ (\ref{ansatz}) ) is 
\begin{equation*}
ds_{Sch}^{2}=g_{\alpha }(u^{\gamma })[du^{\alpha
}]^{2}=g_{1}dr^{2}+g_{2}d\theta ^{2}+g_{3}d\varphi ^{2}+g_{4}dt^{2}.
\end{equation*}%
The Schwarzschild (horizon) radius $r_{s}=2Gm/c^{2}$ in (\ref{sch}) is
determined by the condition that for $r\gg r_{s}$ such a metric defined the
Newton gravitational potentials for a point mass $m.$ This solution
possesses a Killing symmetry on time like vector $\partial _{4}=\partial _{t}
$ because in the chosen coordinate base the coefficients of the
Schwarzschild metric do not depend on the time coordinate $t.$ 

The Friedman-Lema\^{\i}tre-Robertson-Walker, FLRW, diagonal metric is
defined by coefficients%
\begin{equation}
g_{\alpha \beta }=diag[g_{1}(r,t)=a^{2}(t)/(1-\epsilon
r^{2}),g_{2}(r,t)=a^{2}(t)r^{2},g_{3}(r,\theta ,t)=a^{2}(t)r^{2}\sin
^{2}\theta ,g_{4}=-1.  \label{flrw}
\end{equation}%
This is the second important example of physically important solutions in
GR. It is used in constructing homogeneous and isotropic cosmological
models. In (\ref{flrw}), the constant $\epsilon $ represents the curvature
of the space (it can be taken $0,\pm 1$) and the "scale factor" $a(t)$ must
be chosen to define a solution of the Einstein equations (\ref{einstceq1})
reformulated for $\nabla .$ For such equations, the energy-momentum tensor
can be taken in a diagonal form $T_{\alpha \beta }=diag[P,P,P,\rho ]$ which
corresponds to a fluid type matter with pressure $P$ and energy density $%
\rho .$ 

The diagonal ansatz (\ref{sch}) or (\ref{flrw}) were correspondingly
generalized in GR and MGTs with various (extra dimension, or 4-d, spherical,
cylindrical, toroid and with other symmetries) keeping the property to be
diagonalizable by certain coordinate transforms. That allowed to
construction many BH, WH and cosmological solutions in (super) string and
nonassociative/ noncommutative, supersymmetric MGTs and exploited, for
instance, in modern cosmology and astrophysics \cite%
{kramer03,vacaru18,vacaruplb16,vbubuianu17,partner06,partner02}. 

The main property of ansatz of types (\ref{sch}) and/or (\ref{flrw}) is that
they reduce the gravitational field equations in GR and MGTs to some systems
of nonlinear ODEs. In certain systems of coordinates, such equations can be
integrated in certain general or approximate forms determined by integration
constants. The physical interpretation of diagonalizable solutions depends
on the types of assumptions on symmetries, boundary or asymptotic conditions
and (or) how a corresponding Cauchy problem is solved. Such conditions are
stated following certain geometric and important physical considerations.
The cosmological constant and the data for an energy-momentum tensor are
considered effective or matter-generating sources. These types of diagonal
quasi-stationary or cosmological solutions may involve certain singularities
and horizons. During the last 100 years, such solutions were found,
generalized, and studied intensively in GR and MGTs. A series of geometric
and physically important theorems (on BH singularities, cosmic censorships,
conditions of stability, scenarios of geometric evolution, inflation,
acceleration etc.) were proven. The main ideas and principles for
constructing diagonal solutions of (modified) Einstein equations can be
stated in this form:%
\begin{eqnarray*}
&&%
\mbox{\bf Principles 1 (reducing PDEs to ODEs and constructing diagonal
solutions):} \\
&&\left[ 
\begin{array}{c}
\mbox{system of  nonlinear PDEs}, \\ 
\mbox{(modified) Einstein eqs. }%
\end{array}%
\right] \Rightarrow \\
&&\left[ 
\begin{array}{c}
\mbox{frame/coordinate transforms and symmetries:} \\ 
\mbox{ spherical/cylindrical, Killing, Lie algebras,} \\ 
\mbox{diagonal ansatz}\ g_{\alpha }(u^{\gamma })%
\end{array}%
\right] \Rightarrow \left[ 
\begin{array}{c}
\mbox{\bf integrable systems of} \\ 
\mbox{\bf nonlinear ODEs}%
\end{array}%
\right] \\
&&{\qquad }{\qquad }{\qquad }{\qquad }{\qquad }{\qquad }{\qquad }{\qquad }{%
\qquad }{\qquad }\mathbf{\Downarrow } \\
&&\left[ 
\begin{array}{c}
\mbox{special generating functions }f(r),\mbox{or }a(t), \\ 
\begin{array}{c}
\mbox{ non-structure vacuum, or special generating sources}:\ \Lambda
,diag[P,P,P,\rho ];%
\end{array}
\\ 
\begin{array}{c}
\mbox{ integration constants determined by boundary/asymptotic conditions,
Cauchy problems, }%
\end{array}
\\ 
\begin{array}{c}
\mbox{horizons, singularities, BH theorems, hyper-surface  thermodynamics,
cosmic censorship etc.}%
\end{array}%
\end{array}%
\right] .
\end{eqnarray*}

Such methods and most important results are summarized in \cite{kramer03}.
They allow us to elaborate on linear and nonlinear physical models for
gravitational and matter field interactions determined by solutions of some
classes of ODEs. In such cases, only a few diagonal components (maximum 4,
for a 4-d spacetime) of "diagonalizable metrics" are chosen from six
independent components of metrics. For instance, such a diagonal metric is
defined by a function $f(r)$, or $a(t)$, when the presence of other
coordinates is motivated by describing certain spherical/ cylindrical/
ellipsoidal/ toroid systems and considering respective frame and general
coordinate transforms on curved spacetimes. We emphasize that by prescribing
a diagonal ansatz and certain global or local symmetry, we "cut" many other
possibilities to find more general classes of solutions. Other also
physically important solutions may depend on all spacetime coordinates for
corresponding off-diagonal terms. For instance, we may have 6 independent
coefficients for a generic off-diagonal metric. In GR, any metric can be
diagonalized in a point, or along a geodesic, and represented as a standard
diagonal Minkowski metric, $\eta _{\alpha \beta }=diag[1,1,1,-1]$. To
diagonalize a general pseudo-Riemannian metric on a finite spacetime region
is not possible if we use only coordinate transform. We can elaborate on new
physical models with nonlinear interactions and geometric evolution if
certain advanced geometric, analytic and numeric methods are applied for
constructing solutions of (modified) Einstein equations determined by 6
degrees of freedom. 

It is very difficult to construct in explicit forms exact or parametric
off-diagonal solutions of systems of coupled nonlinear PDEs of type (\ref%
{einstceq1})\ or (\ref{cdeq1}). With such purposes, we elaborated the AFCDM
and various generalizations for GR and MGTs \cite%
{vacaru18,partner06,vbubuianu17,partner02,nonassocFinslrev25}. The
parametric off-diagonal solutions may be also exact for a fixed value and
order of a physical parameter. As "small" parameters we can consider the
Planck and string constants, or other ones describing vacuum and non-vacuum
polarizations. The AFCDM was formulated for generalized ansatz adapted to
some nonholonomic 2+2 decompositions. An auxiliary linear connection $%
\widehat{\mathbf{D}}$ (\ref{twocon}) is used; this allows a general
decoupling and constructing new classes of off-diagonal solutions. This is
the price we should pay if we want to solve systems of nonlinear PDEs by not
reducing them to more simple systems of nonlinear ODEs. 

In GR and MGTs, more general classes of solutions were found. Corresponding
geometric and physical properties which are different than those stated for
diagonalizable solutions. For instance, the Kerr solution for rotating BHs
contains odd-diagonal terms defining ellipsoidal ergo-spheres. However such
off-diagonal deformations are induced by rotation frames. In this work, we
elaborate on constructing physically important solutions which are generic
off-diagonal. Such solutions with additional degrees of freedom (for 4-d
gravity theories, being considered 6 independent components of metrics) seem
to be of crucial importance, for instance, in DE and DM physics. We consider
that off-diagonal solutions can be used to solve a series of fundamental
problems in nonlinear physics and elaborate quasi-classical models of
quantum gravity, QG. Generic off-diagonal interactions and nonholonomic
constraints are important in the non-perturbative and nonlinear regimes and
can provide new tests for QG and higher dimension theories. We can also
elaborate on realistic off-diagonal models for inhomogeneous/ anisotropic/
acceleration cosmology; describe DE and DM configurations with
quasi-periodic structure and pattern forming, filaments, vortices, solitons
etc. For such purposes, we consider that to work only with "very simplified"
diagonal ansatz reducing gravitational and matter field equations to certain
systems of nonlinear ODEs is not enough. We have to develop new advanced
geometric methods which allow us to construct generic off-diagonal
solutions, with constraints and generating functions and sources, solving in
direct form respective systems of nonlinear PDEs. 

Having decoupled in a general form the system of nonlinear and nonholonomic
PDEs (\ref{cdeq1}) and (\ref{lccond1}), we can solve it in certain exact or
parametric forms. This way, we can construct generic off-diagonal solutions $%
\mathbf{g}$ (\ref{dm}), or (\ref{cm}), determined by corresponding classes
of generating and integration functions (see next subsections), effective
generating sources $\widehat{\mathbf{Y}}$ (\ref{esourc}), and a
corresponding N-connection splitting. The basic ideas and principles are
stated as 
\begin{eqnarray*}
&&\mbox{\bf Principles 2 - AFCDM: off-diagonal solutions, generalized
connections \& LC-connections} \\
&&\left[ 
\begin{array}{c}
\mbox{system of  nonlinear PDEs}, \\ 
\mbox{distorted Einstein eqs. }(\ref{cdeq1})%
\end{array}%
\right] \Rightarrow \\
&&\left[ 
\begin{array}{c}
\mbox{frame/coordinate transforms, N-adapted }\widehat{\mathbf{D}}[\mathbf{g}%
]\ (\ref{twocon}) \\ 
\mbox{ nonlinear and Killing symmetries,  effective sources } \\ 
\mbox{off-diagonal ansatz}\ g_{\alpha \beta }(u^{\gamma })%
\end{array}%
\right] \Rightarrow \left[ 
\begin{array}{c}
\mbox{\bf decoupling and integrable } \\ 
\mbox{\bf systems of nonlinear PDEs}%
\end{array}%
\right] \\
&&{\qquad }{\qquad }{\qquad }{\qquad }{\qquad }{\qquad }{\qquad }{\qquad }{%
\qquad }{\qquad }\mathbf{\Downarrow } \\
&&\left[ 
\begin{array}{c}
\mbox{ generating and integration functions depending on spacetime
coordinates } \\ 
\begin{array}{c}
\mbox{ nontrivial vacuum, generating sources, effective cosmologial
constants};%
\end{array}
\\ 
\begin{array}{c}
\mbox{ integration functions determined by boundary/asymptotic conditions,
Cauchy problems, }%
\end{array}
\\ 
\begin{array}{c}
\mbox{horizons, singularities,  G. Perelman  thermodynamics, stability and
flow evolution, etc.}%
\end{array}%
\end{array}%
\right] \\
&&{\qquad }{\qquad }{\qquad }{\qquad }{\qquad }{\qquad }{\qquad }{\qquad }{%
\qquad }{\qquad }\mathbf{\Downarrow }\mbox{\ nonholonomic LC-conditions }(%
\ref{lccond1}) \\
&&{\qquad }{\qquad }{\qquad }{\qquad }{\qquad }{\qquad }{\qquad }\left[ 
\begin{array}{c}
\mbox{system of  nonlinear PDEs}, \\ 
\mbox{ standard Einstein eqs for }\nabla%
\end{array}%
\right] .
\end{eqnarray*}

So, one of the main purposes of this article is to show how Principles 2 -
AFCDM can be performed in explicit form for 4-d Lorentz manifolds with
nonholonomic 2+2 splitting and canonical distortion of the LC connection. We
outline a new methodology for constructing generic off-diagonal solutions
and analyzing their possible physical implications in GR. In the next
section, we provide and discuss a series of important physical solutions
related to BH and WH physics and modern cosmology. 

\subsubsection{Decoupling properties of canonical variables}

\label{mainoffdans}Let us consider a d-metric of type (\ref{dm}), defined by
an off-diagonal ansatz not depending on coordinate $y^{4}=t,$ 
\begin{eqnarray}
\mathbf{\hat{g}} &=&g_{i}(x^{k})dx^{i}\otimes dx^{i}+h_{3}(x^{k},y^{3})%
\mathbf{e}^{3}\otimes \mathbf{e}^{3}+h_{4}(x^{k},y^{3})\mathbf{e}^{4}\otimes 
\mathbf{e}^{4},  \notag \\
&&\mathbf{e}^{3}=dy^{3}+w_{i}(x^{k},y^{3})dx^{i},\ \mathbf{e}%
^{4}=dy^{4}+n_{i}(x^{k},y^{3})dx^{i}.  \label{dmq}
\end{eqnarray}%
Such metrics (\ref{ansatz}) with Killing symmetry on the time-like
coordinate $\partial _{4}=\partial _{t}$ are called quasi-stationary. We
denote the N-connection coefficients in a canonical form, $\widehat{N}%
_{i}^{3}=w_{i}(x^{k},y^{3})$ and $\widehat{N}_{i}^{4}=n_{i}(x^{k},y^{3}),$
and consider that the coefficients of d-metric $\widehat{\mathbf{g}}_{\alpha
\beta }=[\widehat{g}_{ij}(x^{\kappa }),\widehat{g}_{ab}(x^{\kappa },y^{3})]$
are functions of necessary smooth class. Such a parametrization can be
obtained using some frame or coordinate transforms even, in general, such a $%
\mathbf{\hat{g}}(u)$ may depend on all spacetime coordinates. We use a hat
label for $\mathbf{\hat{g}}$ to emphasize that (\ref{dmq}) is with a (in
this case, time-like) Killing symmetry. Similarly, we can define ansatz for
d-metrics with other orders of space coordinates $(x^{k},y^{3}),$ for
instance, $(x^{1},y^{3},x^{2})$, when the v-coordinate is $x^{2}$ instead of 
$y^{3}.$ Such space coordinates can be spherical, cylindric, toroid and
other types. 

For generating locally anisotropic cosmological solutions, we can use such
ansatz for d-metrics: 
\begin{eqnarray}
\underline{\mathbf{g}} &=&g_{i}(x^{k})dx^{i}\otimes dx^{i}+\underline{h}%
_{3}(x^{k},t)\underline{\mathbf{e}}^{3}\otimes \underline{\mathbf{e}}^{3}+%
\underline{h}_{4}(x^{k},t)\underline{\mathbf{e}}^{4}\otimes \underline{%
\mathbf{e}}^{4},  \notag \\
&&\underline{\mathbf{e}}^{3}=dy^{3}+\underline{n}_{i}(x^{k},t)dx^{i},\ 
\underline{\mathbf{e}}^{4}=dy^{4}+\underline{w}_{i}(x^{k},t)dx^{i}.
\label{dmc}
\end{eqnarray}%
It involves a Killing symmetry on the space-like coordinate $\partial _{3}$
and emphasizes a generic dependence on $y^{4}=t,$ which is important in
cosmology. Correspondingly, the N-connection coefficients are parameterized $%
\underline{N}_{i}^{3}=\underline{n}_{i}(x^{k},t)$ and $\underline{N}_{i}^{4}=%
\underline{w}_{i}(x^{k},t)$ and the coefficients of d-metrics are of type $%
\underline{\mathbf{g}}_{\alpha \beta }=[g_{ij}(x^{\kappa }),\underline{g}%
_{ab}(x^{\kappa },t)];$ all such coefficients can be functions of a
necessary smooth class. Equivalently, (\ref{dmc}) can be transformed into a
different type of off-diagonal ansatz (\ref{ansatz}), when the coefficients
of metrics do not depend on $y^{3},$ but depend in certain adapted forms on
spacetime coordinates $(x^{i},y^{4}=t).$

In this work, we shall sketch proofs of general decoupling and integration
properties using only the quasi-stationary ansatz (\ref{dmq}). To obtain
solutions for locally anisotropic cosmological d-metrics we can change the
N-adapted coefficients in formal symbolic forms: for instance, $%
h_{3}(x^{k},y^{3})\rightarrow \underline{h}_{4}(x^{k},t),$ $%
h_{4}(x^{k},y^{3})\rightarrow \underline{h}_{3}(x^{k},t)$ and $%
w_{i}(x^{k},y^{3})\rightarrow \underline{n}_{i}(x^{k},t),$ $%
n_{i}(x^{k},y^{3})\rightarrow \underline{w}_{i}(x^{k},t).$ Such "dual" space
and time symmetries can be prescribed only for generic off-diagonal
solutions with respective Killing symmetries on $\partial _{4},$ or $%
\partial _{3}.$ Even in such cases, we can study main geometric and physical
properties of generic off-diagonal metrics. We can consider additional
nonholonomic constraints and deformations generating new classes of
solutions of systems of nonlinear PDEs. The coefficients depend generically
on 3 from 4 spacetime coordinates, and do not reduce the problem to finding
solutions of some "very simplified" systems of nonlinear ODEs. 

We emphasize that similar decoupling properties of (modified) Einstein
equations can be proven for more general ansatz for d-metrics, for instance,
parameterized 
\begin{eqnarray*}
\mathbf{g} &=&g_{i}(x^{k})dx^{i}\otimes dx^{i}+\omega ^{2}(x^{k},y^{a})[%
\underline{h}_{3}(x^{k},y^{4})h_{3}(x^{k},y^{3})\mathbf{e}^{3}\otimes 
\mathbf{e}^{3}+h_{4}(x^{k},y^{3})\underline{h}_{4}(x^{k},y^{4})]\ \mathbf{e}%
^{4}\otimes \mathbf{e}^{4}, \\
&&\mathbf{e}^{3}=dy^{3}+[w_{i}(x^{k},y^{3})+\underline{n}%
_{i}(x^{k},y^{4})]dx^{i},\ \mathbf{e}^{4}=dy^{4}+[n_{i}(x^{k},y^{3})+%
\underline{w}_{i}(x^{k},y^{4})]dx^{i}.
\end{eqnarray*}%
Such an ansatz may not have explicit Killing symmetries but involves
vertical co-space conformal transforms with a factor $\omega (x^{k},y^{a}),$
see discussions and examples in \cite{vacaru18} and references therein. Such
a generalized ansatz (we can consider other types of frame or conformal
transforms) results in more cumbersome formulas and implies additional
technical difficulties for generating exact and parametric solutions. For
simplicity, we omit such generalizations in this work and consider
off-diagonal solutions with at least one Killing symmetry. 

Using formulas (\ref{hcdric}), $\ $(\ref{vhcdric3}), (\ref{vhcdric4}) and (%
\ref{vcdric}), we can write the canonical distorted Einstein equations (\ref%
{cdeq1}) for the ansatz (\ref{dmq}) in the form: 
\begin{eqnarray}
\widehat{R}_{1}^{1} &=&\widehat{R}_{2}^{2}=\frac{1}{2g_{1}g_{2}}[\frac{%
g_{1}^{\bullet }g_{2}^{\bullet }}{2g_{1}}+\frac{(g_{2}^{\bullet })^{2}}{%
2g_{2}}-g_{2}^{\bullet \bullet }+\frac{g_{1}^{\prime }g_{2}^{\prime }}{2g_{2}%
}+\frac{\left( g_{1}^{\prime }\right) ^{2}}{2g_{1}}-g_{1}^{\prime \prime
}]=-\ ^{h}\Upsilon ,  \notag \\
\widehat{R}_{3}^{3} &=&\widehat{R}_{4}^{4}=\frac{1}{2h_{3}h_{4}}[\frac{%
\left( h_{4}^{\ast }\right) ^{2}}{2h_{4}}+\frac{h_{3}^{\ast }h_{4}^{\ast }}{%
2h_{3}}-h_{4}^{\ast \ast }]=-\ ^{v}\Upsilon ,  \label{riccist2} \\
\widehat{R}_{3k} &=&\frac{\ w_{k}}{2h_{4}}[h_{4}^{\ast \ast }-\frac{\left(
h_{4}^{\ast }\right) ^{2}}{2h_{4}}-\frac{(h_{3}^{\ast })(h_{4}^{\ast })}{%
2h_{3}}]+\frac{h_{4}^{\ast }}{4h_{4}}(\frac{\partial _{k}h_{3}}{h_{3}}+\frac{%
\partial _{k}h_{4}}{h_{4}})-\frac{\partial _{k}(h_{3}^{\ast })}{2h_{3}}=0; 
\notag \\
\widehat{R}_{4k} &=&\frac{h_{4}}{2h_{3}}n_{k}^{\ast \ast }+\left( \frac{3}{2}%
h_{4}^{\ast }-\frac{h_{4}}{h_{3}}h_{3}^{\ast }\right) \frac{\ n_{k}^{\ast }}{%
2h_{3}}=0.  \notag
\end{eqnarray}%
These equations can be written in a more compact symbolic form if we express 
$g_{i}=e^{\psi (x^{k})}$ and introduce the coefficients 
\begin{equation}
\alpha _{i}=h_{4}^{\ast }\partial _{i}(\varpi ),\beta =h_{4}^{\ast }(\varpi
)^{\ast }\mbox{  and }\gamma =(\ln \frac{|h_{4}|^{3/2}}{|h_{3}|})^{\ast },
\label{coeff}
\end{equation}%
for $\varpi =\ln |h_{4}^{\ast }/\sqrt{|h_{3}h_{4}}|,$ where $\ \Psi =\exp
(\varpi )$ will be considered in next subsection as a \textbf{generating
function.} This way, we represent the nonlinear system (\ref{riccist2}) in
the form: 
\begin{eqnarray}
\psi ^{\bullet \bullet }+\psi ^{\prime \prime } &=&2\ ^{h}\Upsilon ,
\label{eq1} \\
(\varpi )^{\ast }h_{4}^{\ast } &=&2h_{3}h_{4}\ ^{v}\Upsilon ,  \label{e2a} \\
\beta w_{j}-\alpha _{j} &=&0,  \label{e2b} \\
\ n_{k}^{\ast \ast }+\gamma n_{k}^{\ast } &=&0.  \label{e2c}
\end{eqnarray}%
Any solution of this system of nonlinear PDEs is a solution of (\ref{cdeq1})
parameterized a respective quasi-stationary d-metric ansatz (\ref{dmq}) for
canonically parameterized effective sources (\ref{esourc}). The equations (%
\ref{eq1}) and (\ref{e2a}) involve respectively two \textbf{generating
sources} $\ ^{h}\Upsilon (x^{k})$ and $\ ^{v}\Upsilon (x^{k},y^{3}).$ 

Let us explain the general decoupling property of the above systems of
equations for quasi-stationary configurations: The equation (\ref{eq1}) is a
standard 2-d Poisson equation with source $2\ ^{h}\Upsilon .$ It can be a
2-d wave equation if we consider h-metrics with signature, for instance, $%
(+,-)$ but we shall not analyze such configurations in this work.
Prescribing any data $(h_{3},^{v}\Upsilon ),$ we can search a $h_{4}$ as a
solution of a second order on $\partial _{3}$ nonlinear PDE (\ref{e2a}). We
can consider an inverse problem with prescribed data $(h_{4},^{v}\Upsilon )$
when a $h_{3}$ is a solution of a first-order nonlinear PDE. In the next
subsection, we show how using a generating function $\Psi $, such equations
can be integrated in explicit form. Having defined in some general forms $%
h_{3}(x^{k},y^{3})$ and $h_{4}(x^{k},y^{3}),$ we can compute respective
coefficients $\alpha _{i}$ and $\beta $ for (\ref{e2b}). Such linear
equations for $w_{j}(x^{k},y^{3})$ can be solved in general form. This means
that such equations and respective unknown functions are decoupled from the
rest of the system of equations. At the forth step, we can solve (\ref{e2c})
and find $n_{k}(x^{k},y^{3}).$ We have to perform two general integrations
on $y^{3}$ for any $\gamma (x^{k},y^{3})$ determined by $h_{3}(x^{k},y^{3})$
and $h_{4}(x^{k},y^{3})$ as we described above. So, solving step-by-step
four equations (\ref{eq1}) - (\ref{e2c}), we can generate off-diagonal
solutions of (modified) Einstein equations written in canonical d-connection
variables. This can be done in explicit form by using the general decoupling
property for quasi-stationary off-diagonal metric ansatz (\ref{dmq}) and
respective generating sources. In a similar form, we can prove general
decoupling properties for locally anisotropic cosmological d-metrics (\ref%
{dmc}). In generic form, respective coefficients depend on $y^{4}=t$ \ and
respective symbols are underlined, for instance, in the form $(\underline{h}%
_{3}(x^{k},t),\ ^{v}\underline{\Upsilon }(x^{k},t))$ for $\underline{\Psi }%
(x^{k},t); \underline{\alpha}_{i}$ and $\underline{\beta };$ etc. 

Finally (in this subsection), we note that quasi-stationary d-metrics (\ref%
{dmq}) subjected to above conditions (\ref{eq1}) - (\ref{e2c}) can be
represented local coordinate form as some generic off-diagonal ansatz $%
\mathbf{\hat{g}}=$ $\underline{\widehat{g}}_{\alpha \beta
}(u)du^{\alpha}\otimes du^{\beta },$ (\ref{ansatz}), when 
\begin{eqnarray}
\widehat{\underline{g}}_{\alpha \beta } &=&\left[ 
\begin{array}{cccc}
g_{1}+(N_{1}^{3})^{2}h_{3}+(N_{1}^{4})^{2}h_{4} & 
N_{1}^{3}N_{2}^{3}h_{3}+N_{1}^{4}N_{2}^{4}h_{4} & N_{1}^{3}h_{3} & 
N_{1}^{4}h_{4} \\ 
N_{2}^{3}N_{1}^{3}h_{3}+N_{2}^{4}N_{1}^{4}h_{4} & 
g_{2}+(N_{2}^{3})^{2}h_{3}+(N_{2}^{4})^{2}h_{4} & N_{2}^{3}h_{3} & 
N_{2}^{4}h_{4} \\ 
N_{1}^{3}h_{3} & N_{2}^{3}h_{3} & h_{3} & 0 \\ 
N_{1}^{4}h_{4} & N_{2}^{4}h_{4} & 0 & h_{4}%
\end{array}%
\right]  \notag \\
&=&\left[ 
\begin{array}{cccc}
e^{\psi }+(w_{1})^{2}h_{3}+(n_{1})^{2}h_{4} & w_{1}w_{2}h_{3}+n_{1}n_{2}h_{4}
& w_{1}h_{3} & n_{1}h_{4} \\ 
w_{1}w_{2}h_{3}+n_{1}n_{2}h_{4} & e^{\psi }+(w_{2})^{2}h_{3}+(n_{2})^{2}h_{4}
& w_{2}h_{3} & n_{2}h_{4} \\ 
w_{1}h_{3} & w_{2}h_{3} & h_{3} & 0 \\ 
n_{1}h_{4} & n_{2}h_{4} & 0 & h_{4}%
\end{array}%
\right] .  \label{qeltorsoffd}
\end{eqnarray}%
Constructing exact or parametric solutions for such an ansatz is not
possible if we work directly with the LC connection. The AFCDM prescribes
using the canonical d-connection $\widehat{\mathbf{D}}$ for decoupling and
generating solutions. Then, certain LC configurations can be extracted by
imposing additional nonholonomic constraints (\ref{lccond1}). 

\subsubsection{General quasi-stationary solutions and their nonlinear
symmetries}

Taking the values of adapted N-coefficients constructed in Appendix \ref%
{appendixab}, we construct d-metrics (\ref{dmq}) as general quasi-stationary
solutions of the (modified) Einstein equations (\ref{cdeq1}) with generating
sources (\ref{esourc}). The decoupling properties are determined by $%
g_{i}=e^{\psi (x^{k})}$ as a solution of 2-d Poisson equations (\ref{eq1});
v-coefficients $h_{3}$ (\ref{g3}) and $h_{4}$ (\ref{g4}); and N-connection
coefficients $w_{i}$ (\ref{gw}) and $n_{k}$ (\ref{gn}). The corresponding
quadratic element can be written in the form%
\begin{eqnarray}
d\widehat{s}^{2} &=&e^{\psi (x^{k},\ ^{h}\Upsilon
)}[(dx^{1})^{2}+(dx^{2})^{2}]+\frac{[\Psi ^{\ast }]^{2}}{4(\ ^{v}\Upsilon
)^{2}\{g_{4}^{[0]}-\int dy^{3}[\Psi ^{2}]^{\ast }/4(\ ^{v}\Upsilon )\}}%
(dy^{3}+\frac{\partial _{i}\Psi }{\Psi ^{\ast }}dx^{i})^{2}+  \label{qeltors}
\\
&&\{g_{4}^{[0]}-\int dy^{3}\frac{[\Psi ^{2}]^{\ast }}{4(\ ^{v}\Upsilon )}%
\}\{dt+[\ _{1}n_{k}+\ _{2}n_{k}\int dy^{3}\frac{[(\Psi )^{2}]^{\ast }}{4(\
^{v}\Upsilon )^{2}|g_{4}^{[0]}-\int dy^{3}[\Psi ^{2}]^{\ast }/4(\
^{v}\Upsilon )|^{5/2}}]dx^{k}\}.  \notag
\end{eqnarray}%
We shall provide in the next section some examples of such physically
important quasi-stationary and (after "dualization") locally anisotropic
cosmological solutions. In coordinate bases, a d-metric (\ref{qeltors}) can
be represented by off-diagonal ansatz of type (\ref{qeltorsoffd}). 

\paragraph{Nonlinear symmetries of quasi-stationary configurations \newline
}

A very important and new property of the class of solutions (\ref{qeltors})
is that they are described by some nonlinear shell symmetries which allow
uss to transform generating functions and effective sources into other types
of generating functions with effective cosmological constants. By tedious
computations (see details in \cite%
{vacaru18,vbubuianu17,nonassocFinslrev25,sv11}), we can prove that such
solutions admit a changing of the generating data, $(\Psi ,\ \ ^{v}\Upsilon
)\leftrightarrow (\Phi ,\ ^{v}\Lambda =const\neq 0)$,  when v-cosmological
constant$\ ^{v}\Lambda $ may be different from a h-cosmological constant  $\
^{h}\Lambda $ (in GR, we can consider $\ ^{h}\Lambda =\ ^{v}\Lambda =$ $%
\Lambda $). For such nonlinear transforms, the quasi-stationary solutions $%
\mathbf{\hat{g}}[\Psi ]$ (\ref{qeltors}) of $\widehat{\mathbf{R}}_{\ \ \beta
}^{\alpha }[\Psi ]=\widehat{\mathbf{\Upsilon }}_{\ \ \beta }^{\alpha }$ (\ref%
{cdeq1}) can be expressed in an equivalent class of solutions of 
\begin{equation}
\widehat{\mathbf{R}}_{\ \ \beta }^{\alpha }[\Phi ]=\Lambda \mathbf{\delta }%
_{\ \ \beta }^{\alpha }.  \label{cdeq1a}
\end{equation}%
Such equivalent systems of nonlinear PDEs involve an effective cosmological
constant $\Lambda .$ The generating data $(\Phi ,\ \Lambda ),$ or $(\Psi ,\
\ ^{v}\Upsilon )$ can be chosen, for instance, to describe DE and DM
configurations in accelerating cosmology (we provide an example at the end
of next section). 

The condition that quasi-stationary configurations (\ref{qeltors}) are
transformed into certain quasi-stationary solutions of (\ref{cdeq1a}) result
into such differential or integral equations: 
\begin{eqnarray}
\frac{\lbrack \Psi ^{2}]^{\ast }}{\ ^{v}\Upsilon } &=&\frac{[\Phi
^{2}]^{\ast }}{\Lambda },\mbox{ which can be integrated as  }
\label{ntransf1} \\
\Phi ^{2} &=&\ \Lambda \int dy^{3}(\ ^{v}\Upsilon )^{-1}[\Psi ^{2}]^{\ast }%
\mbox{ and/or }\Psi ^{2}=\Lambda ^{-1}\int dy^{3}(\ ^{v}\Upsilon )[\Phi
^{2}]^{\ast }.  \label{ntransf2}
\end{eqnarray}%
Using (\ref{ntransf1}), we can simplify the formula (\ref{g4}) considered in
appendix and write $h_{4}=h_{4}^{[0]}-\frac{\ \Phi ^{2}}{4\ \Lambda }.$ This
allows us to express the formulas (\ref{g3}) and (\ref{gn}) in terms of new
generating data. For such transforms, we have to write $(\Psi )^{\ast }/\
^{v}\Upsilon $ in terms of such $(\Phi ,\Lambda ).$ Corresponding formulas
in (\ref{ntransf1}) and (\ref{ntransf2}) can be written in a new form:%
\begin{equation*}
\frac{\Psi (\ \Psi )^{\ast }}{\ ^{v}\Upsilon }=\frac{(\Phi ^{2})^{\ast }}{%
2\Lambda }\mbox{ and }\ \Psi =|\Lambda |^{-1/2}\sqrt{|\int dy^{3}\
^{v}\Upsilon \ (\Phi ^{2})^{\ast }|}.
\end{equation*}%
If we introduce $\Psi $ from the above second equation in the first
equation, we re-define $\Psi ^{\ast }$ in terms of generating data $(\
^{v}\Upsilon ,\Phi ,\Lambda ),$ when 
\begin{equation*}
\frac{\Psi ^{\ast }}{\ ^{v}\Upsilon }=\frac{[\Phi ^{2}]^{\ast }}{2\sqrt{|\
\Lambda \int dy^{3}(\ ^{v}\Upsilon )[\Phi ^{2}]^{\ast }|}}.
\end{equation*}

We conclude that any quasi-stationary solution (\ref{qeltors}) possess
important nonlinear symmetries of type (\ref{ntransf1}) and (\ref{ntransf2})
which are trivial or do not exist for diagonal ansatz. 

\paragraph{Quasi-stationary solutions with effective cosmological constant 
\newline
}

The quadratic element for quasi-stationary solutions (\ref{qeltors}) can be
written in an equivalent form using generating data $(\ ^{v}\Upsilon ,\Phi
,\Lambda )$ 
\begin{eqnarray}
d\widehat{s}^{2} &=&\widehat{g}_{\alpha \beta }(x^{k},y^{3},\Phi ,\Lambda
)du^{\alpha }du^{\beta }=e^{\psi (x^{k})}[(dx^{1})^{2}+(dx^{2})^{2}]
\label{offdiagcosmcsh} \\
&&-\frac{\Phi ^{2}[\Phi ^{\ast }]^{2}}{|\Lambda \int dy^{3}\ ^{v}\Upsilon
\lbrack \Phi ^{2}]^{\ast }|[h_{4}^{[0]}-\Phi ^{2}/4\Lambda ]}\{dy^{3}+\frac{%
\partial _{i}\ \int dy^{3}\ ^{v}\Upsilon \ [\Phi ^{2}]^{\ast }}{\
^{v}\Upsilon \ [(\ \Phi )^{2}]^{\ast }}dx^{i}\}^{2}-  \notag \\
&&\{h_{4}^{[0]}-\frac{\Phi ^{2}}{4\Lambda }\}\{dt+[\ _{1}n_{k}+\
_{2}n_{k}\int dy^{3}\frac{\Phi ^{2}[\Phi ^{\ast }]^{2}}{|\Lambda \int
dy^{3}\ ^{v}\Upsilon \lbrack \Phi ^{2}]^{\ast }|[h_{4}^{[0]}-\Phi
^{2}/4\Lambda ]^{5/2}}]\}.  \notag
\end{eqnarray}%
In this formula, indices: $i,j,k,...=1,2;a,b,c,...=3,4;$ the generating
functions are parameterized $\psi (x^{k})$ and $\Phi (x^{k_{1}}y^{3});$ the
generating sources are $^{h}\Upsilon (x^{k})$ and $\ ^{v}\Upsilon
(x^{k},y^{3});$ and the effective cosmological constant is $\ \Lambda .$
Integration functions$\ _{1}n_{k}(x^{j}),\ _{2}n_{k}(x^{j})$ and $%
g_{4}^{[0]}(x^{k}),$ are also considered. We emphasize that the
quasi-stationary solutions represented in the form (\ref{offdiagcosmcsh})
"disperse" into respective off-diagonal forms the prescribed generating data 
$(\Psi ,\ ^{v}\Upsilon )$ into another type ones $(\Phi ,\ \Lambda )$, with
effective cosmological constant. The contributions of a generating source $\
^{v}\Upsilon $ are not completely transformed into a cosmological constant $%
\Lambda .$ The coefficients of d-metrics $\mathbf{\hat{g}}[\Phi ,\
^{v}\Upsilon ,\Lambda ]$ (\ref{offdiagcosmcsh}) keep certain memory about $\
^{v}\Upsilon \,\ $stated in $\mathbf{\hat{g}}[\Psi ,\ \ ^{v}\Upsilon ]$ (\ref%
{qeltors}). Nevertheless, the possibility of introducing an effective $%
\Lambda $ simplifies the method of computing G. Perelman thermodynamic
variables as we show at the end of section \ref{sec04}. 

\paragraph{A d-metric coefficient can be used as a generating function 
\newline
}

Taking the partial derivative on $y^{3}$ of formula (\ref{g4}) allows us to
write $h_{4}^{\ast }=-[\Psi ^{2}]^{\ast }/4\ ^{v}\Upsilon .$ For some
prescribed data for $h_{4}(x^i,y^3)$ and $\ ^{v}\Upsilon (x^i,y^3)$, we can
compute (up to an integration function) a generating function $\ \Psi $
which satisfies $[\Psi ^{2}]^{\ast}=\int dy^{3}\ ^{v}\Upsilon h_{4}^{\ast }$
and work with off-diagonal solutions of type (\ref{qeltors}). But in
equivalent form, we can consider generating data $(h_{4},\ ^{v}\Upsilon )$
and re-write the quadratic elements for a quasi-stationary d-metric (\ref%
{qeltors}) as 
\begin{eqnarray}
d\widehat{s}^{2} &=&\widehat{g}_{\alpha \beta }(x^{k},y^{3};h_{4},\
^{v}\Upsilon )du^{\alpha }du^{\beta }  \label{offdsolgenfgcosmc} \\
&=&e^{\psi (x^{k})}[(dx^{1})^{2}+(dx^{2})^{2}]-\frac{(h_{4}^{\ast })^{2}}{%
|\int dy^{3}[\ \ ^{v}\Upsilon h_{4}]^{\ast }|\ h_{4}}\{dy^{3}+\frac{\partial
_{i}[\int dy^{3}(\ ^{v}\Upsilon )\ h_{4}^{\ast }]}{\ ^{v}\Upsilon \
h_{4}^{\ast }}dx^{i}\}^{2}  \notag \\
&&+h_{4}\{dt+[\ _{1}n_{k}+\ _{2}n_{k}\int dy^{3}\frac{(h_{4}^{\ast })^{2}}{%
|\int dy^{3}[\ ^{v}\Upsilon h_{4}]^{\ast }|\ (h_{4})^{5/2}}]dx^{k}\}.  \notag
\end{eqnarray}

The nonlinear symmetries (\ref{ntransf1}) and (\ref{ntransf2}) allow to
perform similar computations related to (\ref{offdiagcosmcsh}). Expressing $%
\Phi ^{2}=-4\ \Lambda h_{4},$ we can eliminate $\Phi $ from the nonlinear
element and generate a solution of type (\ref{offdsolgenfgcosmc}) which are
determined by the generating data $(h_{4};\Lambda ,\ ^{v}\Upsilon ).$ 

\paragraph{Quasi-stationary gravitational polarizations of prime metrics 
\newline
}

The above-generated off-diagonal solutions and their nonlinear symmetries
can be parameterized in certain forms describing nonholonomic distortions of
certain pseudo-Riemannian prime metrics (which may be or not be solutions of
some gravitational field equations). The main condition is that the target
d-metrics define quasi-stationary configurations as solutions of (\ref{cdeq1}%
) or (\ref{cdeq1a}).

We denote a \textbf{prime} d-metric as 
\begin{equation}
\mathbf{\mathring{g}=}[\mathring{g}_{\alpha },\mathring{N}_{i}^{a}]
\label{offdiagpm}
\end{equation}%
and transform it into a \textbf{target} d-metric $\mathbf{g,}$ 
\begin{equation}
\mathbf{\mathring{g}}\rightarrow \mathbf{g}=[g_{\alpha }=\eta _{\alpha }%
\mathring{g}_{\alpha },N_{i}^{a}=\eta _{i}^{a}\ \mathring{N}_{i}^{a}].
\label{offdiagdefr}
\end{equation}%
is a quasi-stationary d-metric of type (\ref{dmq}). The functions $\eta
_{\alpha }(x^{k},y^{3})$ and $\eta _{i}^{a}(x^{k},y^{3})$ from (\ref%
{offdiagdefr}) are called gravitational polarization ($\eta $-polarization)
functions. To generate solutions of (\ref{cdeq1}) or (\ref{cdeq1a}) we can
consider that the nonlinear symmetries (\ref{ntransf1}) are parameterized in
the form 
\begin{eqnarray}
(\Psi ,\ ^{v}\Upsilon ) &\leftrightarrow &(\mathbf{g},\ ^{v}\Upsilon
)\leftrightarrow (\eta _{\alpha }\ \mathring{g}_{\alpha }\sim (\zeta
_{\alpha }(1+\kappa \chi _{\alpha })\mathring{g}_{\alpha },\ ^{v}\Upsilon
)\leftrightarrow  \label{nonlintrsmalp} \\
(\Phi ,\ \Lambda ) &\leftrightarrow &(\mathbf{g},\ \Lambda )\leftrightarrow
(\eta _{\alpha }\ \mathring{g}_{\alpha }\sim (\zeta _{\alpha }(1+\kappa \chi
_{\alpha })\mathring{g}_{\alpha },\ \Lambda ),  \notag
\end{eqnarray}%
where $\Lambda $ is an effective cosmological and $\kappa $ is a small
parameter $0\leq \kappa <1;$ $\zeta _{\alpha }(x^{k},y^{3})$ and $\chi
_{\alpha }(x^{k},y^{3})$ are respective polarization functions. Such
nonholonomic transforms have to result in a target metric $\mathbf{g}$
defined as a solution of type (\ref{qeltors}) or, equivalently, (\ref%
{offdiagcosmcsh}), if the $\eta $- and/or $\chi $-polarizations are
subjected to the conditions (\ref{ntransf2}) written in the form: 
\begin{eqnarray}
\partial _{3}[\Psi ^{2}] &=&-\int dy^{3}\ ^{v}\Upsilon \partial
_{3}h_{4}\simeq -\int dy^{3}\ ^{v}\Upsilon \partial _{3}(\eta _{4}\ 
\mathring{g}_{4})\simeq -\int dy^{3}\ ^{v}\Upsilon \partial _{3}[\zeta
_{4}(1+\kappa \ \chi _{4})\ \mathring{g}_{4}],  \notag \\
\Phi ^{2} &=&-4\ \Lambda h_{4}\simeq -4\ \Lambda \eta _{4}\mathring{g}%
_{4}\simeq -4\ \Lambda \ \zeta _{4}(1+\kappa \chi _{4})\ \mathring{g}_{4}.
\label{nonlinsymrex}
\end{eqnarray}

Off-diagonal $\eta $-transforms resulting in d-metrics (\ref{offdiagdefr})
can be parameterized to be generated for $\psi$- and $\eta $-polarizations, 
\begin{equation}
\psi \simeq \psi (\kappa ;x^{k}),\eta _{4}\ \simeq \eta _{4}(x^{k},y^{3}),
\label{etapolgen}
\end{equation}%
in a form equivalent to (\ref{offdsolgenfgcosmc}) if the quasi-stationary
quadratic element can be written in the form 
\begin{eqnarray}
d\widehat{s}^{2} &=&\widehat{g}_{\alpha \beta }(x^{k},y^{3};\mathring{g}%
_{\alpha };\psi ,\eta _{4};\ \Lambda ,\ ^{v}\Upsilon )du^{\alpha }du^{\beta
}=e^{\psi }[(dx^{1})^{2}+(dx^{2})^{2}]  \label{offdiagpolfr} \\
&&-\frac{[\partial _{3}(\eta _{4}\ \mathring{g}_{4})]^{2}}{|\int dy^{3}\ \
^{v}\Upsilon \partial _{3}(\eta _{4}\ \mathring{g}_{4})|\ \eta _{4}\mathring{%
g}_{4}}\{dy^{3}+\frac{\partial _{i}[\int dy^{3}\ ^{v}\Upsilon \partial
_{3}(\eta _{4}\mathring{g}_{4})]}{\ \ ^{v}\Upsilon \partial _{3}(\eta _{4}%
\mathring{g}_{4})}dx^{i}\}^{2}  \notag \\
&&+\eta _{4}\mathring{g}_{4}\{dt+[\ _{1}n_{k}+\ _{2}n_{k}\int dy^{3}\frac{%
[\partial _{3}(\eta _{4}\mathring{g}_{4})]^{2}}{|\int dy^{3}\ \ ^{v}\Upsilon
\partial _{3}(\eta _{4}\mathring{g}_{4})|\ (\eta _{4}\mathring{g}_{4})^{5/2}}%
]dx^{k}\}^{2}.  \notag
\end{eqnarray}%
We can relate a solution of type (\ref{offdiagcosmcsh}) to an another one in
the form (\ref{offdiagpolfr}) if $\Phi ^{2}=-4\ \Lambda h_{4}$ and the $\eta 
$-polarizations are determined by the generating data $(h_{4}=\eta _{4}%
\mathring{g}_{4};\Lambda ,\ ^{v}\Upsilon ).$

If a primary d-metric $\mathbf{\mathring{g}}$ (\ref{offdiagpm}) is chosen,
for instance, as a BH solution we can analyze possible embedding into a
quasi-stationary background determined by a target $\mathbf{g}$ (\ref%
{offdiagdefr}). In general, a derived solution (\ref{offdiagpolfr}) may not
define a BH solution. We need additional assumptions to distinguish certain
cases when a target d-metric has 1) an unclear physical meaning; or 2) it
describes a BH embedded in a nontrivial gravitational vacuum determined by $%
\eta $-polarizations, for instance, taken as certain solitonic waves; or 3)
we generate a BH solutions with small $\chi $-polarizations of physical
constants and horizons. An additional analysis reviewed in \cite%
{vacaru18,vbubuianu17} is necessary. In principle, non-stable solutions also
have physical importance in physics. In all cases of solutions of type (\ref%
{offdiagcosmcsh}) or (\ref{offdiagpolfr}), we can compute generalized G.
Perelman variables, see section \ref{sec04}. Only in some special cases when
(\ref{offdiagpolfr}) involve some horizons, we can consider the
Bekenstein-Hawking thermodynamic paradigm. In Appendix \ref{appendixac}, we
provide the formulas for off-diagonal deformations on a $\kappa $-parameter
of d-metrics. The nonholonomic structure of such
solutions can be organized in such a form that the target solutions preserve
the physical meaning of the prime ones with certain differences, for
instance, that BHs are transformed into BEs, some constants became
effectively polarized by a nontrivial gravitational vacuum structure, some
DE effects appear via polarization of the cosmological constant etc. 

\subsubsection{Space and time duality of quasi-stationary and cosmological
solutions}

We mentioned in subsection \ref{mainoffdans} the existence of a specific
space and time duality between ansatz (\ref{dmq}) and (\ref{dmc}). A
corresponding duality principle can be formulated for generic off-diagonal
solutions. It allows us to not repeat all computations presented for
quasi-stationary metrics with nontrivial partial derivatives $\partial _{3}$
for locally anisotropic cosmological solutions with nontrivial partial
derivatives $\partial_{4}=\partial _{t}$. All formulas for quasi-stationary
solutions from the previous subsection and Appendices \ref{appendixab} and %
\ref{appendixac} can be re-defined by constructing locally anisotropic
cosmological solutions.

In abstract symbolic form, the \textbf{principle of space and time duality }
of generic off-diagonal configurations with one Killing symmetry on a
space-like $\partial _{3}$ or time-like $\partial _{t}$ if formulated: 
\begin{eqnarray*}
y^{3} &\longleftrightarrow &y^{4}=t,h_{3}(x^{k},y^{3})\longleftrightarrow 
\underline{h}_{4}(x^{k},t),h_{4}(x^{k},y^{3})\longleftrightarrow \underline{h%
}_{3}(x^{k},t), \\
N_{i}^{3} &=&w_{i}(x^{k},y^{3})\longleftrightarrow N_{i}^{4}=\underline{n}%
_{i}(x^{k},t),N_{i}^{4}=n_{i}(x^{k},y^{3})\longleftrightarrow N_{i}^{3}=%
\underline{w}_{i}(x^{k},t).
\end{eqnarray*}%
For constructing explicit classes of solutions, the above duality conditions
have to be stated also for prime d-metrics and respective generating
functions, generating sources and gravitational polarization functions (and
in certain cases, for the integration functions). Such details on the
duality of generic off-diagonal solutions are given by 
\begin{equation*}
\Upsilon _{~3}^{3}=\Upsilon _{~4}^{4}=~^{v}\Upsilon
(x^{k},y^{3})\longleftrightarrow \underline{\Upsilon }_{~4}^{4}=\underline{%
\Upsilon }_{~3}^{3}=~^{v}\underline{\Upsilon }(x^{k},t),\mbox{ see }(\ref%
{esourc});
\end{equation*}%
\begin{equation}
\begin{array}{ccc}
\begin{array}{c}
(\Psi ,~^{v}\Upsilon )\leftrightarrow (\mathbf{g},\ ~^{v}\Upsilon
)\leftrightarrow \\ 
(\eta _{\alpha }\ \mathring{g}_{\alpha }\sim (\zeta _{\alpha }(1+\kappa \chi
_{\alpha })\mathring{g}_{\alpha },~^{v}\Upsilon )\leftrightarrow%
\end{array}
& \Longleftrightarrow & 
\begin{array}{c}
(\underline{\Psi },\ ~^{v}\underline{\Upsilon })\leftrightarrow (\underline{%
\mathbf{g}},\ ~^{v}\underline{\Upsilon })\leftrightarrow \\ 
(\underline{\eta }_{\alpha }\ \underline{\mathring{g}}_{\alpha }\sim (%
\underline{\zeta }_{\alpha }(1+\kappa \underline{\chi }_{\alpha })\underline{%
\mathring{g}}_{\alpha },\ ~^{v}\underline{\Upsilon })\leftrightarrow%
\end{array}
\\ 
\begin{array}{c}
(\Phi ,\ \Lambda )\leftrightarrow (\mathbf{g},\ \Lambda )\leftrightarrow \\ 
(\eta _{\alpha }\ \mathring{g}_{\alpha }\sim (\zeta _{\alpha }(1+\kappa \chi
_{\alpha })\mathring{g}_{\alpha },\ \Lambda ),%
\end{array}
& \Longleftrightarrow & 
\begin{array}{c}
(\underline{\Phi },\ \underline{\Lambda })\leftrightarrow (\underline{%
\mathbf{g}},\ \underline{\Lambda })\leftrightarrow \\ 
(\underline{\eta }_{\alpha }\ \underline{\mathring{g}}_{\alpha }\sim (%
\underline{\zeta }_{\alpha }(1+\kappa \underline{\chi }_{\alpha })\underline{%
\mathring{g}}_{\alpha },\ \underline{\Lambda }).%
\end{array}%
\end{array}
\label{dualnonltr}
\end{equation}%
The duality conditions are extended also to the corresponding systems of
nonlinear PDE with decoupling (see (\ref{eq1}) - (\ref{e2c}) and respective
coefficients): 
\begin{equation}
\begin{array}{ccc}
\Psi ^{\ast }h_{4}^{\ast }=2h_{3}h_{4}\ ~^{v}\Upsilon \Psi , & 
\longleftrightarrow & \sqrt{|\underline{h}_{3}\underline{h}_{4}|}\underline{%
\Psi }=\underline{h}_{3}^{\diamond }, \\ 
\sqrt{|h_{3}h_{4}|}\Psi =h_{4}^{\ast }, & \longleftrightarrow & \underline{%
\Psi }^{\diamond }\underline{h}_{3}^{\diamond }=2\underline{h}_{3}\underline{%
h}_{4}\ \ ~^{v}\underline{\Upsilon }\underline{\Psi }, \\ 
\Psi ^{\ast }w_{i}-\partial _{i}\Psi =\ 0, & \longleftrightarrow & 
\underline{n}_{i}^{\diamond \diamond }+\left( \ln \frac{|\underline{h}%
_{3}|^{3/2}}{|\underline{h}_{4}|}\right) ^{\diamond }\underline{n}%
_{i}^{\diamond }=0, \\ 
\ n_{i}^{\ast \ast }+\left( \ln \frac{|h_{4}|^{3/2}}{|h_{3}|}\right) ^{\ast
}n_{i}^{\ast }=0 & \longleftrightarrow & \underline{\Psi }^{\diamond }%
\underline{w}_{i}-\partial _{i}\underline{\Psi }=\ 0,%
\end{array}%
\mbox{ see }(\ref{auxa1})-(\ref{aux1ac}).  \label{dualcosm}
\end{equation}

The nonlinear symmetries (\ref{ntransf1}) and (\ref{ntransf2}) are written
in respective dual forms for locally anisotropic cosmological solutions: 
\begin{eqnarray*}
\frac{\lbrack \underline{\Psi }^{2}]^{\diamond }}{~^{v}\underline{\Upsilon }}
&=&\frac{[\underline{\Phi }^{2}]^{\diamond }}{\underline{\Lambda }},%
\mbox{ which can be
integrated as  } \\
\underline{\Phi }^{2} &=&\ \underline{\Lambda }\int dt(~^{v}\underline{%
\Upsilon })^{-1}[\underline{\Psi }^{2}]^{\diamond }\mbox{ and/or
}\underline{\Psi }^{2}=(\underline{\Lambda })^{-1}\int dt(~^{v}\underline{%
\Upsilon })[\underline{\Phi }^{2}]^{\diamond }.
\end{eqnarray*}%
These nonlinear symmetries allow us to re-define for different types of
cosmological models the quasi-stationary d-metrics (\ref{qeltors}), (\ref%
{offdiagcosmcsh}), (\ref{offdsolgenfgcosmc}), (\ref{offdiagpolfr}) and (\ref%
{offdncelepsilon}). The corresponding locally anisotropic cosmological
analogs also define exact or parametric solutions of the nonholonomic
gravitational equations (\ref{cdeq1}) or, respectively, (\ref{cdeq1a}). As
an example of applications of such an abstract symbolic calculus, we provide
the formula for the dualized d-metric (\ref{qeltors}): 
\begin{eqnarray}
d\underline{s}^{2} &=&e^{\psi (x^{k})}[(dx^{1})^{2}+(dx^{2})^{2}]
\label{qeltorsc} \\
&&+\{g_{3}^{[0]}-\int dt\frac{[\underline{\Psi }^{2}]^{\diamond }}{4~^{v}%
\underline{\Upsilon }}\}\{dy^{3}+[\ _{1}n_{k}+\ _{2}n_{k}\int dt\frac{[(%
\underline{\Psi })^{2}]^{\diamond }}{4(\ ~^{v}\underline{\Upsilon }%
)^{2}|g_{3}^{[0]}-\int dt[\underline{\Psi }^{2}]^{\diamond }/4\ ~^{v}%
\underline{\Upsilon }|^{5/2}}]dx^{k}\}  \notag \\
&&+\frac{[\underline{\Psi }^{\diamond }]^{2}}{4(\ ~^{v}\underline{\Upsilon }%
)^{2}\{g_{3}^{[0]}-\int dt[\underline{\Psi }^{2}]^{\diamond }/4\ ~^{v}%
\underline{\Upsilon }\}}(dt+\frac{\partial _{i}\underline{\Psi }}{\underline{%
\Psi }^{\diamond }}dx^{i})^{2}.  \notag
\end{eqnarray}%
To conclude the constructions related to the above-stated space and time
duality principle we state: locally anisotropic d-metrics can be derived in
abstract dual form from some respective quasi-stationary solutions by
changing corresponding indices 3 into 4, 4 into 3. We correspondingly
underline the cosmological generating functions, effective sources and
gravitational polarizations for dependencies on $(x^{i},t);$ the v-partial
derivatives are changed in the form: $\ast \rightarrow \diamond $, i.e. $%
\partial _{3}\rightarrow \partial _{4}$. 

\subsubsection{Constraints on generating functions and sources for
extracting LC configurations}

The generic off--diagonal solutions of (\ref{cdeq1}) or (\ref{cdeq1a}) from
the previous subsections are constructed for an auxiliary canonical
d--connection $\widehat{\mathbf{D}}.$ In general, such solutions are
characterized by nonholonomically induced d--torsion coefficients $\widehat{%
\mathbf{T}}_{\ \alpha \beta }^{\gamma },$ see formulas (\ref{fundgeomc}) and
(\ref{nontrtors}) completely defined by the N--connection and d--metric
structures. To generate exact and parametric solutions in GR we have to
solve additional anholonomic constraints of type (\ref{lccond}) or (\ref%
{lccond1}).

We can extract zero torsion LC configurations in explicit form if we impose
additionally zero conditions for equation (\ref{lcconstr}). Corresponding
computations for quasi-stationary configurations state that all d-torsion
coefficients vanish if the coefficients of the N--adapted frames and the $v$%
--components of d--metrics are subjected to the conditions: 
\begin{eqnarray}
\ w_{i}^{\ast }(x^{i},y^{3}) &=&\mathbf{e}_{i}\ln \sqrt{|\
h_{3}(x^{i},y^{3})|},\mathbf{e}_{i}\ln \sqrt{|\ h_{4}(x^{i},y^{3})|}%
=0,\partial _{i}w_{j}=\partial _{j}w_{i}\mbox{ and }n_{i}^{\ast }=0;  \notag
\\
n_{k}(x^{i}) &=&0\mbox{ and }\partial _{i}n_{j}(x^{k})=\partial
_{j}n_{i}(x^{k}).  \label{zerot1}
\end{eqnarray}%
The solutions for such $w$- and $n$-functions depend on the class of vacuum
or non--vacuum metrics which we are going to generate. To solve this problem
we can follow such steps:

If we prescribe a generating function $\Psi =\check{\Psi}(x^{i},y^{3})$ for
which $[\partial _{i}(\ _{2}\check{\Psi})]^{\ast }=\partial _{i}(\ _{2}%
\check{\Psi})^{\ast },$ we can solve the equations for $w_{j}$ from (\ref%
{zerot1}). This is possible in explicit form if $\ ^{v}\Upsilon =const,$ or
if the effective source is expressed as a functional $\ ^{v}\Upsilon
(x^{i},y^{3})=\ \ ^{v}\Upsilon \lbrack \ _{2}\check{\Psi}].$ Then, we can
solve the third conditions $\partial _{i}w_{j}=\partial _{j}w_{i}$ if we
chose a generating function $\ \check{A}=\check{A}(x^{k},y^{3})$ and define 
\begin{equation*}
w_{i}(x^{i},y^{3})=\check{w}_{i}(x^{i},y^{3})=\partial _{i}\ \check{\Psi}/(%
\check{\Psi})^{\ast }=\partial _{i}\check{A}(x^{i},y^{3}).
\end{equation*}%
The equations for $n$-functions in (\ref{zerot1}) are solved by any $%
n_{i}(x^{k})=\partial _{i}[\ ^{2}n(x^{k})].$ 

The above formulas allow us to write the quadratic element for
quasi-stationary solutions with zero canonical d-torsion in a form similar
to (\ref{qeltors}), 
\begin{eqnarray}
d\check{s}^{2} &=&\check{g}_{\alpha \beta }(x^{k},y^{3})du^{\alpha
}du^{\beta }  \label{qellc} \\
&=&e^{\psi }[(dx^{1})^{2}+(dx^{2})^{2}]+\frac{[\check{\Psi}^{\ast }]^{2}}{%
4(\ ^{v}\Upsilon \lbrack \check{\Psi}])^{2}\{h_{4}^{[0]}-\int dy^{3}[\check{%
\Psi}]^{\ast }/4\ ^{v}\Upsilon \lbrack \check{\Psi}]\}}\{dy^{3}+[\partial
_{i}(\check{A})]dx^{i}\}^{2}  \notag \\
&&+\{h_{4}^{[0]}-\int dy^{3}\frac{[\check{\Psi}^{2}]^{\ast }}{4(\
^{v}\Upsilon \lbrack \check{\Psi}])}\}\{dt+\partial _{i}[\
^{2}n]dx^{i}\}^{2}.  \notag
\end{eqnarray}%
Similar constraints on generation functions as in (\ref{zerot1}), with
re-defined nonlinear symmetries allow us to extract LC configurations for
all classes of quasi-stationary or locally anisotropic cosmologic
constructed in this section. This is always possible if for some generic
off-diagonal metrics with nontrivial canonical d-torsion we chose respective
(more special) conditions for generating data, for instance, of type $(%
\check{\Psi}(x^{i},y^{3}),\ ^{v}\Upsilon \lbrack \ \check{\Psi}],\check{A}%
(x^{k},y^{3})).$ Dualizing the coefficient formulas as in (\ref{dualcosm}),
we transform (\ref{qellc}) into locally anisotropic cosmological solutions
of the Einstein equations in GR. 

\section{Four examples of physically important off-diagonal solutions in GR}

\label{sec03} In this section, we show how the AFCDM \cite%
{vacaru18,partner06,vbubuianu17,partner02,nonassocFinslrev25} can be applied
for constructing four classes of physically important solutions of (\ref%
{cdeq1}). The first three ones are defined by quasi-stationary off-diagonal
metrics and may describe: 1] nonholonomic BH-like solutions with distortions
to BE configurations; 2] locally anisotropic wormhole, WH, solutions; 3]
some systems of black torus solutions. We also provide an example of locally
anisotropic cosmological solutions describing nonholonomic cosmological
solitonic and spheroid deformations involving 2-d vertices. Such solutions
were studied in a series of our work on MGTs and nonholonomic geometric
flows . We prove how choosing generating and integration functions we can
construct certain physically important examples of quasi-stationary generic
off-diagonal solutions. 

\subsection{Nonholonomic deformations of new Kerr de Sitter BHs to
spheroidal configurations}

In MGTs (for instance, involving effective contributions from (non)
associative/ commutative sources in string theory and geometric information
flows), nonholonomic off-diagonal deformations of the Kerr and Schwarzschild
- (a) de Sitter, K(a)dS, BH metrics were studied \cite%
{vi17,vacaru18,vbubuianu17,partner02,partner06,nonassocFinslrev25}. For
spherical rotating configurations of KdS in Einstein gravity, such metrics
can be described by various families of rotating diagonal metrics involving,
or not, certain warping effects of curvature, see details in \cite{ovalle21}%
. The goal of this subsection is to study a new class of solutions in GR
constructed as off-diagonal deformations of some primary KdS metrics. We
show how such rotating BHs can be deformed to parametric quasi-stationary
d-metrics of type (\ref{offdncelepsilon}). Spheroidal rotoid deformations
are computed in explicit forms. 

\subsubsection{Prime new KdS metrics and gravitational polarizations}

We consider a prime quadratic d-metric of type (\ref{offdiagpm}) for
spherical coordinates parameterized in the form $x^{1}=r,x^{2}=%
\varphi,y^{3}=\theta ,y^{4}=t.$ The quadratic line element is written in the
form 
\begin{equation}
d\breve{s}^{2}=\breve{g}_{\alpha }(r,\varphi ,\theta )(\mathbf{\breve{e}}%
^{\alpha })^{2},  \label{offdiagpm1}
\end{equation}%
when the corresponding nontrivial coefficients of the d-metric and
N-connection are 
\begin{eqnarray*}
\breve{g}_{1} &=&\frac{\breve{\rho}^{2}}{\triangle _{\Lambda }},\breve{g}%
_{2}=\frac{\sin ^{2}\theta }{\breve{\rho}^{2}}[\Sigma _{\Lambda }-\frac{%
(r^{2}+a^{2}-\triangle _{\Lambda })^{2}}{a^{2}\sin ^{2}\theta -\triangle
_{\Lambda }}],\breve{g}_{3}=\breve{\rho}^{2},\breve{g}_{4}=\frac{a^{2}\sin
^{2}\theta -\triangle _{\Lambda }}{\breve{\rho}^{2}},\mbox{ and } \\
\breve{N}_{2}^{4} &=&\breve{n}_{2}=-a\sin \theta \frac{r^{2}+a^{2}-\triangle
_{\Lambda }}{a^{2}\sin ^{2}\theta -\triangle _{\Lambda }}.
\end{eqnarray*}
A new KdS solution in GR (see \cite{ovalle21} and references in that work)
is generated if the functions and parameters are chosen in the form 
\begin{eqnarray*}
\Sigma _{\Lambda } &=&(r^{2}+a^{2})^{2}-\triangle _{\Lambda }a^{2}\sin
^{2}\theta ,\triangle _{\Lambda }=r^{2}-2Mr+a^{2}-\frac{\Lambda _{0}}{3}%
r^{4}, \\
\breve{\rho}^{2} &=&r^{2}+a^{2}\cos ^{2}\theta ,\mbox{ for constants  }%
a=J/M=const.
\end{eqnarray*}%
In these formulas, $J$ is the angular momentum, $M$ is the total mass of the
system, and the cosmological constant $\Lambda _{0}>0.$ In this paper, we
use a different system of notations which is stated for another signature of
the metrics. The solution (\ref{offdiagpm1}) is different from the standard
KdS metrics, defining $\Lambda $-vacuum solutions, because the scalar
curvature 
\begin{equation*}
R(r,\theta )=4\widetilde{\Lambda }(r,\theta )=4\Lambda _{0}\frac{r^{2}}{%
\breve{\rho}^{2}}\neq 4\Lambda _{0}.
\end{equation*}%
So, the above formulas define a new KdS solution which posseses a warped
effect when the curvature is warped everywhere excepting the equatorial
plane. This is a rotating configuration of a BH with an effective
polarization $(r,\theta) $ of a cosmological constant $\Lambda _{0}.$ It
shows a rotational effect on the vacuum energy in GR with a cosmological
constant. Such an effect disappears for $r\gg a.$

A d-metric (\ref{offdiagpm1}) can be treated as a rotating version of the
Schwarzschild de Sitter metric and represents a new solution describing the
exterior of a BH with cosmological constant. Certain bond conditions for $%
M(a,\Lambda _{0})$ have to be imposed for the existence of such a BH
solution. The respective upper, $M_{\max }:=M_{+}$ and lower, $M_{\min }:=M,$
bounds are computed 
\begin{equation}
18\Lambda _{0}M_{\pm }^{2}=1+12\Lambda _{0}a^{2}\pm (1-4\Lambda
_{0}a^{2})^{3/2}.  \label{bonds}
\end{equation}%
A(\ref{offdiagpm1})defines a LC-configuration for the standard Einstein
equations with fluid type energy momentum tensor 
\begin{equation}
\breve{T}_{\alpha \beta }(r,\theta )=diag[p_{r},p_{\varphi }=p_{\theta
},p_{\theta }=\rho -2\Lambda _{0}r^{2}/\mathring{\rho}^{2},\rho =-p_{r}=%
\widetilde{\Lambda }^{2}/\Lambda _{0}].  \label{efmt1}
\end{equation}%
There are two physically interpretations of such primary d-metrics: 1) to
consider that they are defined as solutions of some vacuum locally
anisotropic polarizations on $(r,\theta )$ of the cosmological constant, $%
\Lambda _{0}\rightarrow \widetilde{\Lambda }(r,\theta );$ or 2) to consider
that they consist a result of some locally anisotropic energy-momentum
tensors of type $\breve{T}_{\alpha \beta }(r,\theta )$, or more general
(effective) sources. 

In this subsection, we study more general off-diagonal deformations of the
standard Kerr solution when there are involved gravitational polarizations
both of effective cosmological constants. Such target d-metric are defined
by coefficients depending on all space coordinates $(r,\varphi ,\theta )$
not only on $(r,\theta )$ as we considered for above prime d-metrics. Our
new classes of quasi-stationary nonholonomic spacetimes possess nonlinear
symmetries of type (\ref{ntransf1}) and (\ref{ntransf2}), defined by
respective classes of nonholonomic deformations and constraints. We generate
target solutions of type (\ref{dmq}), when $\mathbf{\hat{g}}%
(r,\varphi,\theta )$ is defined equivalently by generating sources of type (%
\ref{esourc}),%
\begin{equation}
\mathbf{\breve{\Upsilon}}_{\ \ \beta }^{\alpha }(r,\varphi ,\theta )=[\
^{h}\Upsilon \delta _{\ \ j}^{i},\ ^{v}\Upsilon \delta _{\ \ b}^{a}]=[\ ^{h}%
\mathbf{\breve{\Upsilon}}(r,\varphi ),\ ^{v}\mathbf{\breve{\Upsilon}}%
(r,\varphi ,\theta )].  \label{sourcbh}
\end{equation}%
Using Table 2 from Appendix \ref{appendixb} for such $\mathbf{\breve{\Upsilon%
}}_{\ \ \beta }^{\alpha },$ we can construct off-diagonal solutions with $%
\eta $-polarization functions as in (\ref{offdiagpolfr}), 
\begin{eqnarray}
d\widehat{s}^{2} &=&\widehat{g}_{\alpha \beta }(x^{k},y^{3};\breve{g}%
_{\alpha };\psi ,\eta _{4};\ \widetilde{\Lambda },\ ^{v}\mathbf{\breve{%
\Upsilon}})du^{\alpha }du^{\beta }=e^{\psi (r,\varphi )}[(dx^{1}(r,\varphi
))^{2}+(dx^{2}(r,\varphi ))^{2}]  \label{nkernew} \\
&&-\frac{[\partial _{\theta }(\eta _{4}\ \breve{g}_{4})]^{2}}{|\int d\theta
\ \ ^{v}\mathbf{\breve{\Upsilon}}\partial _{\theta }(\eta _{4}\ \breve{g}%
_{4})|\ \eta _{4}\breve{g}_{4}}\{dy^{3}+\frac{\partial _{i}[\int d\theta \ \
^{v}\mathbf{\breve{\Upsilon}}\ \partial _{3}(\eta _{4}\breve{g}_{4})]}{\ \
^{v}\mathbf{\breve{\Upsilon}}\partial _{\theta }(\eta _{4}\breve{g}_{4})}%
dx^{i}\}^{2}  \notag \\
&&+\eta _{4}\breve{g}_{4}\{dt+[\ _{1}n_{k}(r,\varphi )+\ _{2}n_{k}(r,\varphi
)\int d\theta \frac{\lbrack \partial _{\theta }(\eta _{4}\breve{g}_{4})]^{2}%
}{|\int d\theta \ \ ^{v}\mathbf{\breve{\Upsilon}}\partial _{3}(\eta _{4}%
\breve{g}_{4})|\ (\eta _{4}\breve{g}_{4})^{5/2}}]dx^{k}\}^{2}.  \notag
\end{eqnarray}%
This family of off-diagonal solutions is determined by a generating function 
$\eta _{4}=\eta _{4}(r,\varphi ,\theta )$ and respective integration
functions like $\ _{1}n_{k}(r,\varphi )$ and $\ _{2}n_{k}(r,\varphi ).$ The
locally anisotropic vacuum effects in such a d-metric with anisotropic
vertical coordinate $\theta $ is very complex and it is difficult to state
well-defined and general conditions when BH configurations. We need
additional assumptions to generate BH solutions in a non-trivial
gravitational vacuum, for instance, of solitonic type. A corresponding
stability analysis is necessary for some explicit generating and integrating
data if we try to construct stable configurations. Here we note that
non-stable solutions may have also certain physical importance, for
instance, describing some evolution, or structure formation, for a period of
time or under certain temperature regimes.

Any quasi-stationary d-metric (\ref{nkernew}) can be characterized by
nonlinear symmetries of type (\ref{nonlinsymrex}), 
\begin{eqnarray}
\partial _{\theta }[\Psi ^{2}] &=&-\int d\theta \ ^{v}\mathbf{\breve{\Upsilon%
}}\partial _{\theta }h_{4}\simeq -\int d\theta \ ^{v}\mathbf{\breve{\Upsilon}%
}\partial _{\theta }(\eta _{4}\ \breve{g}_{4})\simeq -\int d\theta \ ^{v}%
\mathbf{\breve{\Upsilon}}\partial _{\theta }[\zeta _{4}(1+\kappa \ \chi
_{4})\ \breve{g}_{4}],  \label{nlims2} \\
\Psi &=&|\ \ \widetilde{\Lambda }|^{-1/2}\sqrt{|\int d\theta \ ^{v}\mathbf{%
\breve{\Upsilon}}\ (\Phi ^{2})^{\ast }|},\Phi ^{2}=-4\ \widetilde{\Lambda }%
h_{4}\simeq -4\ \ \widetilde{\Lambda }\eta _{4}\breve{g}_{4}\simeq -4\ 
\widetilde{\Lambda }\ \zeta _{4}(1+\kappa \chi _{4})\ \breve{g}_{4}.  \notag
\end{eqnarray}%
In next section, we shall compute respective G. Perelman's thermodynamic
variables.

We note that in a series of our former works on MGTs \cite%
{vi17,vacaru18,partner02,partner06,nonassocFinslrev25} K(a)dS and other type
BH solutions were nonholonomically deformed for $y^{3}=\varphi .$ In those
papers, effective sources are generated by certain extra dimension (super)
string contributions, nonassociative and/ or noncommutative terms,
generalized Finsler or other types of modified dispersion deformations. In
MGTs and GR, we can state explicit conditions when off-diagonal $\varphi $-,
or $\theta $-, deformations may result in black ellipsoid, BE,
configurations. Such solutions are also quasi-stationary and different from
some primary Kerr-Newmann-(a)dS or target configurations. We proved that
such off-diagonal configurations can be stable, or stabilized by imposing
corresponding nonholonomic constraints. Choosing gravitational $\eta $%
-polarizations with $y^{3}=\theta ,$ we generate different classes of
solutions constructed for other types of effective sources. They also are
characterized by nonlinear symmetries and polarized or fixed values of some
prescribed cosmological constants. 

\subsubsection{Off-diagonal solutions with small parametric deformations of
KdS d-metrics}

Considering small parametric decompositions with $\kappa $-linear terms as
in (\ref{epsilongenfdecomp}), we can provide a physical interpretation of
off-diagonal quasi-stationary solutions (\ref{nkernew}). We avoid singular
off-diagonal frame or coordinate deformations if we use a new system of
coordinates with nontrivial terms of a prime N-connection. Respectively, for 
$a=3,$ with some $\breve{N}_{i}^{3}=\breve{w}_{i}(r,\varphi ,\theta ),$
which can be zero in certain rotation frames; and, for $a=4,$ $\breve{N}%
_{i}^{4}=\breve{n}_{i}(r,\varphi ,\theta )$ which may be with a nontrivial $%
\breve{n}_{2}=-a\sin \theta (r^{2}+a^{2}-\triangle _{\Lambda })/(a^{2}\sin
^{2}\theta -\triangle _{\Lambda }).$ So, applying the AFCDM with a small
parameter, we construct a d-metric of type (\ref{offdncelepsilon})
determined by $\chi $-generating functions: 
\begin{equation*}
d\ \widehat{s}^{2}=\widehat{g}_{\alpha \beta }(r,\varphi ,\theta ;\psi
,g_{4};\ ^{v}\mathbf{\breve{\Upsilon}})du^{\alpha }du^{\beta }=e^{\psi
_{0}}(1+\kappa \ ^{\psi }\chi )[(dx^{1}(r,\varphi ))^{2}+(dx^{2}(r,\varphi
))^{2}]
\end{equation*}%
\begin{eqnarray*}
&&-\{\frac{4[\partial _{\theta }(|\zeta _{4}\breve{g}_{4}|^{1/2})]^{2}}{%
\breve{g}_{3}|\int d\theta \lbrack \ \ ^{v}\mathbf{\breve{\Upsilon}}\partial
_{3}(\zeta _{4}\breve{g}_{4})]|}-\kappa \lbrack \frac{\partial _{\theta
}(\chi _{4}|\zeta _{4}\breve{g}_{4}|^{1/2})}{4\partial _{\theta }(|\zeta _{4}%
\breve{g}_{4}|^{1/2})}-\frac{\int d\theta \{\ ^{v}\mathbf{\breve{\Upsilon}}%
\partial _{\theta }[(\zeta _{4}\breve{g}_{4})\chi _{4}]\}}{\int d\theta
\lbrack \ ^{v}\mathbf{\breve{\Upsilon}}\partial _{\theta }(\zeta _{4}\breve{g%
}_{4})]}]\}\breve{g}_{3} \\
&&\{d\theta +[\frac{\partial _{i}\ \int d\theta \ ^{v}\mathbf{\breve{\Upsilon%
}}\ \partial _{\theta }\zeta _{4}}{(\breve{N}_{i}^{3})\ ^{v}\mathbf{\breve{%
\Upsilon}}\partial _{\theta }\zeta _{4}}+\kappa (\frac{\partial _{i}[\int
d\theta \ ^{v}\mathbf{\breve{\Upsilon}}\ \partial _{\theta }(\zeta _{4}\chi
_{4})]}{\partial _{i}\ [\int d\theta \ ^{v}\mathbf{\breve{\Upsilon}}\partial
_{\theta }\zeta _{4}]}-\frac{\partial _{\theta }(\zeta _{4}\chi _{4})}{%
\partial _{\theta }\zeta _{4}})]\breve{N}_{i}^{3}dx^{i}\}^{2}
\end{eqnarray*}%
\begin{eqnarray}
&&+\zeta _{4}(1+\kappa \ \chi _{4})\ \breve{g}_{4}\{dt+[(\breve{N}%
_{k}^{4})^{-1}[\ _{1}n_{k}+16\ _{2}n_{k}[\int d\theta \frac{\left( \partial
_{\theta }[(\zeta _{4}\breve{g}_{4})^{-1/4}]\right) ^{2}}{|\int d\theta
\partial _{\theta }[\ ^{v}\mathbf{\breve{\Upsilon}}(\zeta _{4}\breve{g}%
_{4})]|}]  \label{offdnceleps1} \\
&&+\kappa \frac{16\ _{2}n_{k}\int d\theta \frac{\left( \partial _{\theta
}[(\zeta _{4}\breve{g}_{4})^{-1/4}]\right) ^{2}}{|\int d\theta \partial
_{\theta }[\ ^{v}\mathbf{\breve{\Upsilon}}(\zeta _{4}\breve{g}_{4})]|}(\frac{%
\partial _{\theta }[(\zeta _{4}\breve{g}_{4})^{-1/4}\chi _{4})]}{2\partial
_{\theta }[(\zeta _{4}\breve{g}_{4})^{-1/4}]}+\frac{\int d\theta \partial
_{\theta }[\ ^{v}\mathbf{\breve{\Upsilon}}(\zeta _{4}\chi _{4}\breve{g}_{4})]%
}{\int d\theta \partial _{\theta }[\ ^{v}\mathbf{\breve{\Upsilon}}(\zeta _{4}%
\breve{g}_{4})]})}{\ _{1}n_{k}+16\ _{2}n_{k}[\int d\theta \frac{\left(
\partial _{\theta }[(\zeta _{4}\breve{g}_{4})^{-1/4}]\right) ^{2}}{|\int
d\theta \partial _{\theta }[\ ^{v}\mathbf{\breve{\Upsilon}}(\zeta _{4}\breve{%
g}_{4})]|}]}]\breve{N}_{k}^{4}dx^{k}\}^{2}.  \notag
\end{eqnarray}%
The polarization functions $\zeta _{4}(r,\varphi ,\theta )$ and $\chi
_{4}(r,\varphi ,\theta )$ in (\ref{offdnceleps1}) can be prescribed to be of
a necessary smooth class. Such a d-metric describes small $\kappa $%
-parametric deformations of a new KdS d-metric when the coefficients are
additional anisotropic on the $\varphi $-coordinate.

We can generate additional ellipsoidal deformations on $\theta $ using (\ref%
{offdnceleps1}) if we chose 
\begin{equation}
\chi _{4}(r,\varphi ,\theta )=\underline{\chi }(r,\varphi )\sin (\omega
_{0}\theta +\theta _{0}).  \label{rotoid}
\end{equation}%
In this formula, $\underline{\chi }(r,\varphi )$ is a smooth function and $%
\omega _{0}$ and $\theta _{0}$ are some constants. For such generating
polarization functions and $\zeta _{4}(r,\varphi ,\theta )\neq 0,$ we obtain
that $(1+\kappa \ \chi _{4})\ \breve{g}_{4}\simeq a^{2}\sin ^{2}\theta
-\triangle _{\Lambda }+\kappa \ \chi _{4}=0.$ For small $a$ and $\frac{%
\Lambda _{0}}{3},$ we can approximate $r=2M/(1+\kappa \ \chi _{4}),$ which
is a parametric equation for a rotoid configuration. The parameter with $%
\kappa $ is the eccentricity parameter and generating function (\ref{rotoid}%
). 

We can prescribe polarization functions generating KdS BH embedded into a
nontrivial nonholonomic quasi-stationary background for GR. For small
ellipsoidal deformations of type (\ref{rotoid}), we model black ellipsoid,
BE, objects as generic off-diagonal solutions of the Einstein equations.
Such configurations can be stable, see details and references in \cite%
{vi17,vacaru18} if we impose necessary types of nonholonomic constraints.
Selecting generating and integration functions of type (\ref{zerot1}), we
extract LC-configurations with scalar curvature $R(r,\varphi ,\theta )\simeq
\Lambda (r,\varphi ,\theta ),$ see also the nonlinear symmetries (\ref%
{nlims2}). This modifies (with new terms proportional to $\kappa $) the
boundary conditions (\ref{bonds}) for a effective mass $M$ and cosmological
constant $\Lambda _{0};$ such values are with local anisotropic
polarizations of the vacuum gravitational background. The phenomenon of
warped curvature \cite{ovalle21} can be preserved for some subclasses of
nonholonomic deformations but the gravitational vacuum is with additional
polarizations and the effective matter tensor (\ref{efmt1}) transforms into
an effective source $Y_{\alpha \beta }(r,\theta ,\varphi ).$ 

\subsection{Off-diagonal deformed WHs}

Nonholonomic deformations of WH solutions \cite{morris88,bron20} to locally
anisotropic quasi-stationary configurations were studied in \cite{v13,v14},
with generalizations to MGTs \cite{vacaru18,partner06,nonassocFinslrev25}.
Let us revise those constructions and generate new classes of off-diagonal
quasi-stationary solutions of (modified) Einstein equations (\ref{cdeq1})
derived as gravitational polarizations of primary WH metrics.

\subsubsection{Nonholonomic quasi-stationary gravitational polarizations of
WHs in GR}

Let us begin with the main formulas defining the generic Morris-Thorne WH
solution \cite{morris88}: 
\begin{equation*}
d\mathring{s}^{2}=(1-\frac{b(r)}{r})^{-1}dr^{2}+r^{2}d\theta ^{2}+r^{2}\sin
^{2}\theta d\varphi ^{2}-e^{2\Phi (r)}dt^{2},
\end{equation*}%
In this quadratic line element, $e^{2\Phi (r)}$ is a red-shift function and $%
b(r)$ is the shape function defined in spherically polar coordinates $%
u^{\alpha }=(r,\theta ,\varphi ,t).$ The usual Ellis-Bronnikov, EB, WHs are
defined for $\Phi (r)=0$ and $b(r)=\ _{0}b^{2}/r$ which state a zero tidal
WH with $\ _{0}b$ the throat radius, see \cite{kar94,roy20,souza22} for
details and recent reviews. A generalized EB configuration is characterized
by considering even integers $2k$ (with $k=1,2,...$), where $r(l)=(l^{2k}+\
_{0}b^{2k})^{1/2k}$ is a proper radial distance (tortoise coordinate) and
the cylindrical angular coordinate is $\phi \in \lbrack 0,2\pi ).$ In such
coordinates, $-\infty <l<\infty .$ We can define a prime metric 
\begin{equation*}
d\mathring{s}^{2}=dl^{2}+r^{2}(l)d\theta ^{2}+r^{2}(l)\sin ^{2}\theta
d\varphi ^{2}-dt^{2},
\end{equation*}%
when $dl^{2}=(1-\frac{b(r)}{r})^{-1}dr^{2}$ and $b(r)=r-r^{3(1-k)}(r^{2k}-\
_{0}b^{2k})^{(2-1/k))}.$

To avoid off-diagonal deformations with singularities we can perform some
frame transforms to a parametrization with trivial N-connection coefficients 
$\check{N}_{i}^{a}=$ $\check{N}_{i}^{a}(u^{\alpha }(l,\theta ,\varphi ,t))$
and $\check{g}_{\beta }(u^{j}(l,\theta ,\varphi ),u^{3}(l,\theta ,\varphi
)). $ We can introduce new coordinates $u^{1}=x^{1}=l,u^{2}=\theta ,$ and $%
u^{3}=y^{3}=\varphi +\ ^{3}B(l,\theta ),u^{4}=y^{4}=t+\ ^{4}B(l,\theta ),$
when 
\begin{eqnarray*}
\mathbf{\check{e}}^{3} &=&d\varphi =du^{3}+\check{N}_{i}^{3}(l,\theta
)dx^{i}=du^{3}+\check{N}_{1}^{3}(l,\theta )dl+\check{N}_{2}^{3}(l,\theta
)d\theta , \\
\mathbf{\check{e}}^{4} &=&dt=du^{4}+\check{N}_{i}^{4}(l,\theta
)dx^{i}=du^{4}+\check{N}_{1}^{4}(l,\theta )dl+\check{N}_{2}^{4}(l,\theta
)d\theta ,
\end{eqnarray*}%
are defined for $\mathring{N}_{i}^{3}=-\partial \ ^{3}B/\partial x^{i}$ and $%
\mathring{N}_{i}^{4}=-\partial \ ^{4}B/\partial x^{i}.$ So, the quadratic
line elements for the above WH solutions can be parameterized as a prime
d-metric, 
\begin{equation}
d\mathring{s}^{2}=\check{g}_{\alpha }(l,\theta ,\varphi )[\mathbf{\check{e}}%
^{\alpha }(l,\theta ,\varphi )]^{2},  \label{pmwh}
\end{equation}%
where $\check{g}_{1}=1,\check{g}_{2}=r^{2}(l),\check{g}_{3}=r^{2}(l)\sin
^{2}\theta $ and $\check{g}_{4}=-1.$

We can perform off-diagonal quasi-stationary deformations of WHs (\ref{pmwh}%
) by introducing nontrivial sources $\ ^{v}\Upsilon (l,\theta )$ and $\
^{v}\Upsilon (l,\theta ,\varphi )=\ \ ^{wh}\Upsilon $ related via nonlinear
symmetries (\ref{nonlinsymrex}) to a nonzero (effective) cosmological
constant $\Lambda .$ Considering gravitations $\eta $-polarization
functions, we construct such target quasi-stationary metrics: 
\begin{eqnarray}
d\widehat{s}^{2} &=&\widehat{g}_{\alpha \beta }(l,\theta ,\varphi ;\psi
,\eta _{4};\ \Lambda ,\ \ ^{wh}\Upsilon ,\ \check{g}_{\alpha })du^{\alpha
}du^{\beta }  \notag \\
&=&e^{\psi l,\theta )}[(dx^{1}(l,\theta ))^{2}+(dx^{2}(l,\theta ))^{2}]
\label{whpolf} \\
&&-\frac{[\partial _{\varphi }(\eta _{4}\ \check{g}_{4})]^{2}}{|\int
d\varphi \ \ ^{wh}\Upsilon \partial _{\varphi }(\eta _{4}\ \breve{g}_{4})|\
\eta _{4}\ \check{g}_{4}}\{dy^{3}+\frac{\partial _{i}[\int d\varphi \ \
^{wh}\Upsilon \ \partial _{\varphi }(\eta _{4}\ \check{g}_{4})]}{\ \
^{wh}\Upsilon \partial _{\varphi }(\eta _{4}\ \check{g})}dx^{i}\}^{2}+\eta
_{4}\breve{g}_{4}  \notag \\
&&\{dt+[\ _{1}n_{k}(r,z)+\ _{2}n_{k}(r,z)\int d\varphi \frac{\lbrack
\partial _{\varphi }(\eta _{4}\ \breve{g}_{4})]^{2}}{|\int d\varphi \ \
^{wh}\Upsilon \partial _{\varphi }(\eta _{4}\ \breve{g}_{4})|\ (\eta _{4}\ 
\breve{g}_{4})^{5/2}}]dx^{k}\}.  \notag
\end{eqnarray}%
Such parameterizations can be considered in various 4-d MGTs and GR, when
the physical interpretation is different because of different effective and
matte sources. This quasi-stationary solutions (\ref{whpolf}) \ are
determined by a generating function $\eta _{4}=\eta _{4}(l,\theta ,\varphi )$
and integration functions $\ _{1}n_{k}(l,\theta )$ and $\
_{2}n_{k}(l,\theta).$ The function $\psi (l,\theta )$ is defined as a
solution of 2-d Poisson equation $\partial _{11}^{2}\psi +\partial
_{22}^{2}\psi =2\ \ _{1}\widehat{\Upsilon }(l,\theta ).$

Finally, we emphasize that the target d-metrics (\ref{whpolf}) do not
describe a WH-like d-object for general nonholonomic deforms and general
classes of generating and integrating data. General off-diagonal
deformations may "anihilate or dissipate" a WH object. 

\subsubsection{Small parametric quasi-stationary deformations of WH d-metrics%
}

We can define locally anisotropic WH configurations if we consider for small
parametric off-diagonal deformation of prime metrics of type (\ref{pmwh}).
In terms of $\chi $-polarization functions, the quadratic linear elements
are computed 
\begin{equation*}
d\ \widehat{s}^{2}=\widehat{g}_{\alpha \beta }(l,\theta ,\varphi ;\psi ,\eta
_{4};\Lambda ,\ ^{wh}\Upsilon ,\ \check{g}_{\alpha })du^{\alpha }du^{\beta
}=e^{\psi _{0}(l,\theta )}[1+\kappa \ ^{\psi (l,\theta )}\chi (l,\theta
)][(dx^{1}(l,\theta ))^{2}+(dx^{2}(l,\theta ))^{2}]
\end{equation*}%
\begin{eqnarray*}
&&-\{\frac{4[\partial _{\varphi }(|\zeta _{4}\ \breve{g}_{4}|^{1/2})]^{2}}{%
\breve{g}_{3}|\int d\varphi \{\ \ ^{wh}\Upsilon \partial _{\varphi }(\zeta
_{4}\ \breve{g}_{4})\}|}-\kappa \lbrack \frac{\partial _{\varphi }(\chi
_{4}|\zeta _{4}\breve{g}_{4}|^{1/2})}{4\partial _{\varphi }(|\zeta _{4}\ 
\breve{g}_{4}|^{1/2})}-\frac{\int d\varphi \{\ ^{wh}\Upsilon \partial
_{\varphi }[(\zeta _{4}\ \breve{g}_{4})\chi _{4}]\}}{\int d\varphi \{\
^{wh}\Upsilon \partial _{\varphi }(\zeta _{4}\ \breve{g}_{4})\}}]\}\ \breve{g%
}_{3} \\
&&\{d\varphi +[\frac{\partial _{i}\ \int d\varphi \ ^{wh}\Upsilon \ \partial
_{\varphi }\zeta _{4}}{(\check{N}_{i}^{3})\ \ ^{wh}\Upsilon \partial
_{\varphi }\zeta _{4}}+\kappa (\frac{\partial _{i}[\int d\varphi \ \
^{wh}\Upsilon \ \partial _{\varphi }(\zeta _{4}\chi _{4})]}{\partial _{i}\
[\int d\varphi \ \ ^{wh}\Upsilon \partial _{\varphi }\zeta _{4}]}-\frac{%
\partial _{\varphi }(\zeta _{4}\chi _{4})}{\partial _{\varphi }\zeta _{4}})]%
\check{N}_{i}^{3}dx^{i}\}^{2}
\end{eqnarray*}%
\begin{eqnarray}
&&+\zeta _{4}(1+\kappa \ \chi _{4})\ \breve{g}_{4}\{dt+[(\check{N}%
_{k}^{4})^{-1}[\ _{1}n_{k}+16\ _{2}n_{k}[\int d\varphi \frac{\left( \partial
_{\varphi }[(\zeta _{4}\ \breve{g}_{4})^{-1/4}]\right) ^{2}}{|\int d\varphi
\partial _{\varphi }[\ ^{wh}\Upsilon (\zeta _{4}\ \breve{g}_{4})]|}]  \notag
\\
&&+\kappa \frac{16\ _{2}n_{k}\int d\varphi \frac{\left( \partial _{\varphi
}[(\zeta _{4}\ \breve{g}_{4})^{-1/4}]\right) ^{2}}{|\int d\varphi \partial
_{\varphi }[\ ^{wh}\Upsilon (\zeta _{4}\ \breve{g}_{4})]|}(\frac{\partial
_{\varphi }[(\zeta _{4}\ \breve{g}_{4})^{-1/4}\chi _{4})]}{2\partial
_{\varphi }[(\zeta _{4}\ ^{cy}g)^{-1/4}]}+\frac{\int d\varphi \partial
_{\varphi }[\ ^{wh}\Upsilon (\zeta _{4}\chi _{4}\ \breve{g}_{4})]}{\int
d\varphi \partial _{\varphi }[\ ^{wh}\Upsilon (\zeta _{4}\ \breve{g}_{4})]})%
}{\ _{1}n_{k}+16\ _{2}n_{k}[\int d\varphi \frac{\left( \partial _{\varphi
}[(\zeta _{4}\ \breve{g}_{4})^{-1/4}]\right) ^{2}}{|\int d\varphi \partial
_{\varphi }[\ ^{wh}\Upsilon (\zeta _{4}\ \breve{g}_{4})]|}]}]\check{N}%
_{k}^{4}dx^{k}\}^{2}.  \label{whpolf1}
\end{eqnarray}

In details, the formula (\ref{whpolf1}) was derived in part I of \cite%
{nonassocFinslrev25} and generalized for nonassociative 8-d WHs in the part
II of that work. In this subsection, we analyze 4-d configurations which
define 4-d off-diagonal modifications in GR. We can can model elliptic
deformations of WHs in GR in a particular case of target d-metrics
determined by generating functions of type $\chi _{4}(l,\theta ,\varphi )=%
\underline{\chi }(l,\theta )\sin (\omega _{0}\varphi +\varphi _{0}).$ These
are cylindric-elliptic configurations with $\varphi $-anisotropy. 

\subsection{Nonholonomic toroid configurations and black torus, BT}

Various classes of black torus, BT, or black ring solutions were constructed
in GR and MGTs \cite{lemos01,peca98,emparan02,emparan08}, see \cite%
{astorino17} for a recent review of results. Nonholonomic and off-diagonal
deformations of toroidal BHs were studied in \cite{v01t,v01q}. In this
subsection, we study such an example when the AFCDM is applied for
generating quasi-stationary locally anisotropic solutions using prime BT
metrics considered in \cite{astorino17}. We consider only small parametric
deformations when the physical interpretation of new classes of solutions is
very similar to some holonomic/ diagonalizable metric ansatz.

We consider a quadratic line element 
\begin{eqnarray}
d\tilde{s}^{2} &=&f^{-1}(\tilde{r})d\tilde{r}^{2}+\tilde{r}^{2}(\tilde{k}%
_{1}^{2}dx^{2}+\tilde{k}_{2}^{2}dy^{2})-f(\tilde{r})d\tilde{t}^{2}
\label{prmtor1} \\
&=&\tilde{g}_{\alpha }(\tilde{x}^{1})(d\tilde{u}^{\alpha })^{2},\mbox{ for }%
f(\tilde{r})=-\epsilon ^{2}b^{2}-\tilde{\mu}/\tilde{r}-\Lambda \tilde{r}%
^{2}/3  \notag
\end{eqnarray}%
as in section 3.1 of \cite{astorino17}. The coordinates in this metric are
related via re-scaling parameter $\epsilon $ to standard toroid "normalized"
coordinates, when $r$ is a radial coordinate, with $\theta =2\pi k_{1}x$ and 
$\varphi =2\pi k_{2}y$ (when $x,y\in \lbrack 0,1]$) and re-scaling $%
k_{1}=\epsilon \tilde{k}_{1},k_{2}=\epsilon \tilde{k}_{2},\mu \rightarrow 
\frac{\mu }{(2\pi )^{3}}=\tilde{\mu}/\epsilon ^{3};r\rightarrow \frac{r}{%
2\pi }=\tilde{r}/\epsilon ,t\rightarrow 2\pi t=\epsilon \tilde{r}.$ In (\ref%
{prmtor1}), the parameter $b$ is a coupling constant as in the
energy-momentum tensor for the nonlinear SU(2) sigma model, 
\begin{equation}
T_{\mu \nu }=\frac{b^{2}\epsilon ^{2}}{8\pi G\tilde{r}^{2}}[f(\tilde{r}%
)\delta _{\mu }^{4}\delta _{\nu }^{4}-f^{-1}(\tilde{r})\delta _{\mu
}^{1}\delta _{\nu }^{1}].  \label{emtsm}
\end{equation}%
In this formula, $\mu $ is an integration constant which can be fixed as a
mass parameter. The value $\epsilon =0$ allows to recover in a formal way
certain toroid vacuum solution \cite{lemos01,peca98}. The toroid metric (\ref%
{prmtor1}) defines an exact static solution of the Einstein equations for
the LC connection and energy-momentum tensor (\ref{emtsm}). The above
formulas generate an AdS BH with a toroidal horizon in 4-d Einstein gravity
and nonlinear $\sigma $-model. 

For our purposes, we consider frame transforms to an off-diagonal
parametrization of (\ref{prmtor1}) to a form with trivial N-connection
coefficients $\tilde{N}_{i}^{a}=$ $\tilde{N}_{i}^{a}(u^{\alpha }(\tilde{r}%
,x,y,t))$ and $\tilde{g}_{\alpha \beta }(u^{j}(\tilde{r},x,y),u^{3}(\tilde{r}%
,x,y)).$ Such transforms are defined in any form which do not involve
singular frame transforms and off-diagonal deformations. For instance, we
can introduce new coordinates $u^{1}=x^{1}=\tilde{r},u^{2}=x,$ and $%
u^{3}=y^{3}=y+\ ^{3}B(\tilde{r},x),u^{4}=y^{4}=t+\ ^{4}B(\tilde{r},x),$ when
for $\tilde{N}_{i}^{3}=-\partial \ ^{3}B/\partial x^{i}$ and $\tilde{N}%
_{i}^{4}=-\partial \ ^{4}B/\partial x^{i}:$ 
\begin{eqnarray*}
\mathbf{\tilde{e}}^{3} &=&dy=du^{3}+\tilde{N}_{i}^{3}(\tilde{r}%
,x)dx^{i}=du^{3}+\tilde{N}_{1}^{3}(\tilde{r},x)dr+\tilde{N}_{2}^{3}(\tilde{r}%
,x)dz, \\
\mathbf{\tilde{e}}^{4} &=&dt=du^{4}+\tilde{N}_{i}^{4}(\tilde{r}%
,x)dx^{i}=du^{4}+\tilde{N}_{1}^{4}(\tilde{r},x)dr+\tilde{N}_{2}^{4}(\tilde{r}%
,x)dz.
\end{eqnarray*}
In such nonlinear coordinates, the diagonal toroid metric (\ref{prmtor1})
transforms into a off-diagonal toroid d-metric 
\begin{equation}
d\tilde{s}^{2}=\tilde{g}_{\alpha }(\tilde{r},x,y)[\mathbf{\tilde{e}}^{\alpha
}(\tilde{r},x,y)]^{2},  \label{prmtor2}
\end{equation}%
where $\tilde{g}_{1}=f^{-1}(x^{1}),\tilde{g}_{2}=(x^{1})^{2}\tilde{k}%
_{1}^{2},\tilde{g}_{3}=(x^{2})^{2}\tilde{k}_{2}^{2}$ and $\tilde{g}%
_{4}=f(x^{1}).$

We can generate new classes of locally anisotropic toroid solutions if we
construct small parametric quasi-stationary deformations of prime metrics of
type (\ref{prmtor2}) defined by an effective source $\ ^{tor}\mathbf{Y}[%
\mathbf{g,}\widehat{\mathbf{D}}]\simeq \{-\Lambda \mathbf{g}_{\alpha \beta }+%
\mathbf{T}_{\alpha \beta }\},$ for (\ref{emtsm}), see also formulas (\ref%
{emdt}). The left label "tor" is used for toroid configurations.
Corresponding generating sources (\ref{esourc}) are parameterized as $\
_{h}^{tor}\Upsilon (\tilde{r},x)$ and $^{tor}\Upsilon =\ _{v}^{tar}\Upsilon (%
\tilde{r},x,y).$

Any off-diagonal (\ref{prmtor2}) can be nonholonomically deformed to a class
of solutions of type of (\ref{qeltors}), (\ref{offdiagcosmcsh}), (\ref%
{offdsolgenfgcosmc}), (\ref{offdiagpolfr}) or (\ref{offdncelepsilon}). The
corresponding nonlinear symmetries of type (\ref{nonlinsymrex}) are written: 
\begin{eqnarray}
\partial _{y}[\Psi ^{2}(\tilde{r},x,y)] &=&-\int dy\ \ ^{tor}\Upsilon
\partial _{y}g_{4}\simeq -\int dy\ \ ^{tor}\Upsilon (\tilde{r},x,y)\partial
_{y}[\eta _{4}(\tilde{r},x,y)\ \tilde{g}_{4}(\tilde{r})]  \label{nsymtor} \\
&\simeq &-\int dy \ \ ^{tor}\Upsilon (\tilde{r},x,y)\partial _{y}[\zeta _{4}(%
\tilde{r},x,y)(1+\kappa \ \chi _{4}(\tilde{r},x,y)) \ \tilde{g}_{4}(\tilde{r}%
)],  \notag \\
\Psi (\tilde{r},x,y) &=&|\ \ \Lambda +\ ^{tor}\Lambda |^{-1/2}\sqrt{|\int
dy\ \ \ ^{tor}\Upsilon (\tilde{r},x,y)\ \partial _{y}(\Phi ^{2})|},(\Phi (%
\tilde{r},x,y))^{2}=-4\ \Lambda \tilde{g}_{4}(\tilde{r},x,y)  \notag \\
&\simeq &-4\ (\ \Lambda +\ ^{tor}\Lambda )\eta _{4}(\tilde{r},x,y)\ \ \tilde{%
g}_{4}(\tilde{r})\simeq -4(\ \Lambda +\ ^{tor}\Lambda )\ \zeta _{4}(\tilde{r}%
,x,y)(1+\kappa \chi _{4}(\tilde{r},x,y))\ \tilde{g}_{4}(\tilde{r}).  \notag
\end{eqnarray}%
In these formulas, we use $\ ^{tor}\Lambda $ as an effective cosmological
constant related via nonlinear symmetries to and energy-momentum tensor (\ref%
{emtsm}). Such a $\ ^{tor}\Lambda $ can be different from a prescribed
cosmological constant associated to other types of gravitational and matter
interactions. In this subsection, we write $\widetilde{\Lambda }=\Lambda +\
^{tor}\Lambda .$ 

For small parametric deformations with $\chi$-polarization functions, the
quadratic linear elements for nonholonomic toroid solutions are computed: 
\begin{eqnarray*}
d\ \widehat{s}^{2} &=&\widehat{g}_{\alpha \beta }(\tilde{r},x,y;\psi ,\eta
_{4};\widetilde{\Lambda }=\Lambda +\ ^{tor}\Lambda ,\ ^{tor}\Upsilon ,\ 
\tilde{g}_{\alpha })du^{\alpha }du^{\beta } \\
&=&e^{\psi _{0}(\tilde{r},x)}[1+\kappa \ ^{\psi (\tilde{r},x)}\chi (\tilde{r}%
,x)][(dx^{1}(\tilde{r},x))^{2}+(dx^{2}(\tilde{r},x))^{2}]-
\end{eqnarray*}%
\begin{eqnarray*}
&&\{\frac{4[\partial _{y}(|\zeta _{4}\ \tilde{g}_{4}|^{1/2})]^{2}}{\tilde{g}%
_{3}|\int dy\{\ ^{tor}\Upsilon \partial _{y}(\zeta _{4}\ \tilde{g}_{4})\}|}%
-\kappa \lbrack \frac{\partial _{y}(\chi _{4}|\zeta _{4}\tilde{g}_{4}|^{1/2})%
}{4\partial _{y}(|\zeta _{4}\tilde{g}_{4}|^{1/2})}-\frac{\int dy\{\
^{tor}\Upsilon \partial _{y}[(\zeta _{4}\ \tilde{g}_{4})\chi _{4}]\}}{\int
dy\{\ ^{tor}\Upsilon \partial _{y}(\zeta _{4}\ \tilde{g}_{4})\}}]\}\ \tilde{g%
}_{3} \\
&&\{dy+[\frac{\partial _{i}\ \int dy\ ^{tor}\Upsilon \ \partial _{y}\zeta
_{4}}{(\tilde{N}_{i}^{3})\ \ ^{tor}\Upsilon \partial _{y}\zeta _{4}}+\kappa (%
\frac{\partial _{i}[\int dy\ \ ^{tor}\Upsilon \ \partial _{y}(\zeta _{4}\chi
_{4})]}{\partial _{i}\ [\int dy\ \ ^{tor}\Upsilon \partial _{y}\zeta _{4}]}-%
\frac{\partial _{y}(\zeta _{4}\chi _{4})}{\partial _{y}\zeta _{4}})]\tilde{N}%
_{i}^{3}dx^{i}\}^{2}+
\end{eqnarray*}%
\begin{eqnarray}
&&\zeta _{4}(1+\kappa \ \chi _{4})\ \tilde{g}_{4}\{dt+[(\tilde{N}%
_{k}^{4})^{-1}[\ _{1}n_{k}+16\ _{2}n_{k}[\int dy\frac{\left( \partial
_{y}[(\zeta _{4}\ \tilde{g}_{4})^{-1/4}]\right) ^{2}}{|\int dy\partial
_{y}[\ ^{tor}\Upsilon (\zeta _{4}\ \tilde{g}_{4})]|}]+  \notag \\
&&\kappa \frac{16\ _{2}n_{k}\int dy\frac{\left( \partial _{y}[(\zeta _{4}\ 
\tilde{g}_{4})^{-1/4}]\right) ^{2}}{|\int dy\partial _{y}[\ ^{tor}\Upsilon
(\zeta _{4}\ \tilde{g}_{4})]|}(\frac{\partial _{y}[(\zeta _{4}\ \tilde{g}%
_{4})^{-1/4}\chi _{4})]}{2\partial _{y}[(\zeta _{4}\ \tilde{g}_{4})^{-1/4}]}+%
\frac{\int dy\partial _{y}[\ ^{tor}\Upsilon (\zeta _{4}\chi _{4}\ \tilde{g}%
_{4})]}{\int dy\partial _{y}[\ ^{tor}\Upsilon (\zeta _{4}\ \tilde{g}_{4})]})%
}{\ _{1}n_{k}+16\ _{2}n_{k}[\int dy\frac{\left( \partial _{y}[(\zeta _{4}\ 
\tilde{g}_{4})^{-1/4}]\right) ^{2}}{|\int dy\partial _{y}[\ ^{tor}\Upsilon
(\zeta _{4}\ \tilde{g}_{4})]|}]}]\tilde{N}_{k}^{4}dx^{k}\}^{2}.
\label{tortpol2}
\end{eqnarray}%
This formula involves elliptic deformations if we chose a generating
function $\chi _{4}(\tilde{r},x,y)=\underline{\chi }(\tilde{r},x)\sin
(\omega _{0}y+y_{0}).$ In such cases, we generate a family of toroid
configurations with ellipsoidal deformations on $y$ coordinate. In a similar
way, we can construct solutions with ellipsoidal deformations on $x$%
-coordinate.

Finally, we note that we can consider more sophisticate classes of
nonholonomic deformations which transform a toroid prime d-metric into
"spagetti" quasi-stationary configurations (with different sections, curved
and waved, possible interruptions, singularities etc.) embedded into locally
anisotropic gravitational vacuum media. Such solutions can be used for
modelling DM quasi-stationary configurations when the effective cosmological
constant $\widetilde{\Lambda }=\Lambda +\ ^{tor}\Lambda $ can be considered
as a phenomenological sum of parameters to be related to DE and encoding
toroid configurations. Such parameters have to be fixed by certain
observational data. 

\subsection{Off-diagonal cosmological solitonic and spheroid cosmological
involving 2-d vertices}

The goal of this subsection is to study some physically important examples
of locally anisotropic cosmological solutions (\ref{qeltorsc}) and their
equivalents when the gravitational $\eta $- and $\chi $-polarizations depend
on a time like coordinate. Such solutions can be generic off-diagonal and
characterized by nonlinear symmetries of type. 

\subsubsection{Off-diagonal transforms of cosmological models with
spheroidal symmetry and voids}

The Minkowski spacetime can be written in \textbf{prolate} spheroidal
coordinates $u^{\alpha }=(r,\theta ,\phi ,t),$ when certain Cartesian
coordinates can be defined in the form $u^{\alpha }=(x=r\sin \theta \cos
\phi ,y=r\sin \theta \sin \phi ,z=\sqrt{r^{2}+r_{\lozenge }^{2}}\cos \theta
,t).$ For such coordinates, the constant parameter $r_{\lozenge }$ has the
meaning of the distance of the foci from the origin of the coordinate
system. This allows us to define a prolate spheroid (i.e. a rotoid, or
ellipsoid) with the foci along the $z$-axis, when 
\begin{equation*}
\frac{x^{2}+y^{2}}{(\ _{0}r)^{2}}+\frac{z^{2}}{(\ _{0}r)^{2}+r_{\lozenge
}^{2}}=1,
\end{equation*}%
for any fixed $r=\ _{0}r.$ Such a $\ _{0}r$ corresponds to the length of its
minor radius and the size of its major radius is $\sqrt{(\
_{0}r)^{2}+r_{\lozenge }^{2}}.$ Using prolate coordinates, the flat
Minkowski spacetime metric can be written 
\begin{equation*}
ds^{2}=(r^{2}+r_{\lozenge }^{2}\sin ^{2}\theta )(\frac{dr^{2}}{%
r^{2}+r_{\lozenge }^{2}}+d\theta ^{2})+r^{2}\sin ^{2}\theta d\phi -dt^{2}.
\end{equation*}

In a similar form, we can introduce \textbf{oblate} coordinates, when $x=%
\sqrt{r^{2}+r_{\lozenge }^{2}}\sin \theta \cos \phi ,$\newline
$y=\sqrt{r^{2}+r_{\lozenge }^{2}}\sin \theta \sin \phi ,z=r\cos \theta .$
So, for a fixed $r=\ _{0}r,$ an oblate spheroid with a $z$ symmetric axis, 
\begin{equation*}
\frac{x^{2}+y^{2}}{(\ _{0}r)^{2}+r_{\lozenge }^{2}}+\frac{z^{2}}{(\
_{0}r)^{2}}=1,
\end{equation*}%
can be defined. For such a hypersurface, the value $\sqrt{%
r^{2}+r_{\lozenge}^{2}}$ corresponds to the major radius and $_{0}r$ is the
minor one. Correspondingly, the flat Minkowki spacetime metric can be
written in the form 
\begin{equation*}
ds^{2}=(r^{2}+r_{\lozenge }^{2}\cos ^{2}\theta )(\frac{dr^{2}}{%
r^{2}+r_{\lozenge }^{2}}+d\theta ^{2})+r^{2}\sin ^{2}\theta d\phi -dt^{2}.
\end{equation*}

We consider a quadratic element introduced in \cite{boero16}: 
\begin{eqnarray}
d\underline{s}^{2} &=&\frac{a^{2}(t)}{[1+\frac{\varsigma }{4}%
(r^{2}+r_{\lozenge }^{2}\cos ^{2}\theta )]^{2}}[(r^{2}+r_{\lozenge }^{2}\sin
^{2}\theta )(\frac{dr^{2}}{r^{2}-\frac{M(r)}{r}(r^{2}+r_{\lozenge }^{2}\sin
^{2}\theta )+r_{\lozenge }^{2}}+d\theta ^{2})  \notag \\
&&+r^{2}\sin ^{2}\theta d\phi ]-B(r)dt^{2},%
\mbox{ with prolate spheroidal
symmetry};  \label{cosmvoidm1}
\end{eqnarray}%
\begin{eqnarray*}
d\underline{s}^{2} &=&\frac{a^{2}(t)}{[1+\frac{\varsigma }{4}%
(r^{2}+r_{\lozenge }^{2}\sin ^{2}\theta )]^{2}}[(r^{2}+r_{\lozenge }^{2}\sin
^{2}\theta )(\frac{dr^{2}}{r^{2}-\frac{M(r)}{r}(r^{2}+r_{\lozenge }^{2}\cos
^{2}\theta )+r_{\lozenge }^{2}}+d\theta ^{2}) \\
&&+(r^{2}+r_{\lozenge }^{2})\sin ^{2}\theta d\phi ]-B(r)dt^{2},\mbox{with
oblate spheroidal symmetry}.
\end{eqnarray*}%
The conditions For $B(r)=1$ and $M(r)=0$ are used if the above formulas
define respective FLRW cosmological quadratic line elements, when $\varsigma
=1,0,-1 $ refer respectively to a positive curved, flat, hyperbolic space
geometry, see also explanations for formulas (\ref{flrw}). 

For elaborating cosmological models, the mass profile function $M(r)$ from (%
\ref{cosmvoidm1}) can be specified as in \cite{amendola98} (in a simple
choice, $B(r)=1$),%
\begin{equation*}
M(r)=\left\{ 
\begin{array}{cc}
\frac{4\pi }{3}\rho _{int}r^{3}, & \mbox{ for }r<\ _{v}r; \\ 
M(\ _{v}r)+\frac{4\pi }{3}\rho _{bor}(r^{3}-\ _{v}r^{3}), & \mbox{ for }\
_{v}r\leq r<\ _{v}r+\ _{w}r; \\ 
0 & \mbox{ for }\ _{v}r+\ _{w}r\leq r.%
\end{array}%
\right.
\end{equation*}%
Let us explain the meaning of two important constants: $\ _{v}r$ is
associated with the radius of the void, and the parameter $\ _{w}r$ is
related to the size of the wall. For instance, such a profile is modelled in
a form that the border compensates for the amount missing in the void (i.e.
it models a compensated void) by choosing the spherical symmetry.
Respectively, the internal density of the matter, $\rho _{int},$ and border
density of matter, $\rho _{bor},$ are related to the mean density outside
the void, $\rho _{0};$ we use the formulas%
\begin{equation}
\rho _{int}=-\rho _{0}\xi \mbox{ and }\rho _{bor}=\rho _{0}\xi /[(1+\
_{w}r/\ _{v}r)^{3}-1],  \label{emtvoid}
\end{equation}%
for a constant parameter $\xi <1.$ For such conditions, a cosmological
metric (\ref{cosmvoidm1}) is a solution of the Einstein equations in GR if $%
a(t)$ is defined by the Friedman equations, $\frac{3}{a^{2}(t)}[\frac{da}{dt}%
+\varsigma ]=8\pi \rho _{0}.$ The function $B(r)$ can be parameterized in
the form $\ B(r)=B_{0}[B_{1}+\ln (\frac{r}{r_{\lozenge }})]^{2},$ for some
constants $B_{0}$ and $B_{1}$ if we try to use such solutions to explain
certain phenomenology for astrophysical systems with DM as in \cite{galo12}.
For such configurations, the value of $B_{1}$ can be fixed in a form that
the component $T_{r}^{r}=T_{1}^{1}$ of the energy-momentum tensor remains of
the same order as $\rho _{0}$ (they fix $B_{1}=10^{7}$). Other
phenomenological parameters take values like $\ _{w}r=0.3\ _{v}r,$ $\xi
=0.1, $ $r_{\lozenge }=0.1\ _{v}r$ when a radius $\ _{v}r$ corresponds to a
physical size of 22Mpc. 

We can re-write (\ref{cosmvoidm1}) in some appropriate curved coordinates,
when non-trivial N-connection coefficients $\underline{\mathring{N}}%
_{i}^{a}= $ $\underline{\mathring{N}}_{i}^{a}(u^{\alpha }(r,\theta ,\phi
,t)) $ and \underline{$\mathring{g}$}$_{\alpha \beta }(u^{j}(r,\theta ,\phi
,t),u^{4}(r,\theta ,\phi ,t))$ which are defined in any form not involving
singular frame transforms and off-diagonal deformations. This allows us to
apply the AFCDM to generate new classes of locally anisotropic cosmological
solutions. Such new coordinates are defined $u^{1}=x^{1}=r,u^{2}=\theta ,$
and $u^{3}=y^{3}=y^{3}(r,\theta ,\phi )$ and $u^{4}=y^{4}=t+\
^{4}B(r,\theta),$ when (for $\underline{\mathring{N}}_{i}^{3}=-\partial \
y^{3}/\partial x^{i}$ and $\underline{\mathring{N}}_{i}^{4}=-\partial \
^{4}B/\partial x^{i}):$ 
\begin{eqnarray*}
\underline{\mathbf{\mathring{e}}}^{3} &=&du^{3}+\underline{\mathring{N}}%
_{i}^{3}(r,\theta )dx^{i}=du^{3}+\underline{\mathring{N}}_{1}^{3}(r,\theta
)dr+\underline{\mathring{N}}_{2}^{3}(r,\theta )d\theta , \\
\underline{\mathbf{\mathring{e}}}^{4} &=&du^{4}+\underline{\mathring{N}}%
_{i}^{4}(r,\theta )dx^{i}=du^{4}+\underline{\mathring{N}}_{1}^{4}(r,\theta
)dr+\underline{\mathring{N}}_{2}^{4}(r,\theta )dz.
\end{eqnarray*}
This way, we obtain an off-diagonal spheroid-type cosmological metric
parameterized as a d-metric, 
\begin{eqnarray}
d\underline{\mathring{s}}^{2} &=&\underline{\mathring{g}}_{\alpha }(r,\theta
,t)[\underline{\mathbf{\mathring{e}}}^{\alpha }(r,\theta ,t)]^{2},%
\mbox{where for }\left\{ 
\begin{array}{c}
\mbox{ prolate }: \\ 
\mbox{ oblate }:%
\end{array}%
\right.  \label{cosmvoidm2} \\
\underline{\mathring{g}}_{1}(r,\theta ,t) &=&\left\{ 
\begin{array}{c}
\frac{a^{2}(t)(r^{2}+r_{\lozenge }^{2}\sin ^{2}\theta )}{[1+\frac{\varsigma 
}{4}(r^{2}+r_{\lozenge }^{2}\cos ^{2}\theta )]^{2}[r^{2}-\frac{M(r)}{r}%
(r^{2}+r_{\lozenge }^{2}\sin ^{2}\theta )+r_{\lozenge }^{2}]} \\ 
\frac{a^{2}(t)(r^{2}+r_{\lozenge }^{2}\sin ^{2}\theta )}{[1+\frac{\varsigma 
}{4}(r^{2}+r_{\lozenge }^{2}\sin ^{2}\theta )]^{2}[r^{2}-\frac{M(r)}{r}%
(r^{2}+r_{\lozenge }^{2}\cos ^{2}\theta )+r_{\lozenge }^{2}]}%
\end{array}%
\right. ,  \notag \\
\underline{\mathring{g}}_{2}(r,\theta ,t) &=&\left\{ 
\begin{array}{c}
\frac{a^{2}(t)}{[1+\frac{\varsigma }{4}(r^{2}+r_{\lozenge }^{2}\cos
^{2}\theta )]^{2}} \\ 
\frac{a^{2}(t)}{[1+\frac{\varsigma }{4}(r^{2}+r_{\lozenge }^{2}\sin
^{2}\theta )]^{2}}%
\end{array}%
\right. ,\ \underline{\mathring{g}}_{3}(r,\theta ,t)=\left\{ 
\begin{array}{c}
\frac{a^{2}(t)r^{2}\sin ^{2}\theta }{[1+\frac{\varsigma }{4}%
(r^{2}+r_{\lozenge }^{2}\cos ^{2}\theta )]^{2}} \\ 
\frac{a^{2}(t)(r^{2}+r_{\lozenge }^{2})\sin ^{2}\theta }{[1+\frac{\varsigma 
}{4}(r^{2}+r_{\lozenge }^{2}\sin ^{2}\theta )]^{2}}%
\end{array}%
\right. ,\underline{\ \mathring{g}}_{4}(r)=-B(r).  \notag
\end{eqnarray}

In the next subsection, prime cosmological metrics (\ref{cosmvoidm2}) will
be nonholonomically deformed to locally anisotropic cosmological d-metrics (%
\ref{dmc}). We shall use gravitational $\eta $-polarization functions in
some forms which allow us to generate exact and parametric solutions of
nonholonomic Einstein equations (\ref{cdeq1}). 

\subsubsection{Off-diagonal cosmological solitonic evolution encoding 2-d
vertices}

We consider nonholonomic deformations of data $(\underline{\mathring{g}}%
_{\alpha},\underline{\mathring{N}}_{i}^{a})\rightarrow (\underline{g}%
_{\alpha }=\underline{\eta }_{\alpha }\underline{\mathring{g}}_{\alpha },%
\underline{N}_{i}^{a}=\underline{\eta }_{i}^{a}\underline{\mathring{N}}%
_{i}^{a})$ using underlined versions of formulas (\ref{offdiagpm}), (\ref%
{offdiagdefr}) with nonlinear symmetries (\ref{dualnonltr}). The
gravitational polarizations $\underline{\eta }_{i}(r,\theta
,t)=a^{-2}(t)\eta _{i}(r,\theta ),\underline{\eta }_{3}(r,\theta
,t)=a^{-2}(t)\underline{\eta } (r,\theta ,t)$ and $\underline{\eta }%
_{4}(r,\theta ,t)$ will be prescribed or computed in such forms that 
\begin{eqnarray}
\underline{\mathbf{g}} &=&(g_{i},g_{b},\underline{N}_{i}^{3}=\underline{n}%
_{i},\underline{N}_{i}^{4}=\underline{w}_{i})=g_{i}(r,\theta )dx^{i}\otimes
dx^{i}+\underline{h}_{3}(r,\theta ,t)\underline{\mathbf{e}}^{3}\otimes 
\underline{\mathbf{e}}^{3}+\underline{h}_{4}(r,\theta ,t)\underline{\mathbf{e%
}}^{4}\otimes \underline{\mathbf{e}}^{4},  \label{lacosm1} \\
&&\underline{\mathbf{e}}^{3}=d\phi +\underline{n}_{i}(r,\theta ,t)dx^{i},\ 
\underline{\mathbf{e}}^{4}=dt+\underline{w}_{i}(r,\theta ,t)dx^{i},  \notag
\end{eqnarray}%
with Killing symmetry on the angular coordinate $\varphi ,$ when $\partial
_{\varphi }$ transforms into zero the N-adapted coefficients of such a
d-metric.

In terms of $\eta $-polarization functions, a (\ref{cdeq1}) can be written
in a $t$-dual form to (\ref{offdiagpolfr}), when 
\begin{eqnarray}
d\widehat{s}^{2} &=&\widehat{g}_{\alpha \beta }(r,\theta ,t;\underline{%
\mathring{g}}_{\alpha };\psi ,\eta _{3};\ \underline{\Lambda },\ ^{v}%
\underline{\Upsilon })du^{\alpha }du^{\beta }=e^{\psi
}[(dx^{1})^{2}+(dx^{2})^{2}]  \label{lacosm2} \\
&&+ (\underline{\eta }\underline{\mathring{g}}_{3})\{d\phi +[\ _{1}n_{k}+\
_{2}n_{k}\int dt\frac{[\partial _{t}(\underline{\eta }\underline{\mathring{g}%
}_{3})]^{2}}{|\int dt\ ^{v}\underline{\Upsilon }\partial _{t}(\underline{%
\eta }\underline{\mathring{g}}_{3})|\ (\underline{\eta }\underline{\mathring{%
g}}_{3})^{5/2}}]dx^{k}\}^{2}  \notag \\
&& -\frac{[\partial _{t}(\underline{\eta }\ \underline{\mathring{g}}%
_{3})]^{2}}{|\int dt\ ^{v}\underline{\Upsilon }\partial _{t}(\underline{\eta 
}\underline{\mathring{g}}_{3})|\ \eta \mathring{g}_{3}}\{dt+\frac{\partial
_{i}[\int dt\ ^{v}\underline{\Upsilon }\ \partial _{t}(\underline{\eta }%
\underline{\mathring{g}}_{3})]}{\ ^{v}\underline{\Upsilon }\partial _{t}(%
\underline{\eta }\underline{\mathring{g}}_{3})}dx^{i}\}^{2}.  \notag
\end{eqnarray}%
For $\underline{\Phi }^{2}=-4\ \underline{\Lambda }\underline{g}_{3},$ we
can transform (\ref{lacosm2}) in a variant of (\ref{qeltorsc}) with
underlined $\eta $-polarizations determined by the generating data $(%
\underline{g}_{3};\ \underline{\Lambda },\ ^{v}\underline{\Upsilon }).$ The
effective cosmological constant $\underline{\Lambda }$ is chosen as
effective ones which correspond via nonlinear symmetries (\ref{dualnonltr})
to an energy-momentum tensor (\ref{emtvoid}) in a fluid type form.
Respective generating sources $(\ ^{h}\underline{\Upsilon },\ ^{v}\underline{%
\Upsilon }) $ are related to a $T_{\alpha \beta }$ via respective frame or
coordinate transforms. Locally anisotropic cosmological scenarios with
nonholonomic evolution from a primary void configuration (\ref{cosmvoidm2})
are determined by generating polarization functions $\psi \simeq \psi (x^{k})
$ and $\underline{\eta }\ \simeq \underline{\eta }(x^{k},t).$ 

In explicit form, we consider such a variant: The h-part of the d-metric (%
\ref{lacosm2}) is prescribed to satisfy instead of a 2-d Poisson equation
the generalized Taubes equation for vortices on a curved background 2-d
surface, 
\begin{equation}
_{h}\nabla ^{2}\psi =\Omega _{0}(C_{0}-C_{1}e^{2\psi }).  \label{taubeq}
\end{equation}%
In this formula, the position-dependent conformal factor $\Omega _{0}$ and
effective source $(C_{0}-C_{1}e^{2\psi })$ are prescribed as respective
generating h-function $\psi (x^{k})$ and generating h-source $\ ^{h}%
\underline{\Upsilon }(x^{k}).$ By re-scaling, both constants $C_{0}$ and $%
C_{1}$ take standard values $-1,0,$ or 1, but there are only five
combinations of these values allow vortex solutions $\psi \lbrack vortex]$
without singularities \cite{manton17}, see also \cite{jaffe80}.

The v-part of (\ref{lacosm2}) can be modelled in a solitonic wave form for 
\begin{equation}
\underline{\eta }\ \simeq \left\{ 
\begin{array}{ccc}
_{r}^{sol}\underline{\eta }(r,t) & 
\mbox{ as a solution of the modified KdV
equation }\frac{\partial \underline{\eta }}{\partial t}-6\underline{\eta }%
^{2}\frac{\partial \underline{\eta }}{\partial r}+\frac{\partial ^{3}%
\underline{\eta }}{\partial r^{3}}=0, & \mbox{  radial solitons}; \\ 
_{\theta }^{sol}\underline{\eta }(\theta ,t) & 
\mbox{ as a solution of the
modified KdV equation }\frac{\partial \underline{\eta }}{\partial \theta }-6%
\underline{\eta }^{2}\frac{\partial \underline{\eta }}{\partial \theta }+%
\frac{\partial ^{3}\underline{\eta }}{\partial \,^{3}}=0, & 
\mbox{ angular
solitons}.%
\end{array}%
\right.  \label{solitonw}
\end{equation}%
We cite \cite{doikou20} and reference therein on such types of solitonic wave equations.

The generic off-diagonal metrics (\ref{lacosm2}) describe nonholonomic cosmological evolution scenarios with conventional h- and v-splitting, when the gravitational polarization $\underline{\eta }$ is prescribed to be a solitonic wave of type (\ref{solitonw}). In principle, we can prescribe
other types of nonlinear wave polarizations, nonholonomic distributions, etc. It is like for BH solutions in GR, when we have integration constants and we can prescribe them as respective mass, rotation momenta, zero or nonzero cosmological constants etc. (that is possible for diagonal ansatz
transforming the Einstein equations into a system of nonlinear ODE of second order). The AFCDM allows us to construct (\ref{lacosm2}) as a very general off-diagonal cosmological solutions of nonholonomic deformed Einstein equations (\ref{cdeq1}) (or their equivalents (\ref{cdeq1a}), with nonlinear symmetries and properties described by formulas (\ref{dualnonltr}) and (\ref{dualcosm})). Such locally anisotropic cosmological models can be elaborated in explicit form if we prescribe certain physically important configurations defined by generating and integration functions (because we integrated more PDEs). The above cosmological model describes a scenario when a primary metric with prolate/oblate rotoid void transforms into a vertex h-configuration (\ref{taubeq}) and the v-part describes solitonic wave evolution of type (\ref{solitonw}). This holds for a geometric evolution with gravitational polarizations and for the respective generating sources. Such a cosmological evolution results also in solitonic configurations for the N-connection coefficients. Corresponding solitonic waves on $t$-variable can be with a radial space variable, $r,$ or with an angular variable, $\theta .$ 

In a series of our and co-authors' works, there were constructed more general classes of generic off-diagonal cosmological and quasi-stationary solutions with 3-d solitonic waves and solitonic hierarchies in GR and MGTs, see  \cite{vacaruplb16,vbubuianu17,vacaru18,partner06} and references therein. In \cite{nonassocFinslrev25} such solitonic cosmological solutions are considered for nonassociative Finsler-like MGTs for modelling DM quasi-periodic and pattern-forming structures. In this subsection, two (void and solitonic) examples of those results were modified in such forms that analogous cosmological solutions can be constructed in the framework of the Einstein gravity theory. Let us finally explain how vertex (\ref{taubeq}) and solitonic (\ref{solitonw}) configurations can be modelled in GR.  For this, we have to consider off-diagonal cosmological metrics  (\ref{lacosm2}) when the coefficients (defined respectively by generating and integration functions and generating sources) are subjected to the LC-conditions  (\ref{zerot1}) and (\ref{qellc}). In such cases, the nonholonomic torsion is zero, as imposed by (\ref{lccond1}), and the system of nonlinear PDEs (\ref{cdeq1}) transforms into standard Einstein equations (\ref{einstceq1}).    

\subsubsection{Small parametric cosmological deformations with off-diagonal solitonic vacua for voids}

We can generate off-diagonal cosmological solutions with small $\kappa $%
--parametric deformations of (\ref{cosmvoidm2}) using nonlinear symmetries
and transforms (\ref{dualnonltr}), when the d-metrics are written in terms
of $\chi $-polarization functions, 
\begin{equation*}
d\ \widehat{s}^{2}=\widehat{g}_{\alpha \beta }(r,\theta ,t;\psi ,\underline{%
\Lambda },\ ^{v}\underline{\Upsilon })du^{\alpha }du^{\beta }=e^{\psi
_{0}(r,\theta )}[1+\kappa \ ^{\psi }\chi (r,\theta )][(dx^{1}(r,\theta
))^{2}+(dx^{2}(r,\theta ))^{2}]
\end{equation*}%
\begin{eqnarray}
&&+\zeta _{3}(1+\kappa \ \underline{\chi })\underline{\mathring{g}}%
_{3}\{d\phi +[(\underline{\mathring{N}}_{k}^{3})^{-1}[\ _{1}n_{k}+16\
_{2}n_{k}[\int dt\frac{\left( \partial _{t}[(\underline{\zeta }_{3}%
\underline{\mathring{g}}_{3})^{-1/4}]\right) ^{2}}{|\int d\varphi \partial
_{\varphi }[\ \ ^{v}\underline{\Upsilon }(\underline{\zeta }_{3}\ \underline{%
\mathring{g}}_{3})]|}]  \notag \\
&&+\kappa \frac{16\ _{2}n_{k}\int dt\frac{\left( \partial _{t}[(\underline{%
\zeta }_{3}\ \underline{\mathring{g}}_{3})^{-1/4}]\right) ^{2}}{|\int
dt\partial _{t}[\ ^{v}\underline{\Upsilon }(\underline{\zeta }_{3}\ 
\underline{\mathring{g}}_{3})]|}(\frac{\partial _{t}[(\underline{\zeta }%
_{3}\ \underline{\mathring{g}}_{3})^{-1/4}\underline{\chi })]}{2\partial
_{t}[(\underline{\zeta }_{3}\ \underline{\mathring{g}}_{3})^{-1/4}]}+\frac{%
\int dt\partial _{t}[\ ^{v}\underline{\Upsilon }(\underline{\zeta }_{3}%
\underline{\chi }\ \underline{\mathring{g}}_{3})]}{\int dt\partial _{t}[\
^{v}\underline{\Upsilon }(\underline{\zeta }_{3}\ \underline{\mathring{g}}%
_{3})]})}{\ _{1}n_{k}+16\ _{2}n_{k}[\int dt\frac{\left( \partial _{t}[(%
\underline{\zeta }_{3}\ \underline{\mathring{g}}_{3})^{-1/4}]\right) ^{2}}{%
|\int dt\partial _{t}[\ \ ^{v}\underline{\Upsilon }(\underline{\zeta }_{3}\ 
\underline{\mathring{g}}_{3})]|}]}]\underline{\mathring{N}}%
_{k}^{3}dx^{k}\}^{2}.  \label{paramsoliton}
\end{eqnarray}%
\begin{eqnarray*}
&&-\{\frac{4[\partial _{t}(|\underline{\zeta }_{3}\ \underline{\mathring{g}}%
_{3}|^{1/2})]^{2}}{\ \underline{\mathring{g}}_{4}|\int dt\{\ ^{v}\underline{%
\Upsilon }\partial _{t}(\underline{\zeta }_{3}\ \underline{\mathring{g}}%
_{3})\}|}-\kappa \lbrack \frac{\partial _{t}(\underline{\chi }|\underline{%
\zeta }_{3}\ \underline{\mathring{g}}_{3}|^{1/2})}{4\partial _{t}(|%
\underline{\zeta }_{3}\ \underline{\mathring{g}}_{3}|^{1/2})}-\frac{\int
dt\{\ ^{v}\underline{\Upsilon }\partial _{t}[(\underline{\zeta }_{3}\ 
\underline{\mathring{g}}_{3})\underline{\chi }]\}}{\int dt\{\ ^{v}\underline{%
\Upsilon }\partial _{t}(\underline{\zeta }_{3}\ \underline{\mathring{g}}%
_{3})\}}]\}\ \ \underline{\mathring{g}}_{4} \\
&&\{dt+[\frac{\partial _{i}\ \int dt\ ^{v}\underline{\Upsilon }\ \partial
_{t}\underline{\zeta }_{3}}{(\underline{\mathring{N}}_{i}^{3})\ ^{v}%
\underline{\Upsilon }\partial _{t}\underline{\zeta }_{3}}+\kappa (\frac{%
\partial _{i}[\int dt\ ^{v}\underline{\Upsilon }\ \partial _{t}(\underline{%
\zeta }_{3}\ \underline{\mathring{g}}_{3})]}{\partial _{i}\ [\int dt\ \ ^{v}%
\underline{\Upsilon }\partial _{t}\zeta _{4}]}-\frac{\partial _{t}(%
\underline{\zeta }_{3}\ \underline{\mathring{g}}_{3})}{\partial _{t}%
\underline{\zeta }_{3}})]\underline{\mathring{N}}_{i}^{4}dx^{i}\}^{2}
\end{eqnarray*}%
In these formulas, $\psi _{0}(r,\theta )$ and $\ ^{\psi }\chi (r,\theta )$
are solutions of 2-d Poisson equations (we can consider $\kappa $-parametric
solutions of (\ref{taubeq}) with some small parametric generated vortices).
For vertical configurations, the generating function $\underline{\chi }=%
\underline{\chi }_{3}(r,\theta ,t)$ can be taken as a solution of solitonic
wave equation (\ref{solitonw}), $\underline{\eta }\ \longleftrightarrow 
\underline{\chi },$ when $\underline{\zeta }_{3}(r,\theta ,t)$ is also
prescribed in a form to satisfy observational data for $\kappa ^{0}.$ Such
cosmological solutions describe v-solitonic gravitational structures of
voids with $\kappa $--parametric and $t$-evolution. We can model
configurations when a parametric gravitational void vacuum possesses a
nontrivial solitonic energy. The solutions can be characterized by nonlinear
symmetries relating the effective generating source $\ ^{v}\underline{%
\Upsilon }$ to a respective cosmological constant $\ \underline{\Lambda }.$
In a more general context, the vertex-solitonic wave cosmological d-metrics
with prime prolate/oblate symmetry encode a nonholonomic vacuum structure
with nontrivial canonical d-torsion $\widehat{\mathbf{T}}_{ \alpha \beta
}^{\gamma }$. Imposing additional constraints of any type (\ref{lccond}), (%
\ref{zerot1}) or (\ref{lccond1}), we can extract LC configurations for
elaborating locally anisotropic cosmological models in the framework of GR.
Similar off-diagonal cosmological solutions were studied in \cite%
{vacaruplb16,vbubuianu17} for ekpyrotic scenarios, in general, with
quasi-periodic and pattern-forming structures in MGTs and GR. 

\section{A general relativistic formulation of G. Perelman's thermodynamics}

\label{sec04} The generic off-diagonal solutions constructed in the previous section are not characterized, in general, by certain hypersurface or holographic conditions. This means that the Bekenstein--Hawking thermodynamic paradigm does not apply to characterizing important physical properties of such classes of quasi-stationary or locally anisotropic solutions in GR. Such solutions with generic off-diagonal terms and additional degrees of freedom can be used for characterising properties of locally anisotropic cosmological models and elaborating on dark matter and dark energy scenarios remaining in the framework of the Einstein theory. But in such cases, we have to develop in a relativistic nonholonomic form the Ricci flow theory, which allows us to define a new geometric and statistical thermodynamic paradigm (due to G. Perelman). 

Perhaps, a most simple setting why the Ricci flow theory is relevant to GR and string gravity exists in G. Perelman’s famous preprint [26] (see, respectively, sub-sections 1.4*, 2.1, 5.1, 5.2, 5.3).\footnote{Here, we note that, in mathematical preprints, the style of numbering sections and subsections is different.} In modern literature on GR, MGTs, and various models of QG and QFTs, the importance of the geometric flow methods is considered already a well-known fact.  In \cite{svnonh08,gheorghiuap16,partner06,vacaruplb16,vbubuianu17}, see a review and recent developments for nonassociative gravity \cite{vacaru18,nonassocFinslrev25,bsvv24,v25c}, we elaborated on applications of the theory of geometric flows \cite{hamilton82,perelman1} in GR and various types of MGTs, quantum information flows etc. Here, we note that in our works on nonholonomic metric or nonmetric geometric flows, relativistic generalizations of the famous Poincar\'{e}--Thurston conjecture are not formulated/ proved. This is a very difficult mathematical problem. For instance, rigorous mathematical proofs of the results for Ricci flows of Riemannian metrics consist of some hundred pages as in monographs \cite{monogrrf1,monogrrf2,monogrrf3}. 

It is not a goal of this work to develop those topological and geometric analysis methods for Ricci flows, including the Einstein equations in GR as a particular case, i.e. as a relativistic Ricci soliton. Here we note that the term Ricci soliton (used by G. Perelman in section 2.1 of \cite{perelman1} for self-similar geometric flow configurations is different from the concepts used in solitonic analysis, nonlinear waves in gravity and hydrodynamics, etc. Even there were announced certain results on numerical evolution of the Einstein equations when the meaning of certain so-called Z4 terms was related to a cryptic connection to a Ricci flow \cite{alic11}, we do not consider or use the results of numerical methods in this work. 

Our AFCDM is a geometric and analytic one, allowing an integration of respective systems of nonlinear PDEs in certain general/parametric forms. We use the concept of W-entropy \cite{perelman1}, which is useful for formulating his statistical and geometric thermodynamic model (see section 5 of that work, for Riemannian geometric flow; in this section, those results are generalized in relativistic form on Lorentz manifolds). Corresponding thermodynamic variables involve families of Ricci tensors and respective Ricci scalars defined by $\tau $-families of metrics $g_{\alpha \beta }(\tau ):=g_{\alpha \beta }(\tau ,u^{\gamma }),$ where $\tau $ is a positive flow parameter (treated as a conventional temperature). In abstract geometric form, we can consider families of pseudo-Riemannian metrics $g_{\alpha \beta }(\tau )$ and derive respective relativistic geometric flow equations and Perelman's thermodynamic variables which are important for elaborating various applications in GR and modern cosmology. This is possible if we apply the AFCDM, which allows us to decouple and solve in general off-diagonal forms respective systems of nonlinear PDEs. 

\subsection{Relativistic geometric flows of nonholonomic Einstein equations}

In this subsection, we elaborate on relativistic geometric flow models
formulated in canonical dyadic variables which describe, for self-similar
configurations, i. e. as Ricci solitons, the Einstein gravity theory. We
consider $\tau $-families of Lorentzian canonical geometric data 
\begin{equation}
(\mathbf{V,N}(\tau ),\mathbf{g}(\tau ),\widehat{\mathbf{D}}(\tau )=\nabla
(\tau )+\widehat{\mathbf{Z}}(\tau ),\widehat{\mathcal{L}}(\tau )=\ ^{g}%
\widehat{\mathcal{L}}(\tau )+\ ^{m}\widehat{\mathcal{L}}(\tau )),
\label{canongflodata}
\end{equation}%
which are parameterized by a real parameter $\tau ,0\leq \tau \leq \tau
_{1}. $ In this paper, $\tau $ is a temperature like parameter as in \cite%
{perelman1} but the families of d-metric $\mathbf{g}(\tau )$ are of
pseudo-Riemannian signatures. We work with $\widehat{\mathbf{D}}(\tau)$
instead of $\nabla (\tau )$ which allows us to apply the AFCDM and construct
off-diagonal solutions for various classes of geometric flow equations and
related gravitational models. The corresponding left labels are used in (\ref%
{canongflodata}) for respective families of gravitational fields, $g,$ for
mater fields, $m.$ The Lagrange densities can be defined in canonical or
non-canonical nonholonomic variables when the condition 
\begin{equation}
\ \ ^{g}\widehat{\mathcal{L}}(\tau )+\ ^{m}\widehat{\mathcal{L}}(\tau )=\
^{g}\mathcal{L}(\tau )+\ ^{m}\mathcal{L}(\tau )  \label{mlagd}
\end{equation}%
is necessary if we want to obtain the Einstein equations for certain $%
\tau=\tau _{0}=const.$ For any $\tau _{0},\ ^{g}\mathcal{L}$ and $\ ^{m}%
\mathcal{L}$ can be postulated as in GR \cite{hawrking73,misner73,wald82}
for various classes of relativistic matter fields.

The canonical distortion relation (\ref{canondistrel}) re-defines 
\begin{equation*}
\ ^{g}\mathcal{L}[\mathbf{g},\nabla ]\mathcal{\rightarrow \ }\ ^{g}\widehat{%
\mathcal{L}}[\mathbf{g,}\widehat{\mathbf{D}}]\mbox{ and }\ ^{m}\mathcal{L}[%
\mathbf{g},\nabla ,\ ^{m}\phi ]\mathcal{\rightarrow \ }\ ^{m}\widehat{%
\mathcal{L}}[\mathbf{g,}\widehat{\mathbf{D}},\ ^{m}\phi ]+\ ^{e}\widehat{%
\mathcal{L}}[\mathbf{g,}\widehat{\mathbf{D}},\widehat{\mathbf{Z}},\
^{m}\phi].
\end{equation*}%
In these formulas, $\ ^{m}\phi $ is a general symbol for the matter field
(with various tensor, spinor, group and other necessary type indices). The
effective Lagrange density $\ ^{e}\widehat{\mathcal{L}}$ contains
non-trivial contributions by coefficients of distortion d-tensor $\widehat{%
\mathbf{Z}}.$ We can consider nonholonomic distributions when $\ ^{e}%
\widehat{\mathcal{L}}=0$ if $\widehat{\mathbf{Z}}=0.$ The formulas (\ref%
{canongflodata}) and (\ref{mlagd}) can be postulated in such forms that (for
any $\tau _{0})$ using a respective N-adapted variational calculus on $%
\mathbf{V}$ we can derive the Einstein equations (\ref{cdeq1}) and (\ref%
{lccond1}) in canonical nonholonomic variables. In an equivalent form, we
can consider an abstract geometric formalism as in \cite{misner73,vacaru18}.
The main assumption in this work is that the nonholonomic frame structures
can be formulated in necessary N-adapted forms when geometric d-objects are
transformed into respective $\tau $-families of similar geometric and
physical objects as we considered above. Hereafter, we shall omit to write
an explicit dependence on $\tau $ if that will not result in ambiguities.

The main goal of this subsection is to elaborate on a model of relativistic
geometric flows of\textit{\ nonholonomic Einstein systems} (NES) (\ref%
{canongflodata}) and (\ref{mlagd}) which can be integrated in certain
general forms using the AFCDM. 

\subsubsection{Canonical nonholonomic deformations of F- and W-functionals}

An important condition for elaborating various types of geometric flow
theories and applications in modern physics consists in the definition and
generalization of the so-called $F$- and $W$-functionals 
\cite{perelman1,svnonh08,gheorghiuap16,vacaru18,nonassocFinslrev25} from
which geometric flow equations can be proved in variational N-adapted forms.

For relativistic geometric flows of NES we postulate such modified G.
Perelman's functionals: 
\begin{eqnarray}
\widehat{{F}}(\tau ) &=&\int_{t_{1}}^{t_{2}}\int_{\Xi _{t}}e^{-%
\widehat{\zeta }(\tau )}\sqrt{|\mathbf{g}(\tau )|}\delta ^{4}u[\widehat{%
\mathbf{R}}sc(\tau )+\ \widehat{\mathcal{L}}(\tau )+|\widehat{\mathbf{D}}%
(\tau )\widehat{\zeta }(\tau )|^{2}],  \label{fperelmNES} \\
\ \widehat{{W}}(\tau ) &=&\int_{t_{1}}^{t_{2}}\int_{\Xi _{t}}\left(
4\pi \tau \right) ^{-2}e^{-\widehat{\zeta }(\tau )}\sqrt{|\mathbf{g}(\tau )|}%
\delta ^{4}u[\tau (\widehat{\mathbf{R}}sc(\tau )+\widehat{\mathcal{L}}(\tau
)+|\widehat{\mathbf{D}}(\tau )\widehat{\zeta }(\tau )|^{2})+\widehat{\zeta }%
(\tau )-4].  \label{wfperelmNES}
\end{eqnarray}%
To elaborate geometric and physical models normalization conditions%
\begin{equation}
\int_{t_{1}}^{t_{2}}\int_{\Xi _{t}}\left( 4\pi \tau \right) ^{-2}e^{-%
\widehat{\zeta }(\tau )}\sqrt{|\mathbf{g}|}d^{4}u=1  \label{normcond}
\end{equation}%
are imposed, where $\widehat{\zeta }(\tau )=\widehat{\zeta }(\tau ,u)$ is a
family normalizing functions. In our approach, the F- and W- functionals are
postulated in hat variables to derive further geometric flow evolution
equations which can be decoupled and integrated in certain general forms. We
can extract geometric flows of LC configurations if we consider $\tau $%
-families of additional nonholonomic constraints (\ref{lccond1}) or (\ref%
{lccond}).

We can redefine the normalizing function, $\widehat{\zeta }(\tau)\rightarrow 
$ $\zeta (\tau ),$ from (\ref{fperelmNES}) and (\ref{wfperelmNES}), in
different forms which allow to generate, or absorb, respective distortions
of connections and work with respective functionals for other geometric data
labelled, for instance, ${F}(\tau ,\zeta ,\mathbf{g},\mathbf{D},%
\mathbf{R}sc,...)$ and ${W}(\tau ,\zeta ,\mathbf{g},\mathbf{D},%
\mathbf{R}sc,...).$ The difference of above nonholonomic functionals from
the original F- and W-functionals \cite{perelman1} introduced for 3-d
Riemannian $\tau $-flows $(g(\tau ),\nabla (\tau ))$ is that in this work we
study relativistic geometric flows and using canonical geometric data $(%
\mathbf{g}(\tau ),\mathbf{N}(\tau ),\widehat{\mathbf{D}}(\tau )).$ For
canonical nonholonomic geometric flow deformations of the Einstein systems
on nonholonomic Lorentz manifolds, this result in physically important
systems of nonlinear PDEs. Such $\tau $-flows of NES can be decoupled and
integrated in general forms using the same nonholonomic geometric methods as
in the previous section.

\subsubsection{Geometric flow equations and nonholonomic Ricci solitons as
Einstein spaces}

There are two possibilities to derive geometric flow equations from a
functional $\widehat{{F}}(\tau )$ or $\widehat{{W}}(\tau). $
In the first case \cite{svnonh08,gheorghiuap16,vacaru18,nonassocFinslrev25},
we can use $\widehat{\mathbf{D}}(\tau )$ instead of $\nabla (\tau )$ and
reproduce in N-adapted and distorted forms all covariant differential and
integral formulas from \cite{perelman1,monogrrf1,monogrrf2,monogrrf3}. Such
proofs can be written on hundred of pages and involve many problems for
extension to various types of MGT.

In the second case, we can use an abstract (nonholonomic) geometric
formalism \cite{misner73,vacaru18} involving necessary types of distortions.
In canonical nonholonomic dyadic variables, we postulate such
generalizations of the Hamilton-Friedan equations: 
\begin{eqnarray}
\partial _{\tau }g_{ij}(\tau ) &=&-2[\widehat{\mathbf{R}}_{ij}(\tau
)-\Upsilon _{ij}(\tau )];\ \partial _{\tau }g_{ab}(\tau )=-2[\widehat{%
\mathbf{R}}_{ab}(\tau )-\ \Upsilon _{ab}(\tau )];  \label{ricciflowr2} \\
\widehat{\mathbf{R}}_{ia}(\tau ) &=&\widehat{\mathbf{R}}_{ai}(\tau )=0;%
\widehat{\mathbf{R}}_{ij}(\tau )=\widehat{\mathbf{R}}_{ji}(\tau );\widehat{%
\mathbf{R}}_{ab}(\tau )=\widehat{\mathbf{R}}_{ba}(\tau );\   \notag \\
\partial _{\tau }\widehat{\zeta }(\tau ) &=&-\widehat{\square }(\tau )[%
\widehat{\zeta }(\tau )]+\left\vert \widehat{\mathbf{D}}(\tau )[\widehat{%
\zeta }(\tau )]\right\vert ^{2}-\ \widehat{R}sc(\tau )+\ \Upsilon _{\ \alpha
}^{\alpha }(\tau ).  \notag
\end{eqnarray}%
In these formulas, we use a $\tau $-family of effective sources as for (\ref%
{esourc}), when $\widehat{\mathbf{\Upsilon }}_{\ \ \beta }^{\alpha
}(\tau)=[\ ^{h}\Upsilon (\tau )\delta _{\ \ j}^{i},\ ^{v}\Upsilon (\tau
)\delta _{\ \ b}^{a}].$

In (\ref{ricciflowr2}), a generalized Laplace operator $\widehat{\square }%
(\tau)= \widehat{\mathbf{D}}^{\alpha }(\tau )\widehat{\mathbf{D}}_{\alpha
}(\tau )$ is used. Here we note that the conditions $\widehat{\mathbf{R}}%
_{ia}=\widehat{\mathbf{R}}_{ai}=0$ for the Ricci tensor $\widehat{R}ic[%
\widehat{\mathbf{D}}]=\{\widehat{\mathbf{R}}_{\alpha \beta }= [\widehat{R}%
_{ij},\widehat{R}_{ia},\widehat{R}_{ai},\widehat{R}_{ab,}]\}$ have to be
imposed if we want to keep the metrics $\mathbf{g}(\tau )$ to be symmetric.
In general, nonholonomic and nonmetric Ricci flow evolution scenarios can be
elaborated if such conditions are dropped, see a review in \cite{vacaru18}.

We can reproduce all solutions from the previous section by introducing a
formal dependence on and re-defining the effective sources in the form 
\begin{eqnarray}
\widehat{\mathbf{J}}(\tau ) &=&\widehat{\mathbf{\Upsilon }}(\tau )-\frac{1}{2%
}\partial _{\tau }\mathbf{g}(\tau )=[\ ^{h}\widehat{\mathbf{J}}(\tau ),\ ^{v}%
\widehat{\mathbf{J}}(\tau )]  \label{effrfs} \\
&=&[\ J_{i}(\tau )=\ \Upsilon _{i}(\tau )-\frac{1}{2}\partial _{\tau
}g_{i}(\tau ),\ \ J_{a}(\tau )= \Upsilon _{a}(\tau )-\frac{1}{2}\partial
_{\tau }g_{a}(\tau )],  \notag
\end{eqnarray}%
where $\mathbf{g}(\tau )=[g_{i}(\tau ),g_{a}(\tau ),N_{i}^{a}(\tau )]$ (\ref%
{dm}). The nonholonomic variables can be adapted in such forms that the
off-diagonal solutions of (\ref{cdeq1}) are transformed into $\tau $%
-depending solutions of (\ref{ricciflowr2}). For instance, $\tau $-families
of d-metrics , can be generated for ansatz 
\begin{equation}
g_{i}(\tau )=e^{\psi {(\tau ,x^{j})}},g_{a}(\tau )=h_{a}(\tau
,x^{k},y^{3}),\ N_{i}^{3}=w_{i}(\tau
,x^{k},y^{3}),\,\,\,\,N_{i}^{4}=n_{i}(\tau ,x^{k},y^{3})
\label{stationarydm}
\end{equation}%
and effective sources $\widehat{\mathbf{J}}(\tau )$ (\ref{effrfs}).

Nonholonomic Ricci solitons were defined and studied in \cite%
{svnonh08,gheorghiuap16,vacaru18} as self-similar configurations for the
corresponding nonholonomic geometric flow equations. Fixing $\tau =\tau _{0}$
in (\ref{ricciflowr2}), we obtain the equations for the nonholonomic
Einstein equations with effective sources (\ref{effrfs}) and (\ref{esourc}), 
\begin{align}
\widehat{\mathbf{R}}_{ij}& =\ ^{h}\widehat{\mathbf{J}}(\tau _{0},{x}^{k}),\ 
\widehat{\mathbf{R}}_{ab}=\ ^{v}\widehat{\mathbf{J}}(\tau _{0},x^{k},y^{c}),
\label{canriccisolda} \\
\widehat{\mathbf{R}}_{ia}& =\widehat{\mathbf{R}}_{ai}=0;\widehat{\mathbf{R}}%
_{ij}=\widehat{\mathbf{R}}_{ji};\widehat{\mathbf{R}}_{ab}=\widehat{\mathbf{R}%
}_{ba}.  \notag
\end{align}%
For additional constraints (\ref{lccond1}) or (\ref{lccond}) for zero
nonholonomic torsion, such equations transform into standard Einstein
equations $\nabla (\tau _{0}).$ 

\subsubsection{Generalizing G. Perelman's thermodynamics for the Einstein
gravity theory}

We can consider different types of normalization functions for the
functionals (\ref{fperelmNES}) and (\ref{wfperelmNES}), $\zeta
(\tau)\rightarrow \widehat{\zeta }(\tau ).$ For our purposes, we consider a $%
\widehat{\zeta }(\tau )$ defined by 
\begin{equation}
\partial _{\tau }\zeta (\tau )+\widehat{\square }(\tau )[\zeta (\tau
)]-\left\vert \widehat{\mathbf{D}}\zeta (\tau )\right\vert ^{2}-\ \widehat{%
\mathcal{L}}(\tau )=\partial _{\tau }\ \widehat{\zeta }(\tau )+\widehat{%
\square }(\tau )[\widehat{\zeta }(\tau )]-\left\vert \widehat{\mathbf{D}}%
\widehat{\zeta }(\tau )\right\vert ^{2}.  \label{normcondc1}
\end{equation}%
This is possible if we use (\ref{normcond}) and a $\widehat{\zeta }(\tau )$
as a solution of 
\begin{equation}
\partial _{\tau }\widehat{\zeta }(\tau )=-\widehat{\square }(\tau )[\widehat{%
\zeta }(\tau )]+\left\vert \widehat{\mathbf{D}}\widehat{\zeta }(\tau
)\right\vert ^{2}-\widehat{\mathbf{R}}sc(\tau ).  \label{normcondc}
\end{equation}%
Such normalization functions, $\zeta (\tau )$ or $\widehat{\zeta }(\tau ),$
define different integration measures in topological type theories. In our
approach with geometric evolution or gravitational and matter field
equations, we use different types of transforms $\zeta (\tau )\rightarrow 
\widehat{\zeta }(\tau )$ which allow us to absorb, or (inversely)
distinguish different types of (effective) $\tau $-running Lagrange
densities. This simplifies the geometric methods for finding exact and
parametric solutions. 

In this work, we elaborate on nonholonomic geometric flow evolution
scenarios of NES. In terms of an integration measure with a $\widehat{\zeta }%
(\tau )$ (\ref{normcondc1}), the W-functional (\ref{wfperelmNES}) can be
written in the form 
\begin{equation}
\ \widehat{{W}}(\tau )=\int_{t_{1}}^{t_{2}}\int_{\Xi _{t}}\left(4\pi
\tau \right) ^{-2}e^{-\widehat{\zeta }(\tau )}\sqrt{|\ \mathbf{g}(\tau)|}%
\delta ^{4}u[\tau (\widehat{\mathbf{R}}sc(\tau )+|\widehat{\mathbf{D}}(\tau )%
\widehat{\zeta }(\tau )|^{2}+\widehat{\zeta }(\tau )-4].  \label{wf1}
\end{equation}%
In this formula, the effective nonholonomic and matter fields sources $%
\widehat{\mathcal{L}}(\tau )$ can be encoded into geometric data. We have to
consider an explicit class of solutions of (\ref{ricciflowr2}) related for $%
\tau =\tau _{0}$ to solutions of the ENS equations (\ref{cdeq1}), or (\ref%
{canriccisolda}), and extended to (\ref{ricciflowr2}). The W-functional was introduced by G. Perelman as an “inverse entropy” for the (nonrelativistic) Ricci flows of Riemannian metrics and used for elaborating his statistical thermodynamic models in section 5 of [26], with an intuitive content related to statistical thermodynamics in subsection 5.1. In our works \cite{svnonh08,gheorghiuap16,partner06,vacaruplb16,vbubuianu17,vacaru18,
nonassocFinslrev25,bsvv24,v25c}, respective geometric constructions  were generalized for hat-variables which allows to introduce into respective F- and W-functionals the nonholonomic Ricci tensors and scalars determined by off-diagonal metrics and canonical d-connections with $\tau$ – evolution of exact and parametric solutions for ENS. 

On a nonholonomic Lorentz manifold $\mathbf{V}$ a canonical nonholonomic 2+2
decomposition is essential for generating off-diagonal solutions as involved in hat-variables. To
elaborate on thermodynamical models, we must consider an additional
nonholonomic (3+1) splitting. Such double nonholonomic spacetime fibrations
allow us to introduce such a statistical partition function: 
\begin{equation}
\ \widehat{Z}(\tau )=\exp [{\int_{\widehat{\Xi }}[-\widehat{\zeta }+2]\
\left( 4\pi \tau \right) ^{-2}e^{-\widehat{\zeta }}\ \delta \widehat{%
\mathcal{V}}(\tau )}],  \label{spf}
\end{equation}%
where the volume element is defined and computed as 
\begin{equation}
\delta \widehat{V}(\tau )=\sqrt{|\mathbf{g}(\tau )|}\ dx^{1}dx^{2}\delta
y^{3}\delta y^{4}\ .  \label{volume}
\end{equation}
If we consider Riemannian signatures of metrics and LC-configurations, these formulas transforms into 4-d analogs of the values considered in sections 5.1 and 5.2 of  \cite{perelman1}. In this work, the constructions are performed in relativistic form which allows to introduce into the scheme off-diagonal solutions. $\widehat{Z}(\tau )$ can be used as an analog of  "free Gibbs energy", but defined in geometric variables on (pseudo) Riemannian manifolds.

Using $\widehat{Z}(\tau )$ (\ref{spf}), we can modify the variational procedure provided in section 5 of \cite{perelman1}) (in original form, for geometric flows of Riemannian metrics, but all constructions can be performed in abstract geometric form or in N-adapted variations). We have also to canonical nonholonomic deformations for $\widehat{{W}}(\tau )$ (\ref{wf1}) on a closed region in $\mathbf{V}.$ Such abstract geometric calculations allow us to define and compute nonholonomic canonical distortions of G. Perelman thermodynamic variables: 
\begin{align}
\ \widehat{{E}}\ (\tau )& =-\tau ^{2}\int_{\widehat{\Xi }}\ \left(
4\pi \tau \right) ^{-2}\left( \widehat{\mathbf{R}}sc+|\ \widehat{\mathbf{D}}%
\ \widehat{\zeta }|^{2}-\frac{2}{\tau }\right) e^{-\widehat{\zeta }}\ \delta 
\widehat{V}(\tau ),  \label{qthermvar} \\
\ \ \ \widehat{S}(\tau )& =-\ \widehat{{W}}(\tau )=-\int_{\widehat{%
\Xi }}\left( 4\pi \tau \right) ^{-2}\left( \tau (\widehat{\mathbf{R}}sc+|%
\widehat{\mathbf{D}}\widehat{\zeta }|^{2})+\widehat{\zeta }-4\right) e^{-%
\widehat{\zeta }}\ \delta \widehat{V}(\tau ),  \notag \\
\ \ \ \widehat{\sigma }(\tau )& =2\ \tau ^{4}\int_{\widehat{\Xi }}\left(
4\pi \tau \right) ^{-2}|\ \widehat{\mathbf{R}}_{\alpha \beta }+\widehat{%
\mathbf{D}}_{\alpha }\ \widehat{\mathbf{D}}_{\beta }\widehat{\zeta }_{[1]}-%
\frac{1}{2\tau }\mathbf{g}_{\alpha \beta }|^{2}e^{-\widehat{\zeta }}\ \delta 
\widehat{V}(\tau ).  \notag
\end{align}%
The nonholonomic geometric thermodynamic variables (\ref{qthermvar}) can be
considered for characterizing different classes of physically important
solutions in relativistic geometric flow and the Einstein gravity theories.
The quadratic fluctuation thermodynamic variable $\widehat{\sigma }(\tau )$
can be written as a functional of $\widehat{\mathbf{R}}_{\alpha \beta }$
even $\widehat{{E}}\ (\tau )$ and $\widehat{S}(\tau )$ are
functionals of $\widehat{\mathbf{R}}sc$ if we correspondingly re-define the
normalizing functions $\widehat{\zeta }\rightarrow \widehat{\zeta }_{[1]}$.
We omit such details in this work because we shall not compute $\widehat{%
\sigma }(\tau )$ for certain classes of solutions.

 So, we conclude that the nonholonomic G. Perelman thermodynamic variables (\ref{qthermvar})  can be computed in explicit form for any solution of systems of nonlinear PDEs of type (\ref{cdeq1}), or (\ref{canriccisolda}), and (\ref{ricciflowr2}). The procedure of computing such physical values is simplified substantially if we consider solutions with nonlinear symmetries and $\tau $-families of generating functions $\Phi (\tau )$ resulting in 
 $\tau $-evolution of nonholonomic Einstein equations  (\ref{cdeq1a}). For such models, certain $\tau $-running effective cosmological constants $\Lambda (\tau )$ are introduced by considering respective nonlinear symmetries. They can be determined in a form to be compatible with observational data in accelerating cosmology and DE and DM physics. Let us speculate about this. The conservative view is that physical importance in GR has only some classes solutions generated by diagonal ansatz with spherical/ cylindric symmetries. Typically, these are certain BH, WH, of FLRW metrics, found after the Einstein equations were transformed into certain systems of nonlinear ODEs. 
  
 The AFCDM allows us to construct general off-diagonal solutions depending, in principle, on all spacetime coordinates. This can be done in GR and various MGTs as we explained in previous sections. Such classes of solutions were considered for many years as unphysical  and not supported by experimental/ observational data in cosmology and astrophysics. The situation changed substantially in accelerating cosmology when various filament, pattern forming, locally anisotropic and other type structures were observed. We elaborated on such quasi-crystal, quasi-periodic, solitonic scenarios in \cite{vacaruplb16,vbubuianu17,vacaru18}. A part of the scientific community preferred to modify the Einstein gravity in various forms (elaborating hundreds of MGTs with various classes of BH, WHs, $\Lambda$CDM scenarios) instead of considering new classes of off-diagonal solutions in GR, which may used for modelling DE and DM effects. As we proved in previous sections, see also \cite{v25c}, the AFCDM can be applied for  solving nonlinear  PDEs in general off-diagonal forms, but the  Bekenstein-Hawking thermodynamic paradigm \cite{bek2,haw2} is not applicable to study physical properties of such solutions. In our approach, the Ricci flow theory can be formulated in relativistic form as a thermo-field gravitational theory with dependence on a temperature-like $\tau$ parameter and when a statistical thermodynamic model (based on the concept of Perelman's W-entropy) can be elaborated. Corresponding locally anisotropic thermodynamic phase space models defining phase space transitions can be elaborated and restricted in nonholonomic form for LC-configurations, in particular, in GR (see \cite{vacaruplb16,vbubuianu17,vacaru18} and references therein). For such nonlinear models with  off-diagonal interactions or geometric evolution, we can study a new DE and DM physics etc.

Finally, in this subsection, we emphasize that G. Perelman's thermodynamic paradigm (\ref{qthermvar}) is different from that of Bekenstein-Hawking thermodynamics \cite{bek2,haw2}. The second approach is motivated for certain special classes of solutions in GR which possess respective hypersurface horizons, duality, or holographic conditions. We have to consider relativistic generalizations of G. Perelman thermodynamics of Ricci flows and connect such constructions to the AFCDM if we want understand the physical importance of many classes of generic off-diagonal solutions in GR or MGTs as we motivated in \cite{gheorghiuap16,partner06,vacaruplb16,
vbubuianu17,vacaru18,nonassocFinslrev25,bsvv24,v25c}.


\subsubsection{Computing thermodynamic variables for quasi-stationary NES}

Let us consider $\tau $-families of quasi-stationary d-metrics $\mathbf{g}%
_{\alpha }[\Phi (\tau )]\simeq \ \mathbf{g}_{\alpha }[\eta _{4}(\tau )],$
which for any fixed $\tau _{0}$ define off-diagonal solutions of type (\ref%
{offdiagcosmcsh}) or (\ref{offdiagpolfr}). The formulas for nonlinear
symmetries (\ref{nonlintrsmalp}) and (\ref{nonlinsymrex}) are generalized
for $\tau $-parametric dependencies when, for instance, $\widehat{\mathbf{%
\Upsilon }}_{\ \ \beta }^{\alpha }(\tau )=[\ ^{h}\Upsilon (\tau ),\
^{v}\Upsilon (\tau )]$ (\ref{esourc}) are substituted by $\widehat{\mathbf{J}%
}(\tau )=[\ ^{h}\widehat{\mathbf{J}}(\tau ),\ ^{v}\widehat{\mathbf{J}}(\tau
)]$ (\ref{effrfs}); the gravitational polarization $\eta _{4}(\tau )$ is
extended to $\eta _{4}(\tau ,x^{i},y^{3})=\eta _{4}(\tau ,x^{i},y^{3});$ and 
$\Lambda $ is changed into $\Lambda (\tau ).$ Such assumptions allow us to
express the nonholonomic geometric flow equations (\ref{ricciflowr2}) as a $%
\tau $-family of nonholonomic Einstein equations (\ref{cdeq1a}). We write
them in such a functional form: 
\begin{equation}
\widehat{\mathbf{R}}_{\ \ \beta }^{\alpha }[\Phi (\tau ),\widehat{\mathbf{J}}%
(\tau )]=\Lambda (\tau )\mathbf{\delta }_{\ \ \beta }^{\alpha }.
\label{cdeq1b}
\end{equation}

For the nonholonomic canonical geometric data considered for the system of
nonlinear PDEs (\ref{cdeq1b}), we can compute the canonical thermodynamic
variables $\widehat{{E}}(\tau )$ and $\widehat{{S}}(\tau )$ (%
\ref{qthermvar}) by expressing $\widehat{\mathbf{R}}sc(\tau )=4\Lambda (\tau
).$ We do not present in this work more cumbersome technical results for
computing $\widehat{\sigma }(\tau )$ involving the canonical Ricci d-tensor.
Such assumptions allow us to write the statistical partition function and
thermodynamic variables in the form: 
\begin{eqnarray}
\ ^{q}\widehat{Z}(\tau ) &=&\exp [\int_{\widehat{\Xi }}\frac{1}{8\left( \pi
\tau \right) ^{2}}\ \delta \ ^{q}\mathcal{V}(\tau )],\ ^{q}\widehat{E}\ (\tau )=-\tau ^{2}\int_{\widehat{\Xi }}\ \frac{1}{8\left( \pi \tau
\right) ^{2}}[2\Lambda (\tau )-\frac{1}{\tau }]\ \delta \ ^{q}\mathcal{V}%
(\tau ),  \notag \\
\ ^{q}\widehat{S}(\tau ) &=&-\ \ ^{q}\widehat{W}(\tau )=-\int_{\widehat{\Xi }%
}\frac{1}{4\left( \pi \tau \right) ^{2}}[\tau \Lambda (\tau )\ -1]\delta \
^{q}\mathcal{V}(\tau ).  \label{thermvar1}
\end{eqnarray}%
Here we note that the volume element (\ref{volume}) $\delta \ ^{q}\mathcal{V}
(\tau )= \sqrt{|\ ^{q}\mathbf{g}(\tau )|}\ dx^{1}dx^{2}\delta y^{3}\delta
y^{4}$ (the label $q$ is used for quasi-stationary $\tau $-running
configurations) is determined by a respective class of quasi-stationary
solutions. The formulas and computations simplify substantially if we chose
such frame/ coordinate systems when the normalizing functions have the
properties $\widehat{\mathbf{D}}_{\alpha }\ {\widehat{\zeta }}=0$ and ${%
\widehat{\zeta }}\approx 0.$ To simplify further computations and all
formulas in (\ref{thermvar1}) we can consider trivial integration functions $%
\ _{1}n_{k}=0$ and $\ _{2}n_{k}=0$ (such conditions change for arbitrary
frame and coordinate transforms). 

Using the formulas (\ref{nonlinsymrex}) for nonlinear symmetries, we express 
\begin{equation}
\ \Phi (\tau )=2\sqrt{|\ \Lambda (\tau )\ g_{4}(\tau )|}=\ 2\sqrt{|\ \Lambda
(\tau )\ \eta _{4}(\tau )\ \mathring{g}_{4}(\tau )|}\simeq 2\sqrt{|\ \Lambda
(\tau )\ \zeta _{4}(\tau )\ \mathring{g}_{4}|}[1-\frac{\varepsilon }{2}\chi
_{4}(\tau )],  \notag
\end{equation}%
for $\ [\ \Psi (\tau ),\ ^{v}\widehat{\mathbf{J}}(\tau )]\rightarrow \lbrack
\Phi (\tau ),\Lambda (\tau )].$ This allows us to compute and write 
\begin{eqnarray*}
\ \delta \ \ ^{q}\mathcal{V} &=&\delta \mathcal{V}[\tau ,\ \widehat{\mathbf{J%
}}(\tau ),\ \Lambda (\tau );\psi (\tau ),\ g_{4}(\tau )]=\delta \mathcal{V}%
(\ \widehat{\mathbf{J}}(\tau ),_{Q}\Lambda (\tau ),\eta _{4}(\tau )\ 
\mathring{g}_{4}) \\
&=&\frac{1}{|\ \Lambda (\tau )|}\ \delta \ _{\eta }\mathcal{V},\mbox{ where }%
\ \delta \ _{\eta }\mathcal{V}=\ \delta \ _{\eta }^{h}\mathcal{V}\times
\delta \ _{\eta }^{v}\mathcal{V}, \mbox{ for }
\end{eqnarray*}%
\begin{eqnarray}
\delta \ _{\eta }^{h}\mathcal{V} &=&\delta \ _{\eta }^{h}\mathcal{V}[\ ^{h}%
\widehat{\mathbf{J}}(\tau ),\eta _{1}(\tau )\ \mathring{g}_{1}]
\label{volumfuncts} \\
&=&e^{\widetilde{\psi }(\tau )}dx^{1}dx^{2}=\sqrt{|\ ^{h}\widehat{\mathbf{J}}%
(\tau )|}e^{\psi (\tau )}dx^{1}dx^{2},\mbox{ for }\psi (\tau )%
\mbox{ being a
solution of  }(\ref{eq1}),\mbox{ with sources }\ ^{h}\widehat{\mathbf{J}}%
(\tau );  \notag \\
\delta \ _{\eta }^{v}\mathcal{V} &=&\delta \ _{\eta }^{v}\mathcal{V}[\ ^{v}%
\widehat{\mathbf{J}}(\tau ),\eta _{4}(\tau ),\ \mathring{g}_{4}]  \notag \\
&=&\frac{\partial _{3}|\ \eta _{4}(\tau )\ \mathring{g}_{4}|^{3/2}}{\ \sqrt{%
|\int dy^{3}\ \ _{{}}^{v}\widehat{\mathbf{J}}(\tau )\{\partial _{3}|\ \eta
_{4}(\tau )\ \mathring{g}_{4}|\}^{2}|}}[dy^{3}+\frac{\partial _{i}\left(
\int dy^{3}\ \ ^{v}\widehat{\mathbf{J}}(\tau )\partial _{3}|\ \eta _{4}(\tau
)\ \mathring{g}_{4}|\right) dx^{i}}{\ \ \ ^{v}\widehat{\mathbf{J}}(\tau
)\partial _{3}|\ \eta _{4}(\tau )\mathring{g}_{4}|}]dt.  \notag
\end{eqnarray}%
Integrating such products of forms from (\ref{volumfuncts}) on a closed
hypersurface $\widehat{\Xi },$ we obtain a running spacetime volume
functional 
\begin{equation}
\ _{\eta }^{\mathbf{J}}\mathcal{V[}\ ^{q}\mathbf{g}(\tau )]=\int_{\ \widehat{%
\Xi }}\delta \ _{\eta }\mathcal{V}(\ ^{v}\widehat{\mathbf{J}}(\tau ),\ \eta
_{\alpha }(\tau ),\mathring{g}_{\alpha }).  \label{volumf1}
\end{equation}%
This functional depends on the type of quasi-stationary solutions and
geometric data on $\widehat{\Xi };$ on effective sources for NES; primary
d-metrics and generating functions $\eta _{4}(\tau )$ and effective
cosmological constants $\Lambda (\tau ).$ It is not possible to compute $\
_{\eta }^{\shortmid }\mathcal{\mathring{V}[}\ ^{q}\mathbf{g}(\tau )]$ in a
general form for all classes of quasi-stationary solutions. Nevertheless, we
can separate in the thermodynamic variables the terms depending only on $%
\Lambda (\tau )$ and some coefficients depending only on temperature $\tau $%
. This can be used for analyzing possible thermodynamic implications, for
instance, of different types of DE models with $\tau $-running cosmological
constants, or two alternative DE configurations with different $\Lambda $%
-constants. 

Using the volume functional (\ref{volumf1}), we obtain such formulas for
nonholonomic thermodynamic variables (\ref{thermvar1}): 
\begin{eqnarray}
\ ^{q}\widehat{Z}(\tau ) &=&\exp \left[ \frac{1}{8\pi ^{2}\tau ^{2}}\ \
_{\eta }^{\mathbf{J}}\mathcal{V}[\ ^{q}\mathbf{g}(\tau )]\right] ,\ ^{q}%
\widehat{\mathcal{E}}\ (\tau )=\ \frac{1-2\tau \ \Lambda (\tau )}{8\pi
^{2}\tau }\ \ \ _{\eta }^{\mathbf{J}}\mathcal{V[}\ ^{q}\mathbf{g}(\tau )],
\label{thermvar2} \\
\ \ \ \ ^{q}\widehat{S}(\tau ) &=&-\ ^{q}\widehat{W}(\tau )=\frac{1-\Lambda
(\tau )}{4\pi ^{2}\tau ^{2}}\ \ _{\eta }^{\mathbf{J}}\mathcal{V}[\ ^{q}%
\mathbf{g}(\tau )].  \notag
\end{eqnarray}

Finally, we emphasize that for $\tau =\tau _{0},$ the formulas (\ref%
{thermvar2}) can be used for defining thermodynamic characteristics of
nonholonomic Ricci soliton quasi-stationary configurations, i.e.
off-diagonal quasi-stationary solutions of nonholonomic Einstein equations,
stated by (\ref{canriccisolda}). All formulas derived in the above
subsections, can be re-defined for $\tau $-families of locally cosmological
solutions (\ref{dmc}) with underlined symbols of the geometric and physical
d-objects. Respective nonlinear symmetries and duality properties are
written in formulas with underlined symbols as in (\ref{dualnonltr}) and (%
\ref{dualcosm}). 

\subsection{Geometric flow and Ricci soliton thermodynamics for off-diagonal
solutions in GR}

In this subsection, we show how to compute in explicit form G. Perelman's
thermodynamic variables for four classes of physically important
off-diagonal solutions constructed in section \ref{sec03}. Only in a special
case of rotoid deformations of KdS BHs (for instance, with an ellipsoid
generating function (\ref{rotoid})), we can introduce hypersurface
(ellipsoid type) configurations. This allows us to apply the
Bekenstein-Hawking thermodynamic paradigm \cite{bek2,haw2}. Such examples
are studied and reviewed in \cite{vacaru18,nonassocFinslrev25}. For more
general off-diagonal deformations of KdS BHs, WHs, BTs and locally
anisotropic cosmological solutions, respective thermodynamic
characterizations are possible if we consider a relativistic generalization
of the concept of W-entropy \cite%
{perelman1,svnonh08,gheorghiuap16,vacaru18,nonassocFinslrev25}.

\subsubsection{Perelman thermodynamic variables for general off-diagonal
deformed KdS BHs}

Classes of $\tau $-running nonholonomically deformed KdS BH configurations
do not possess closed horizons and do not involve any duality/ holographic
properties. To characterize the physical properties of corresponding
off-diagonal we have to change the thermodynamic paradigm and (for
respective quasi-stationary configurations) use thermodynamic variables of
type (\ref{thermvar2}). In explicit form, we have to compute the respective
volume forms $\ _{\eta }^{\mathbf{J}}\mathcal{V}[\ ^{q}\mathbf{g}(\tau )]$ (%
\ref{volumf1}) when $\ ^{q}\mathbf{g}(\tau )$ is defined by corresponding $%
\tau $-families of quasi-stationary solutions. We can consider two geometric
flow thermodynamic models:

\begin{enumerate}
\item[{a]}] $\ ^{q}\mathbf{g}(\tau )\rightarrow \ ^{KdS}\mathbf{g}(\tau ),$
when the gravitational $\eta $-polarizations are defined by $\tau $-running
d-metrics of type (\ref{nkernew}) when formulas $\mathbf{\breve{\Upsilon}}%
_{\ \ \beta }^{\alpha }(r,\varphi ,\theta )=[\ ^{h}\Upsilon ,\ ^{v}\Upsilon ]
$ are substituted by $\mathbf{\breve{J}}(\tau ,r,\varphi ,\theta )=[\ ^{h}%
\mathbf{\breve{J}}(\tau ), \ ^{v}\mathbf{\breve{J}}(\tau )]$ as in (\ref%
{effrfs}). Such values are used in the formulas for nonlinear symmetries (%
\ref{nlims2}), when $\tilde{\Lambda}\simeq \Lambda _{0},$ which can be
considered by re-defining the generating functions. The corresponding volume
form is written as a functional $\ _{\eta }^{J}\mathcal{V}[\ \ ^{KdS}\mathbf{%
g}(\tau )]$ of type (\ref{volumf1}).

\item[{b]}] $\ ^{q}\mathbf{g}(\tau )\rightarrow \ _{\kappa }^{KdS}\mathbf{g}%
(\tau ),$ when the gravitational $\chi $-polarizations are defined by $\tau $%
-running d-metrics of type (\ref{offdnceleps1}) involving a small parameter $%
\kappa .$The effective sources and nonlinear symmetries are $\tau $-extended
as in a]. The corresponding volume functional  $\ _{\chi }^{J}\mathcal{V}[\
_{\kappa }^{KdS}\mathbf{g}(\tau )]$ can be computed in $\kappa $-parametric
form.
\end{enumerate}

For general $\eta $-polarizations as in a], the thermodynamic variables (\ref%
{thermvar2}) are defined and computed: 
\begin{eqnarray}
\ \ _{\eta }^{J}\widehat{Z}(\tau ) &=&\exp \left[ \frac{1}{8\pi ^{2}\tau ^{2}%
}\ \ _{\eta }^{J}\mathcal{V[}\ \ ^{KdS}\mathbf{g}(\tau )]\right] ,\ \ _{\eta
}^{J}\widehat{{E}}\ (\tau )=\ \frac{1-2\tau \ \Lambda (\tau )}{8\pi
^{2}\tau }\ \ _{\eta }^{J}\mathcal{V[}\ \ ^{KdS}\mathbf{g}(\tau )],  \notag
\\
\ \ \ \ \ _{\eta }^{J}\widehat{S}(\tau ) &=&-\ \ _{\eta }^{J}\widehat{W}%
(\tau )=\frac{1-\Lambda (\tau )}{4\pi ^{2}\tau ^{2}}\ _{\eta }^{J}\mathcal{V[%
}\ \ ^{KdS}\mathbf{g}(\tau )]\ .  \label{thermvarkds}
\end{eqnarray}

In a similar form, using the variant b], we can compute $\kappa $-parametric
decompositions of (\ref{thermvarkds}) with a respective changing of the left
labels. For instance, we have 
\begin{equation}
..,\ \ _{\chi }^{J}\widehat{S}(\tau )=-\ \ _{\chi }^{J}\widehat{W}(\tau )=%
\frac{1-\Lambda (\tau )}{4\pi ^{2}\tau ^{2}}\ _{\chi }^{J}\mathcal{V[}\ \
_{\kappa }^{KdS}\mathbf{g}(\tau )]\ .  \label{thermvarkdsk}
\end{equation}

Both types of quasi-stationary configurations (\ref{thermvarkds}) and (\ref%
{thermvarkdsk}) are described by a similar behaviour under a $\tau $-running
cosmological constant $\Lambda (\tau ).$ So, such configurations can exist
in an off-diagonal gravitational background determined by the DE
distribution with evolution on effective temperature $\tau .$ For a fixed $%
\tau _{0},$ we obtain thermodynamic models of certain nonholonomic Ricci
soliton configurations. The class a] of thermodynamic models (\ref%
{thermvarkds}) is appropriate for describing general $\eta $-deformations
(for instance, for certain solitionic hierarchies) of KdS BHs. In such
models, a BH can be stable if certain stability conditions are satisfied
(see as a review \cite{vacaru18}), or moved away by some nonlinear waves. A
KdS BH can be "anihilated", i.e. transformed in another type of
quasi-stationary solution which can be more convenient for a lower entropy $%
\ _{\eta }^{J}\widehat{S}(\tau )$ from (\ref{thermvarkds}). In a different
form, we can elaborate on a scenarios by creating KdS BHs using $\eta $%
-deformations under $\tau $-evolution. All such geometric thermodynamic
models depend on the type of $\ _{\eta }^{J}\mathcal{V}[ \ ^{KdS}\mathbf{g}%
(\tau )]$ which can be computed in explicit form by prescribing
corresponding classes of generating functions and (effective) sources and
integration functions.

For small $\kappa $-parametric deformations of KdS BHs, we can speculate on $%
\chi $-polarizations of the gravitational vacuum or of certain $\chi $%
-polarizations of physical constants. In more special cases, we can generate
rotoid deformations of horizons (\ref{rotoid}) and consider alternatively
the Bekenstein-Hawking thermodynamics. The G. Perelman thermodynamic
variables (\ref{thermvarkdsk}) can be defined and computed in all cases.
This is possible even a KdS BHs can be unstable and "slow dissipate" into
another types of quasi-stationary solutions with less known physical
properties. In many cases, we can prescribe certain nonholonomic
distributions when BHs transform into BEs, with some polarized horizons and
effective constants. In such cases, the physical interpretation of
off-diagonal target solutions is quite similar to the prime BH ones.
Corresponding relativistic geometric flow, or nonholononmic Ricci flow
models depend on the type of volume form $\ _{\chi }^{J}\mathcal{V[}\ \
_{\kappa }^{KdS}\mathbf{g}(\tau )]$ we define for respective generating and
integrating data. 

\subsubsection{Perelman thermodynamic variables for off-diagonal deformed WHs%
}

The quasi-stationary Ricci flow thermodynamic variables (\ref{thermvar2})
can be computed in similar forms for nonholonomic deformations of WHs, for
instance, of type (\ref{whpolf}) or (\ref{whpolf1}). We can follow similar
steps a] or b] as described above in the previous subsection.

For $\tau $-families of nonholonomic distorted WHs, we consider $\ ^{q}%
\mathbf{g}(\tau )\rightarrow \ ^{wh}\mathbf{g}(\tau ),$ when the
gravitational $\eta $-polarizations are defined by $\tau $-running d-metrics
of type (\ref{whpolf}). The effective sources are re-defined in a form when $%
\ ^{wh}\Upsilon _{\ \ \beta }^{\alpha }(r,\varphi ,\theta )=[\
_{h}^{wh}\Upsilon ,\ _{v}^{wh}\Upsilon ]$ are substituted by $\ ^{wh}\mathbf{%
J}(\tau ,r,\varphi ,\theta )=[\ _{h}^{wh}\mathbf{J}(\tau ),\ _{v}^{wh}%
\mathbf{J}(\tau )]$ as in (\ref{effrfs}). The formulas for nonlinear
symmetries (\ref{nlims2}) are modified respectively for $\Lambda (\tau ),$
and the volume form (\ref{volumf1}) is labelled in abstract geometric form
as $\ _{\eta }^{J}\mathcal{V}[ \ ^{wh}\mathbf{g}(\tau )].$ This allows to
define and compute for $\tau $-running WH configurations such relativistic
forms of G. Perelman's variables: 
\begin{eqnarray}
\ \ _{\eta }^{J}\widehat{Z}(\tau ) &=&\exp \left[ \frac{1}{8\pi ^{2}\tau ^{2}%
}\ \ _{\eta }^{J}\mathcal{V[}\ \ ^{wh}\mathbf{g}(\tau )]\right] ,\ \ _{\eta
}^{J}\widehat{{E}}\ (\tau )=\ \frac{1-2\tau \ \Lambda (\tau )}{8\pi
^{2}\tau }\ \ _{\eta }^{J}\mathcal{V[}\ \ ^{wh}\mathbf{g}(\tau )],  \notag \\
\ \ \ \ \ _{\eta }^{J}\widehat{S}(\tau ) &=&-\ \ _{\eta }^{J}\widehat{W}%
(\tau )=\frac{1-\Lambda (\tau )}{4\pi ^{2}\tau ^{2}}\ _{\eta }^{J}\mathcal{V[%
}\ \ ^{wh}\mathbf{g}(\tau )]\ .  \label{thermvarwh}
\end{eqnarray}%
These thermodynamic variables can be modified to describe $\tau $-families
of $\kappa $-parametric deformations of prime WH solutions in GR, defined by 
$\chi $-polarizations as in (\ref{whpolf1}). For instance, the entropy
functional from (\ref{thermvarwh}) is written%
\begin{equation}
...,\ _{\chi }^{J}\widehat{S}(\tau )=-\ \ _{\chi }^{J}\widehat{W}(\tau )=%
\frac{1-\Lambda (\tau )}{4\pi ^{2}\tau ^{2}}\ _{\chi }^{J}\mathcal{V[}\ \
_{\kappa }^{wh}\mathbf{g}(\tau )].  \label{thermvarwhk}
\end{equation}

The formulas (\ref{thermvarwh}) \ and (\ref{thermvarwhk}) for off-diagonal
modifications of WHs in GR are respective analogues of (\ref{thermvarkds})
and (\ref{thermvarkdsk}) for off-diagonal deformed KdS BHs. In a general
form, (\ref{thermvarwh}) is not for an "off-diagonal" WH configuration. It
may describe (for certain nonholonomic distributions) some WHs which are
self-consistently embedded in a nontrivial off-diagonal quasi-stationary
gravitational vacuum. For other type conditions, WHs "dissipate"
off-diagonally into a quasi-stationary gravitational vacuum. Such a vacuum
can be modeled as certain solitonic configurations etc. We can transform WHs
into some "spaghetti" configurations or other types off-diagonal
quasi-stationary solutions with less clear physical interpretation.
Inversely, we can elaborate on creating nonholonomic deformed WHs from a
quasi-stationary gravitational vacuum when respective (effective) matter
satisfies necessary type conditions. All such models depend on the type of
generating and integration data are prescribed in $\ _{\eta}^{J}\mathcal{V}%
[\ ^{wh}\mathbf{g}(\tau )]$ and which types of nonlinear symmetries we
consider to relate $\ ^{wh}\mathbf{J}$ to a $\tau $-running effective
cosmological constant $\Lambda (\tau ).$ Such $\tau $-families, or with a
fixed $\tau _{0},$ can be considered for modelling certain DE configurations
where WHs are embedded, or supposed to dissipate.

The $\kappa $-parametric models with geometric entropy $\ _{\chi }^{J}%
\widehat{S}(\tau )$ (\ref{thermvarkdsk}) allow us to construct off-diagonal
quasi-stationary metrics which really describe WH configurations. For
instance, such nonholonomic WHs can be with "small" polarization of physical
constants (in particular, with rotoid-type throats etc.). Certain
off-diagonal deformations may open, or close, certain WH throats. We can
impose LC conditions and generate such locally anisotropic WHs in the
framework of GR. They can be characterized thermodynamically using G.
Perelman's W-entropy but not in the frameworks of Bekenstein-Hawking
paradigm.

\subsubsection{Relativistic Ricci flow thermodynamic variables for
nonholonomic BTs}

The geometric thermodynamic models studied in the previous sections for
quasi-stationary off-diagonal generalizations of KdS BH and WH solutions can
be re-defined nonholonomically in an abstract geometric language to describe
physically important properties of $\tau $-families of nonholonomic BT
solutions (\ref{tortpol2}). Such solutions were constructed in $\kappa $%
-parametric form for nonlinear symmetries (\ref{nsymtor}) and respective
nonholonmic Ricci solitons. For relativistic flows, such flows are defined
for $\tau $-families of nonholonomic distorted BTs, we consider $\ ^{q}%
\mathbf{g}(\tau )\rightarrow \ _{\kappa }^{tor}\mathbf{g}(\tau ),$ with
respective gravitational $\chi $-polarizations. The effective sources are
re-defined in a form when $\ ^{tor}\Upsilon _{\ \ \beta }^{\alpha }=[\
_{h}^{tor}\Upsilon , \ _{v}^{tor}\Upsilon ]$ are substituted by $\ ^{tor}%
\mathbf{J}(\tau )=[\ _{h}^{tor}\mathbf{J}(\tau ),\ _{v}^{tor}\mathbf{J}%
(\tau)]$ as in (\ref{effrfs}). The $\tau $-running cosmological constants
are parameterized into $\widetilde{\Lambda }(\tau )=\Lambda (\tau )+\
^{tor}\Lambda (\tau ).$

The quasi-stationary Ricci flow thermodynamic variables (\ref{thermvar2})
computed for $\tau $-families of $\kappa $-parametric deformations of BTs
can be written in such a form: 
\begin{eqnarray}
\ \ _{\chi }^{J}\widehat{Z}(\tau ) &=&\exp \left[ \frac{1}{8\pi ^{2}\tau ^{2}%
}\ \ _{\chi }^{J}\mathcal{V[}\ \ _{\kappa }^{tor}\mathbf{g}(\tau )]\right]
,\ \ _{\chi }^{J}\widehat{{E}}\ (\tau )=\ \frac{1-2\tau \ \Lambda
(\tau )}{8\pi ^{2}\tau }\ \ _{\chi }^{J}\mathcal{V[}\ \ _{\kappa }^{tor}%
\mathbf{g}(\tau )],  \notag \\
\ \ \ \ \ _{\chi }^{J}\widehat{S}(\tau ) &=&-\ \ _{\chi }^{J}\widehat{W}%
(\tau )=\frac{1-\Lambda (\tau )}{4\pi ^{2}\tau ^{2}}\ _{\chi }^{J}\mathcal{V[%
}\ \ _{\kappa }^{tor}\mathbf{g}(\tau )]\ .  \label{thermvartork}
\end{eqnarray}%
These geometric thermodynamic formulas characterize small parametric
deformations of BT solutions in GR. Such objects can be defined by
off-diagonal solutions of nonholonomic Ricci solitons if $\tau _{0}.$ If the
target d-metrics involve certain $\chi $-polarizations of physical
constants, we can choose such nonholonomic distributions which positively
keep a BT character. We can similarly generate quasi-stationary solutions
for BT and BE systems, see \cite{vacaru18,nonassocFinslrev25} and references
therein.

\subsubsection{Geometric flow thermodynamics for off-diagonal cosmological
solutions}

Locally anisotropic cosmological solutions in GR defined by off-diagonal
metrics with conventional underlined coefficients can be generated as we
summarized in Table 3 from Appendix \ref{appendixb}. A typical ansatz of
d-metrics (\ref{lacosm2}) in a dual on a time-variant of (\ref{qeltorsc}) as
stated by formulas (\ref{dualnonltr}). For instance, the generating
functions are related by formulas $\underline{\Phi }^{2}=-4\ \underline{%
\Lambda }\underline{g}_{3},$ with underlined $\eta $-polarizations
determined by the generating data $(\underline{g}_{3}=\underline{\eta }_{3}%
\underline{\mathring{g}}_{3};\ \underline{\Lambda },\ ^{v}\underline{%
\Upsilon }).$ For relativistic geometric flows, such formulas are
generalized for $\tau $-families, for instance, written as $\underline{\Phi }%
^{2}(\tau )=-4\ \underline{\Lambda }(\tau )\underline{g}_{3}(\tau )$ and $(%
\underline{g}_{3}(\tau )=\underline{\eta }_{3}(\tau) \underline{\mathring{g}}%
_{3};\ \underline{\Lambda }(\tau ),\ ^{v}\underline{J}(\tau )).$

For time dual transforms,$\ ^{q}\mathbf{g}(\tau )\rightarrow \ ^{c}%
\underline{\mathbf{g}}(\tau ),$ of (\ref{thermvar2}) and (\ref{volumf1}), we
obtain such formulas for geometric thermodynamic variables of locally
anisotropic cosmological solutions (\ref{lacosm2}): 
\begin{eqnarray}
\ ^{c}\widehat{\underline{Z}}(\tau ) &=&\exp \left[ \frac{1}{8\pi ^{2}\tau
^{2}}\ \ _{\eta }^{\mathbf{J}}\underline{\mathcal{V}}[\ ^{c}\underline{%
\mathbf{g}}(\tau )]\right] ,\ ^{c}\widehat{\underline{{E}}}\ (\tau
)=\ \frac{1-2\tau \ \underline{\Lambda }(\tau )}{8\pi ^{2}\tau }\ \ \ _{\eta
}^{\mathbf{J}}\underline{\mathcal{V}}[\ ^{c}\underline{\mathbf{g}}%
(\tau )],  \label{thermovar2c} \\
\ \ \ \ ^{c}\widehat{\underline{S}}(\tau ) &=&-\ ^{c}\widehat{\underline{W}}%
(\tau )=\frac{1-\underline{\Lambda }(\tau )}{4\pi ^{2}\tau ^{2}}\ \ _{\eta
}^{\mathbf{J}}\underline{\mathcal{V}}[\ ^{c}\underline{\mathbf{g}}%
(\tau )],\mbox{ where }  \notag \\
\ _{\eta }^{\mathbf{J}}\underline{\mathcal{V}}[\ ^{c}\underline{%
\mathbf{g}}(\tau )] &=&\int_{\ \widehat{\Xi }}\delta \ _{\eta }\underline{%
\mathcal{V}}(\ ^{v}\widehat{\underline{\mathbf{J}}}(\tau ),\ \underline{\eta 
}_{\alpha }(\tau ),\underline{\mathring{g}}_{\alpha }).  \label{volumf1c}
\end{eqnarray}
These formulas can be used for defining thermodynamic characteristics of
nonholonomic Ricci soliton cosmological configurations constructed for $\tau
=\tau _{0}$.

In the previous section, we constructed in the explicit form an off-diagonal
cosmological solution (\ref{paramsoliton}) with a small parameter $\kappa .$
For relativistic geometric flows, such families of cosmological solutions
are denoted $\ _{\kappa }^{c}\underline{\mathbf{g}}(\tau );$ they allow us
to define and compute parametrically the volume forms (\ref{volumf1c}) (we
write $\ _{\chi }^{\mathbf{J}}\underline{\mathcal{V}}[\ _{\kappa}^{c}%
\underline{\mathbf{g}}(\tau )]).$ Respectively, the geometric cosmological
thermodynamic variables (\ref{thermovar2c}) are computed 
\begin{eqnarray}
\ ^{c}\widehat{\underline{Z}}(\tau ) &=&\exp \left[ \frac{1}{8\pi ^{2}\tau
^{2}}\ \ _{\chi }^{\mathbf{J}}\underline{\mathcal{V}}[\ _{\kappa }^{c}%
\underline{\mathbf{g}}(\tau )]\right] ,\ \ ^{c}\widehat{{E}}\ (\tau
)=\ \frac{1-2\tau \ \underline{\Lambda }(\tau )}{8\pi ^{2}\tau }\ \ \ _{\chi
}^{\mathbf{J}}\underline{\mathcal{V}}[ \ _{\kappa }^{c}\underline{\mathbf{g}}%
(\tau )],  \notag \\
\ \ \ \ \ ^{c}\widehat{\underline{S}}(\tau ) &=&-\ \ ^{c}\widehat{\underline{%
W}}(\tau )=\frac{1-\underline{\Lambda }(\tau )}{4\pi ^{2}\tau ^{2}}\ \
_{\chi }^{\mathbf{J}}\underline{\mathcal{V}}[ \ _{\kappa }^{c}\underline{%
\mathbf{g}}(\tau )].  \label{thermvarcosmpar}
\end{eqnarray}%
For fixed $\tau =\tau _{0},$ such thermodynamic variables characterize $%
\kappa $-parametric off-diagonal deformations of cosmological solutions in
GR.

The geometric thermodynamic (\ref{thermvarcosmpar}) characterizes certain
h-terms defined by certain $\tau $-families of solutions of 2-d Poisson
equations. We can consider some particular cases with $\kappa $-parametric
generated vortices (\ref{taubeq}), defined by $\psi _{0}(\tau ,r,\theta )$
and $\ ^{\psi }\chi (\tau ,r,\theta ).$ Conventional v-flows are included
into the thermodynamic model by respective $\tau $-families of generating
functions $\underline{\chi }(\tau)=\underline{\chi }_{3}(\tau ,r,\theta ,t)$%
. They can be considered as some families of solutions of solitonic wave
equation (\ref{solitonw}), $\underline{\eta }(\tau )\ \longleftrightarrow 
\underline{\chi }(\tau ),$ when $\underline{\zeta }_{3}(\tau ,r,\theta ,t)$
are prescribed to satisfy observational data for $\kappa ^{0}.$ Such
geometric thermodynamic models also characterize cosmological solutions by a
family of v-solitonic gravitational structures of voids with $\kappa $%
--parametric deformations and $t$-evolution. The thermodynamic energy for
respective cosmological configurations with parametric gravitational void
vacua may possess nontrivial solitonic energies. In a more general
cosmological context, G. Perelman's thermodynamic variables may encode
vertex - solitonic wave cosmological d-metrics with prime prolate or oblate
symmetry. Such cosmological configurations can describe nonholonomic vacuum
structures with nontrivial canonical d-torsion $\widehat{\mathbf{T}}_{\
\alpha \beta }^{\gamma }(\tau )$. G. Perelman's thermodynamic variables can
be subjected to additional constraints of type (\ref{lccond}), (\ref{zerot1}%
) or (\ref{lccond1}) to extract LC configurations.

\section{Outlook and conclusions}

\label{sec05}The first objective of this work on mathematical relativity is
to review new advanced geometric techniques (the anholonomic frame and
connection deformation method, AFCDM) for constructing generic off-diagonal
exact and parametric solutions in general relativity, GR. As the second main
objective of this paper, new classes of off-diagonal deformed black hole
(BH), wormhole (WH), black torus (BT) and locally anisotropic cosmological
solutions are constructed and analyzed. To characterize and study the
geometric thermodynamic properties of such solutions a relativistic
generalization of the Ricci flow thermodynamics \cite{perelman1} is
elaborated and applied (the third main objective) following our works \cite%
{svnonh08,gheorghiuap16,vacaru18,nonassocFinslrev25,bsvv24}.

\vskip4pt

The bulk of physically important solutions in GR and many modified gravity
theories, MGTs, were constructed and investigated for diagonal ansatz of
metrics and, typically, using the Levi-Civita, LC, connection $\nabla$. The
standard methods of constructing solutions in GR are outlined in \cite%
{hawrking73,misner73,wald82,kramer03}. Most important applications in
astrophysics and cosmology were based on solutions with certain spherical or
cylindrical symmetries of spacetime metrics. Such physically important
solutions were derived from certain systems of nonlinear ordinary
differential equations, ODEs, to which the Einstein equations are
transformed for certain diagonalizable ansatz (for corresponding frame or
coordinate transforms). This type of solution is determined by some
integration constants, a possible nonzero cosmological constant $\Lambda $,
and special forms of energy-momentum tensors. The integration constants and
sources are defined or chosen to satisfy certain compatibility with
observational data, experimental data, stated in theoretical forms by some
boundary or asymptotic conditions, or Cauchy data. A corresponding
thermodynamic paradigm, due to Bekenstein and Hawking \cite{bek2,haw2}, was
formulated to characterize BH solutions or other types of solutions
involving certain hypersurface horizons, duality conditions, or holographic
configurations.

\vskip4pt An "Orthodox" view which dominated for many years in gravity
theory was that only certain classes of diagonal and "higher symmetry"
solutions in GR are physically motivated for applications in astrophysics
and cosmology with inflation. The paradigm has to be changed in many
unstated ways after the discovery of late-time cosmic acceleration \cite%
{riess98,perlmutter99}. In a sense, many researchers are lost in a labyrinth
of MGTs \cite{sotiriou10,nojiri11,capo11,clifton12,harko14} and a plethora
of dark energy (DE) and dark matter (DM) models \cite{copeland06}.
Well-defined, motivated and meaningful motivations of GR and explanations
for accelerating cosmology are not straightforward and many elaborated
directions are considered ambiguous. The existing observational and
experimental data are still not enough to reduce substantially the spaces of
options. In most research articles, the typical assumption is that "we try
any interesting for us MGTs, then use generalizations of some important
solutions based on some systems of nonlinear ODEs, and watch what comes
out". Then, researchers elaborate on new scenarios in modern cosmology,
speculate on possible applications in quantum and classical information
theories etc. If any phenomenological problems, other generalizations of
already generalized versions of GR are proposed.

\vskip4pt In a series of our works on GR and MGTs \cite%
{vacaru18,vacaruplb16,vbubuianu17,partner06,partner02}, we emphasized that
using geometric methods we can construct exact and parametric solutions of
(modified) Einstein equations which are defined by generic off-diagonal
ansatz for metrics (\ref{ansatz}) depending, in principle, on all spacetime
coordinates. We can consider 4-d spacetimes, or extra dimension/phase space
coordinates by using similar nonholonomic geometric methods. In GR, a
generic off-diagonal metric has 6 degrees of freedom (from 10 coefficients
of a symmetric metric tensor, we can make zero 4 of them by using coordinate
transforms). A diagonal ansatz allows us to model gravitational and matter
field dynamics (maximum) only for 4 degrees of freedom. From a mathematical
point of view, the priority to use diagonal ansatz depending on one space,
or time, coordinate (other ones can be introduced, for instance, as rotation
frame effects) is that it allows to work with the LC connection $\nabla $
and obtain exactly/ parametric solvable systems of nonlinear ODE. Even our
AFCDM allowed us to construct various types of generic off-diagonal
solutions in GR and MGTs, a part of "the very Orthodox community" of
researchers in gravity considered them as "not physically motivated" because
observable BH and cosmological data emphasized large-scale spherical
symmetries and homogeneity. Those conservative arguments are not more
supported rigorously by recent observational data on accelerating cosmology
and DE and DM physics, with anisotropies and inhomogeneities. We have to
analyze such a fundamental issue: maybe it is enough to provide a ground for
such new cosmology theories using off-diagonal solutions (with 6 degrees of
freedom) in GR? Or do we have really to work with MGTs when the off-diagonal
solutions are also important?

\vskip4pt The core of the AFCDM is to find exact and parametric solutions of
physically important systems of nonlinear partial differential equations,
PDEs, see former reviews of results and methods in \cite%
{sv11,vacaru18,vbubuianu17,nonassocFinslrev25}. For GR, the key idea is to
consider an auxiliary linear connection $\widehat{\mathbf{D}}[\mathbf{g}]$
defined for the same metric structure $\mathbf{g}$ which is used for the
LC-connection $\nabla \lbrack \mathbf{g}],$ see formulas (\ref{twocon}).
Using corresponding nonholonomic dyadic variables on Lorentz manifolds, the
Einstein equations for $\widehat{\mathbf{D}}$ can be decoupled and solved in
generic off-diagonal form in terms of integration functions depending, in
principle, on all spacetime coordinates. Having defined a general class of
quasi-stationary or locally anisotropic solutions, we can always impose
additional constraints on generating functions and generating sources which
allow us to extract LC configurations. To integrate (modified, or not)
Einstein equations for generic off-diagonal ansatz is not possible if we
work from the very beginning with $\nabla .$ The main point is to distort
correspondingly the necessary types of systems of nonlinear PDEs to certain
decoupled equivalent systems and then to return to zero nonholonomic torsion
if necessary.

\vskip4pt We constructed and analyzed the physical properties of four
examples of off-diagonal deformations of BH, WH, BT, and locally anisotropic
cosmological solutions in section \ref{sec03}. For certain nonholonomic
conditions, such solutions can be embedded self-consistently in off-diagonal
vacuum configurations; they can be defined with polarizations of physical
constants; with deformation of horizons, or without any horizon or
holographic conditions. Such generic off-diagonal solutions do not have a
thermodynamic description using the Bekenstein-Hawking BH paradigm.
Nevertheless, they exist in GR if we "relax" in a way certain topological or
BH uniqueness criteria and theorems based on certain 'strong' assumptions on
asymptotic conditions, higher symmetries, and pre-accelerating cosmology
observational data. DM models with filaments, complex hierarchic and
quasi-periodic structures, and various types of halos and cosmological
anisotropies can be described if we consider more general classes of
off-diagonal solutions.

\vskip4pt Our approach to the theory of relativistic nonholonomic geometric
flows \cite{svnonh08,gheorghiuap16,vacaru18,nonassocFinslrev25,bsvv24} has
both natural and pragmatic motivations in GR because it allows us to provide
a geometric thermodynamic formalism for all classes of diagonal or
off-diagonal solutions. It provides a new paradigm due to G. Perelman's
concept of W-entropy \cite{perelman1} as we motivated in section \ref{sec04}%
. Even though we do not propose in this work to formulate/ prove any
relativistic versions of the Thorston-Poincar\'{e} conjecture, we show that
the thermodynamic part of the G. Perelman work \cite{perelman1} can be
extended in certain nonholonomic-relativistic-ways. Applying the AFCDM, the
corresponding relativistic geometric flow equations (in particular, the
nonholonomic Ricci soliton equations including the Einstein equations), can
be decoupled and solved in certain general off-diagonal forms. For such
models, relativistic extensions of G. Perelman's thermodynamic variables can
be computed in certain general forms in terms of effective cosmological
constants and respective volume forms.

\vskip4pt Finally, we conclude that the results and methods of this work can
be used for constructing quasi-stationary and cosmological off-diagonal
solutions describing Einstein-Yang-Mills-Higgs-Dirac systems with potential
applications for elaborating elaborating on DE and DM models, in classical
and quantum flow information theories and quantum gravity, see recent
partner works \cite{partner06,bsvv24,vacaru25,nonassocFinslrev25,bsvv24}.

\vskip5pt \textbf{Acknowledgement:}\ SV's work is supported by a visiting
fellowship for the Ko\c{c}aeli University in Turkey and extends former
volunteer research programs at California State University at Fresno, the
USA, and Taras Shevchenko National University of Kyiv, Ukraine.

\newpage \appendix\setcounter{equation}{0} 
\renewcommand{\theequation}
{A.\arabic{equation}} \setcounter{subsection}{0} 
\renewcommand{\thesubsection}
{A.\arabic{subsection}}

\section{A general decoupling and integration property of the Einstein
equations}

\label{appendixa}

In this Appendix, we outline the computations and proofs from sections 3.1
and 3.2\ in \cite{nonassocFinslrev25,sv11}, see also \cite%
{vbubuianu17,vacaru18,partner02} and references therein.

\subsection{Canonical Ricci d-tensors for quasi-stationary configurations}

We compute in explicit form the N-adapted coefficients of the
quasi-stationary canonical d-connections; N-conection curvature; canonical
d-torsion and LC-conditions; and cononical Ricci d-tensor.nary canonical
d-connections. To simplify computations we use brief notations of partial
derivatives, for instance, $\partial _{1}q(u^{\alpha })=q^{\bullet },$ $%
\partial _{2}q(u^{\alpha })=q^{\prime }$, $\partial _{3}q(u^{\alpha
})=q^{\ast }$ and $\partial _{4}q(u^{\alpha })=q^{\diamond }$.

The nontrivial coefficients of $\widehat{\mathbf{\Gamma }}_{\ \alpha \beta
}^{\gamma }$ (\ref{cdc}) computed for quasi-stationary d-metrics (\ref{dmq})
are: 
\begin{eqnarray}
\widehat{L}_{11}^{1} &=&\frac{g_{1}^{\bullet }}{2g_{1}},\ \widehat{L}%
_{12}^{1}=\frac{g_{1}^{\prime }}{2g_{1}},\widehat{L}_{22}^{1}=-\frac{%
g_{2}^{\bullet }}{2g_{1}},\ \widehat{L}_{11}^{2}=\frac{-g_{1}^{\prime }}{%
2g_{2}},\ \widehat{L}_{12}^{2}=\frac{g_{2}^{\bullet }}{2g_{2}},\ \widehat{L}%
_{22}^{2}=\frac{g_{2}^{\prime }}{2g_{2}},  \label{nontrdc} \\
\widehat{L}_{4k}^{4} &=&\frac{\mathbf{\partial }_{k}(h_{4})}{2h_{4}}-\frac{%
w_{k}h_{4}^{\ast }}{2h_{4}},\widehat{L}_{3k}^{3}=\frac{\mathbf{\partial }%
_{k}h_{3}}{2h_{3}}-\frac{w_{k}h_{3}^{\ast }}{2h_{3}},\widehat{L}_{4k}^{3}=-%
\frac{h_{4}}{2h_{3}}n_{k}^{\ast },  \notag \\
\widehat{L}_{3k}^{4} &=&\frac{1}{2}n_{k}^{\ast },\widehat{C}_{33}^{3}=\frac{%
h_{3}^{\ast }}{2h_{3}},\widehat{C}_{44}^{3}=-\frac{h_{4}^{\ast }}{h_{3}},\ 
\widehat{C}_{33}^{4}=0,~\widehat{C}_{34}^{4}=\frac{h_{4}^{\ast }}{2h_{4}},%
\widehat{C}_{44}^{4}=0.  \notag
\end{eqnarray}%
We also compute the values 
\begin{equation}
\ \widehat{C}_{3}=\widehat{C}_{33}^{3}+\widehat{C}_{34}^{4}=\frac{%
h_{3}^{\ast }}{2h_{3}}+\frac{h_{4}^{\ast }}{2h_{4}},\widehat{C}_{4}=\widehat{%
C}_{43}^{3}+\widehat{C}_{44}^{4}=0.  \label{aux3}
\end{equation}%
which are necessary togehter with the set of coefficients (\ref{nontrdc})
for computing in explicit form the N-adapted coefficients of the canonical
d-torsion and canonical Ricci and Einstein d-tensors.

Introducing the N-connection coefficients in (\ref{dmq}), we compute the
coefficients of the N-connection curvature $\widehat{\Omega }_{ij}^{a}=%
\widehat{\mathbf{e}}_{j}\left( \widehat{N}_{i}^{a}\right) -\widehat{\mathbf{e%
}}_{i}(\widehat{N}_{j}^{a}),$ see formulas (\ref{anhcoef}). We obtain 
\begin{equation*}
\widehat{\Omega }_{ij}^{a}=\mathbf{\partial }_{j}\left( \widehat{N}%
_{i}^{a}\right) -\partial _{i}(\widehat{N}_{j}^{a})-w_{i}(\widehat{N}%
_{j}^{a})^{\ast }+w_{j}(\widehat{N}_{i}^{a})^{\ast }.
\end{equation*}%
These formulas result in such nontrivial values: 
\begin{eqnarray}
\widehat{\Omega }_{12}^{3} &=&-\widehat{\Omega }_{21}^{3}=\mathbf{\partial }%
_{2}w_{1}-\partial _{1}w_{2}-w_{1}w_{2}^{\ast }+w_{2}w_{1}^{\ast
}=w_{1}^{\prime }-w_{2}^{\bullet }-w_{1}w_{2}{}^{\ast }+w_{2}w_{1}^{\ast }{};
\notag \\
\widehat{\Omega }_{12}^{4} &=&-\widehat{\Omega }_{21}^{4}=\mathbf{\partial }%
_{2}n_{1}-\partial _{1}n_{2}-w_{1}n_{2}^{\ast }+w_{2}n_{1}^{\ast
}=n_{1}^{\prime }-n_{2}^{\bullet }-w_{1}n_{2}^{\ast }{}+w_{2}n_{1}^{\ast }{}.
\label{omeg}
\end{eqnarray}

Using fomrulas (\ref{omeg}), we can compute the nontrivial coefficients of
the canonical d--torsion. Details on such component formulas are provided in 
\cite{nonassocFinslrev25,sv11}. We have nontrivial coefficent $\widehat{T}%
_{\ ji}^{a}=-\Omega _{\ ji}^{a}$ and $\widehat{T}_{aj}^{c}=\widehat{L}%
_{aj}^{c}-e_{a}(\widehat{N}_{j}^{c}).$ We also compute another subsets of
the nontrivial coefficients%
\begin{eqnarray*}
\widehat{T}_{\ jk}^{i} &=&\widehat{L}_{jk}^{i}-\widehat{L}_{kj}^{i}=0,~%
\widehat{T}_{\ ja}^{i}=\widehat{C}_{jb}^{i}=0,~\widehat{T}_{\ bc}^{a}=\ 
\widehat{C}_{bc}^{a}-\ \widehat{C}_{cb}^{a}=0, \\
\widehat{T}_{3k}^{3} &=&\widehat{L}_{3k}^{3}-e_{3}(\widehat{N}_{k}^{3})=%
\frac{\mathbf{\partial }_{k}h_{3}}{2h_{3}}-w_{k}\frac{h_{3}^{\ast }}{2h_{3}}%
-w_{k}^{\ast }{},\widehat{T}_{4k}^{3}=\widehat{L}_{4k}^{3}-e_{4}(\widehat{N}%
_{k}^{3})=-\frac{h_{4}}{2h_{3}}n_{k}^{\ast },\ 
\end{eqnarray*}%
\begin{eqnarray}
\widehat{T}_{3k}^{4} &=&~\widehat{L}_{3k}^{4}-e_{3}(\widehat{N}_{k}^{4})=%
\frac{1}{2}n_{k}^{\ast }-n_{k}^{\ast }=-\frac{1}{2}n_{k}^{\ast },\widehat{T}%
_{4k}^{4}=\widehat{L}_{4k}^{4}-e_{4}(N_{k}^{4})=\frac{\mathbf{\partial }%
_{k}h_{4}}{2h_{4}}-w_{k}\frac{h_{4}^{\ast }}{2h_{4}},  \notag \\
-\widehat{T}_{12}^{3} &=&w_{1}^{\prime }-w_{2}^{\bullet }-w_{1}w_{2}^{\ast
}{}+w_{2}w_{1}^{\ast },\ -\widehat{T}_{12}^{4}=n_{1}^{\prime
}-n_{2}^{\bullet }-w_{1}n_{2}^{\ast }{}+w_{2}n_{1}^{\ast }{}.
\label{nontrtors}
\end{eqnarray}

The zero canonical d-torsions (\ref{lccond1}) for extracting
LC-configurations are satisfied if 
\begin{equation}
\widehat{L}_{aj}^{c}=e_{a}(\widehat{N}_{j}^{c}),\ \widehat{C}_{jb}^{i}=0,\ 
\widehat{\Omega }_{\ ji}^{a}=0,  \label{lcconstr}
\end{equation}%
when in N-adapted frames we can state $\widehat{\mathbf{\Gamma }}_{\ \alpha
\beta }^{\gamma }=\Gamma _{\ \alpha \beta }^{\gamma }$ even, in general, $%
\widehat{\mathbf{D}}\neq \nabla $. This is possible even two different
linear connections have different transformation laws under general frame/
coordinate transforms. We note that d-connections are not (d) tensor
objects. For LC--configurations, all values (\ref{nontrtors}) must be zero.
Nontrivial off-diagonal solutions can be chosen for $h_{4}^{\ast }\neq 0$
and $w_{k}^{\ast }\neq 0$ but stating for other subsets of N-connection
coefficients: $n_{k}^{\ast }=0,$ for $w_{k}=\mathbf{\partial }%
_{k}h_{4}/h_{4}^{\ast }.$ In an alternative form, we can search for other
types of LC-configurations when $n_{k}^{\ast }\neq 0$ and/or $h_{3}^{\ast
}\neq 0.$ It is difficult to obtain explicit formulas for such classes of
solutions which can be off-diagonal. We note that conditions of type (\ref%
{lcconstr}) can be imposed after a general class of quasi-stationary
off-diagonal metrics is constructed in a general off-diagonal form involving
a nonholonomic d-torsion structure. Here, we also \ emphasize that it is not
possible to decouple in a general form the Einstein equations working from
the very beginning with a $\nabla $ defined by a generic off-diagonal ansatz
with coefficients depending on 2-4 coordinates.

The h-coefficients of a canonical Ricci d-tensor (see N-adapted
prameterizations (\ref{fundgeomc}) and details in \cite%
{vacaru18,vbubuianu17,nonassocFinslrev25,sv11}) are computed for respective
contractions of indices, when $\widehat{R}_{ij}=\widehat{R}_{\ ijk}^{k}$,
for 
\begin{eqnarray}
\widehat{R}_{\ hjk}^{i} &=&\mathbf{e}_{k}\widehat{L}_{.hj}^{i}-\mathbf{e}_{j}%
\widehat{L}_{hk}^{i}+\widehat{L}_{hj}^{m}\widehat{L}_{mk}^{i}-\widehat{L}%
_{hk}^{m}\widehat{L}_{mj}^{i}-\widehat{C}_{ha}^{i}\widehat{\Omega }_{jk}^{a}
\notag \\
&=&\mathbf{\partial }_{k}\widehat{L}_{.hj}^{i}-\partial _{j}\widehat{L}%
_{hk}^{i}+\widehat{L}_{hj}^{m}\widehat{L}_{mk}^{i}-\widehat{L}_{hk}^{m}%
\widehat{L}_{mj}^{i}.  \label{auxcurvh}
\end{eqnarray}%
We note that these formulas are considered for a quasi-stationary ansatz (%
\ref{dmq}) and values (\ref{nontrdc}). The conditions $\widehat{C}_{\
ha}^{i}=0$ and formulas 
\begin{equation*}
e_{k}\widehat{L}_{hj}^{i}=\partial _{k}\widehat{L}_{hj}^{i}+N_{k}^{a}%
\partial _{a}\widehat{L}_{hj}^{i}=\partial _{k}\widehat{L}_{hj}^{i}+w_{k}(%
\widehat{L}_{hj}^{i})^{\ast }+n_{k}(\widehat{L}_{hj}^{i})^{\diamond
}=\partial _{k}\widehat{L}_{hj}^{i}
\end{equation*}%
hold true because $\widehat{L}_{hj}^{i}$ depend only on h-coordinates.
Taking respective derivatives of (\ref{nontrdc}), we obtain 
\begin{eqnarray*}
\partial _{1}\widehat{L}_{\ 11}^{1} &=&(\frac{g_{1}^{\bullet }}{2g_{1}}%
)^{\bullet }=\frac{g_{1}^{\bullet \bullet }}{2g_{1}}-\frac{\left(
g_{1}^{\bullet }\right) ^{2}}{2\left( g_{1}\right) ^{2}},\ \partial _{1}%
\widehat{L}_{\ 12}^{1}=(\frac{g_{1}^{\prime }}{2g_{1}})^{\bullet }=\frac{%
g_{1}^{\prime \bullet }}{2g_{1}}-\frac{g_{1}^{\bullet }g_{1}^{\prime }}{%
2\left( g_{1}\right) ^{2}},\  \\
\partial _{1}\widehat{L}_{\ 22}^{1} &=&(-\frac{g_{2}^{\bullet }}{2g_{1}}%
)^{\bullet }=-\frac{g_{2}^{\bullet \bullet }}{2g_{1}}+\frac{g_{1}^{\bullet
}g_{2}^{\bullet }}{2\left( g_{1}\right) ^{2}},\ \partial _{1}\widehat{L}_{\
11}^{2}=(-\frac{g_{1}^{\prime }}{2g_{2}})^{\bullet }=-\frac{g_{1}^{\prime
\bullet }}{2g_{2}}+\frac{g_{1}^{\bullet }g_{2}^{\prime }}{2\left(
g_{2}\right) ^{2}}, \\
\partial _{1}\widehat{L}_{\ 12}^{2} &=&(\frac{g_{2}^{\bullet }}{2g_{2}}%
)^{\bullet }=\frac{g_{2}^{\bullet \bullet }}{2g_{2}}-\frac{\left(
g_{2}^{\bullet }\right) ^{2}}{2\left( g_{2}\right) ^{2}},\ \partial _{1}%
\widehat{L}_{\ 22}^{2}=(\frac{g_{2}^{\prime }}{2g_{2}})^{\bullet }=\frac{%
g_{2}^{\prime \bullet }}{2g_{2}}-\frac{g_{2}^{\bullet }g_{2}^{\prime }}{%
2\left( g_{2}\right) ^{2}},
\end{eqnarray*}%
\begin{eqnarray*}
\partial _{2}\widehat{L}_{\ 11}^{1} &=&(\frac{g_{1}^{\bullet }}{2g_{1}}%
)^{\prime }=\frac{g_{1}^{\bullet \prime }}{2g_{1}}-\frac{g_{1}^{\bullet
}g_{1}^{\prime }}{2\left( g_{1}\right) ^{2}},~\partial _{2}\widehat{L}_{\
12}^{1}=(\frac{g_{1}^{\prime }}{2g_{1}})^{\prime }=\frac{g_{1}^{\prime
\prime }}{2g_{1}}-\frac{\left( g_{1}^{\prime }\right) ^{2}}{2\left(
g_{1}\right) ^{2}}, \\
\partial _{2}\widehat{L}_{\ 22}^{1} &=&(-\frac{g_{2}^{\bullet }}{2g_{1}}%
)^{\prime }=-\frac{g_{2}^{\bullet ^{\prime }}}{2g_{1}}+\frac{g_{2}^{\bullet
}g_{1}^{^{\prime }}}{2\left( g_{1}\right) ^{2}},\ \partial _{2}\widehat{L}%
_{\ 11}^{2}=(-\frac{g_{1}^{\prime }}{2g_{2}})^{\prime }=-\frac{g_{1}^{\prime
\prime }}{2g_{2}}+\frac{g_{1}^{\bullet }g_{1}^{\prime }}{2\left(
g_{2}\right) ^{2}}, \\
\partial _{2}\widehat{L}_{\ 12}^{2} &=&(\frac{g_{2}^{\bullet }}{2g_{2}}%
)^{\prime }=\frac{g_{2}^{\bullet \prime }}{2g_{2}}-\frac{g_{2}^{\bullet
}g_{2}^{\prime }}{2\left( g_{2}\right) ^{2}},\partial _{2}\widehat{L}_{\
22}^{2}=(\frac{g_{2}^{\prime }}{2g_{2}})^{\prime }=\frac{g_{2}^{\prime
\prime }}{2g_{2}}-\frac{\left( g_{2}^{\prime }\right) ^{2}}{2\left(
g_{2}\right) ^{2}}.
\end{eqnarray*}%
Introducing these values in (\ref{auxcurvh}), we obtain two of nontrivial
components (because of anti-symmery, there are four nontrivia such terms): 
\begin{eqnarray*}
\widehat{R}_{\ 212}^{1} &=&\frac{g_{2}^{\bullet \bullet }}{2g_{1}}-\frac{%
g_{1}^{\bullet }g_{2}^{\bullet }}{4\left( g_{1}\right) ^{2}}-\frac{\left(
g_{2}^{\bullet }\right) ^{2}}{4g_{1}g_{2}}+\frac{g_{1}^{\prime \prime }}{%
2g_{1}}-\frac{g_{1}^{\prime }g_{2}^{\prime }}{4g_{1}g_{2}}-\frac{\left(
g_{1}^{\prime }\right) ^{2}}{4\left( g_{1}\right) ^{2}}, \\
\widehat{R}_{\ 112}^{2} &=&-\frac{g_{2}^{\bullet \bullet }}{2g_{2}}+\frac{%
g_{1}^{\bullet }g_{2}^{\bullet }}{4g_{1}g_{2}}+\frac{\left( g_{2}^{\bullet
}\right) ^{2}}{4(g_{2})^{2}}-\frac{g_{1}^{\prime \prime }}{2g_{2}}+\frac{%
g_{1}^{\prime }g_{2}^{\prime }}{4(g_{2})^{2}}+\frac{\left( g_{1}^{\prime
}\right) ^{2}}{4g_{1}g_{2}}.
\end{eqnarray*}%
By definition, $\widehat{R}_{11}=-\widehat{R}_{\ 112}^{2}$ and $\widehat{R}%
_{22}=\widehat{R}_{\ 212}^{1},$ for $g^{i}=1/g_{i}$ and $\widehat{R}%
_{j}^{j}=g^{j}\widehat{R}_{jj}.$ In these formulas, there is no summarizing
on repeating indices. As a result, we compute 
\begin{equation}
\widehat{R}_{1}^{1}=\widehat{R}_{2}^{2}=-\frac{1}{2g_{1}g_{2}}%
[g_{2}^{\bullet \bullet }-\frac{g_{1}^{\bullet }g_{2}^{\bullet }}{2g_{1}}-%
\frac{\left( g_{2}^{\bullet }\right) ^{2}}{2g_{2}}+g_{1}^{\prime \prime }-%
\frac{g_{1}^{\prime }g_{2}^{\prime }}{2g_{2}}-\frac{(g_{1}^{\prime })^{2}}{%
2g_{1}}].  \label{hcdric}
\end{equation}

We compute also the N-adapted coefficients with mixed h- and v-indices of
the canonical Ricci d-tensor. Considering other groups of coefficients, we
write 
\begin{equation*}
\widehat{R}_{\ bka}^{c}=\frac{\partial \widehat{L}_{bk}^{c}}{\partial y^{a}}-%
\widehat{C}_{~ba|k}^{c}+\widehat{C}_{~bd}^{c}\widehat{T}_{~ka}^{d}=\frac{%
\partial \widehat{L}_{bk}^{c}}{\partial y^{a}}-(\frac{\partial \widehat{C}%
_{ba}^{c}}{\partial x^{k}}+\widehat{L}_{dk}^{c\,}\widehat{C}_{ba}^{d}-%
\widehat{L}_{bk}^{d}\widehat{C}_{da}^{c}-\widehat{L}_{ak}^{d}\widehat{C}%
_{bd}^{c})+\widehat{C}_{bd}^{c}\widehat{T}_{ka}^{d}.
\end{equation*}%
Contracting the indices, we obtain 
\begin{equation*}
\widehat{R}_{bk}=\widehat{R}_{\ bka}^{a}=\frac{\partial L_{bk}^{a}}{\partial
y^{a}}-\widehat{C}_{ba|k}^{a}+\widehat{C}_{bd}^{a}\widehat{T}_{ka}^{d},
\end{equation*}%
where $\widehat{C}_{b}:=\widehat{C}_{ba}^{c}$ are given by formuls (\ref%
{aux3}). \ Respectively,%
\begin{equation*}
\widehat{C}_{b|k}=\mathbf{e}_{k}\widehat{C}_{b}-\widehat{L}_{\ bk}^{d\,}%
\widehat{C}_{d}=\partial _{k}\widehat{C}_{b}-N_{k}^{e}\partial _{e}\widehat{C%
}_{b}-\widehat{L}_{\ bk}^{d\,}\widehat{C}_{d}=\partial _{k}\widehat{C}%
_{b}-w_{k}\widehat{C}_{b}^{\ast }-n_{k}\widehat{C}_{b}^{\diamond }-\widehat{L%
}_{\ bk}^{d\,}\widehat{C}_{d}.
\end{equation*}%
Let us introduce a conventional splitting $\widehat{R}_{bk}=\ _{[1]}R_{bk}+\
_{[2]}R_{bk}+\ _{[3]}R_{bk},$ where%
\begin{eqnarray*}
\ _{[1]}R_{bk} &=&(\widehat{L}_{bk}^{3})^{\ast }+(\widehat{L}%
_{bk}^{4})^{\diamond },\ _{[2]}R_{bk}=-\partial _{k}\widehat{C}_{b}+w_{k}%
\widehat{C}_{b}^{\ast }+n_{k}\widehat{C}_{b}^{\diamond }+\widehat{L}_{\
bk}^{d\,}\widehat{C}_{d}, \\
\ _{[3]}R_{bk} &=&\widehat{C}_{bd}^{a}\widehat{T}_{ka}^{d}=\widehat{C}%
_{b3}^{3}\widehat{T}_{k3}^{3}+\widehat{C}_{b4}^{3}\widehat{T}_{k3}^{4}+%
\widehat{C}_{b3}^{4}\widehat{T}_{k4}^{3}+\widehat{C}_{b4}^{4}\widehat{T}%
_{k4}^{4}.
\end{eqnarray*}%
Further computations simplify if we use formulas (\ref{nontrdc}), (\ref%
{nontrtors}) and (\ref{aux3}): 
\begin{eqnarray*}
\ _{[1]}R_{3k} &=&\left( \widehat{L}_{3k}^{3}\right) ^{\ast }+\left( 
\widehat{L}_{3k}^{4}\right) ^{\diamond }=\left( \frac{\mathbf{\partial }%
_{k}h_{3}}{2h_{3}}-w_{k}\frac{h_{3}^{\ast }}{2h_{3}}\right) ^{\ast
}=-w_{k}^{\ast }\frac{h_{3}^{\ast }}{2h_{3}}-w_{k}\left( \frac{h_{3}^{\ast }%
}{2h_{3}}\right) ^{\ast }+\frac{1}{2}\left( \frac{\mathbf{\partial }_{k}h_{3}%
}{h_{3}}\right) ^{\ast }, \\
\ _{[2]}R_{3k} &=&-\partial _{k}\widehat{C}_{3}+w_{k}\widehat{C}_{3}^{\ast
}+n_{k}\widehat{C}_{3}^{\diamond }+\widehat{L}_{\ 3k}^{3\,}\widehat{C}_{3}+%
\widehat{L}_{\ 3k}^{4\,}\widehat{C}_{4}= \\
&=&w_{k}[\frac{h_{3}^{\ast \ast }}{2h_{3}}-\frac{3}{4}\frac{(h_{3}^{\ast
})^{2}}{(h_{3})^{2}}+\frac{h_{4}^{\ast \ast }}{2h_{4}}-\frac{1}{2}\frac{%
(h_{4}^{\ast })^{2}}{(h_{4})^{2}}-\frac{1}{4}\frac{h_{3}^{\ast }}{h_{3}}%
\frac{h_{4}^{\ast }}{h_{4}}]+\frac{\mathbf{\partial }_{k}h_{3}}{2h_{3}}(%
\frac{h_{3}^{\ast }}{2h_{3}}+\frac{h_{4}^{\ast }}{2h_{4}})-\frac{1}{2}%
\partial _{k}(\frac{h_{3}^{\ast }}{h_{3}}+\frac{h_{4}^{\ast }}{h_{4}}), \\
\ _{[3]}R_{3k} &=&\widehat{C}_{33}^{3}\widehat{T}_{k3}^{3}+\widehat{C}%
_{34}^{3}\widehat{T}_{k3}^{4}+\widehat{C}_{33}^{4}\widehat{T}_{k4}^{3}+%
\widehat{C}_{34}^{4}\widehat{T}_{k4}^{4} \\
&=&w_{k}\left( \frac{(h_{3}^{\ast })^{2}}{4(h_{3})^{2}}+\frac{(h_{4}^{\ast
})^{2}}{4(h_{4})^{2}}\right) +w_{k}^{\ast }\frac{h_{3}^{\ast }}{2h_{3}}-%
\frac{h_{3}^{\ast }}{2h_{3}}\frac{\mathbf{\partial }_{k}h_{3}}{2h_{3}}-\frac{%
h_{4}^{\ast }}{2h_{4}}\frac{\mathbf{\partial }_{k}h_{4}}{2h_{4}}.
\end{eqnarray*}%
Putting together above formulas we express 
\begin{eqnarray}
\ \widehat{R}_{3k} &=&w_{k}[\frac{h_{4}^{\ast \ast }}{2h_{4}}-\frac{1}{4}%
\frac{(h_{4}^{\ast })^{2}}{(h_{4})^{2}}-\frac{1}{4}\frac{h_{3}^{\ast }}{h_{3}%
}\frac{h_{4}^{\ast }}{h_{4}}]+\frac{h_{4}^{\ast }}{2h_{4}}\frac{\mathbf{%
\partial }_{k}h_{3}}{2h_{3}}-\frac{1}{2}\frac{\partial _{k}h_{4}^{\ast }}{%
h_{4}}+\frac{1}{4}\frac{h_{4}^{\ast }\partial _{k}h_{4}}{(h_{4})^{2}}  \notag
\\
&=&\frac{w_{k}}{2h_{4}}[h_{4}^{\ast \ast }-\frac{(h_{4}^{\ast })^{2}}{2h_{4}}%
-\frac{h_{3}^{\ast }h_{4}^{\ast }}{2h_{3}}]+\frac{h_{4}^{\ast }}{4h_{4}}(%
\frac{\mathbf{\partial }_{k}h_{3}}{h_{3}}+\frac{\partial _{k}h_{4}^{\ast }}{%
h_{4}})-\frac{1}{2}\frac{\partial _{k}h_{4}^{\ast }}{h_{4}}.
\label{vhcdric3}
\end{eqnarray}

We write $\ \widehat{R}_{4k}=\ _{[1]}R_{4k}+\ _{[2]}R_{4k}+\ _{[3]}R_{4k},$
where%
\begin{eqnarray*}
\ _{[1]}R_{4k} &=&(\widehat{L}_{4k}^{3})^{\ast }+(\widehat{L}%
_{4k}^{4})^{\diamond },\ _{[2]}R_{4k}=-\partial _{k}\widehat{C}_{4}+w_{k}%
\widehat{C}_{4}^{\ast }+n_{k}\widehat{C}_{4}^{\diamond }+\widehat{L}_{\
4k}^{3\,}\widehat{C}_{3}+\widehat{L}_{\ 4k}^{4\,}\widehat{C}_{4}, \\
_{\lbrack 3]}R_{4k} &=&\widehat{C}_{4d}^{a}\widehat{T}_{ka}^{d}=\widehat{C}%
_{43}^{3}\widehat{T}_{k3}^{3}+\widehat{C}_{44}^{3}\widehat{T}_{k3}^{4}+%
\widehat{C}_{43}^{4}\widehat{T}_{k4}^{3}+\widehat{C}_{44}^{4}\widehat{T}%
_{k4}^{4}.
\end{eqnarray*}%
Using $\widehat{L}_{4k}^{3}$ and $\widehat{L}_{4k}^{4}$ from (\ref{nontrdc}%
), we compute%
\begin{equation*}
\ _{[1]}R_{4k}=(\widehat{L}_{4k}^{3})^{\ast }+(\widehat{L}%
_{4k}^{4})^{\diamond }=(-\frac{h_{4}}{2h_{3}}n_{k}^{\ast })^{\ast
}=-n_{k}^{\ast \ast }\frac{h_{4}}{2h_{3}}-n_{k}^{\ast }\frac{h_{4}^{\ast
}h_{3}-h_{4}h_{3}^{\ast }}{2(h_{3})^{2}}.
\end{equation*}%
In these formulas, the second term follows from $\widehat{C}_{3}$ and $%
\widehat{C}_{4},$ see (\ref{aux3}). Using $\ \widehat{L}_{4k}^{3}$ and $%
\widehat{L}_{4k}^{4}$ (\ref{nontrdc}), we compute 
\begin{equation*}
\ _{[2]}R_{4k}=-\partial _{k}\widehat{C}_{4}+w_{k}\widehat{C}_{4}^{\ast
}+n_{k}\widehat{C}_{4}^{\diamond }+\widehat{L}_{\ 4k}^{3\,}\widehat{C}_{3}+%
\widehat{L}_{\ 4k}^{4\,}\widehat{C}_{4}=-n_{k}^{\ast }{}\frac{h_{4}}{2h_{3}}(%
\frac{h_{3}^{\ast }}{2h_{3}}+\frac{h_{4}^{\ast }}{2h_{4}}).
\end{equation*}%
Then considering $\widehat{C}_{43}^{3},\widehat{C}_{44}^{3},\widehat{C}%
_{43}^{4},\widehat{C}_{44}^{4},$ see (\ref{nontrdc}), and $\widehat{T}%
_{k3}^{3},\widehat{T}_{k3}^{4},\widehat{T}_{k4}^{3},\widehat{T}_{k4}^{4},$
see (\ref{nontrtors}), we compute the third term:%
\begin{equation*}
_{\lbrack 3]}R_{4k}=\widehat{C}_{43}^{3}\widehat{T}_{k3}^{3}+\widehat{C}%
_{44}^{3}\widehat{T}_{k3}^{4}+\widehat{C}_{43}^{4}\widehat{T}_{k4}^{3}+%
\widehat{C}_{44}^{4}\widehat{T}_{k4}^{4}=0.
\end{equation*}%
We summarize above three terms \ and express%
\begin{equation}
\widehat{R}_{4k}=-n_{k}^{\ast \ast }{}\frac{h_{4}}{2h_{3}}+n_{k}^{\ast
}{}\left( -\frac{h_{4}^{\ast }}{2h_{3}}+\frac{h_{4}^{\ast }h_{3}^{\ast }}{%
2(h_{3})^{\ast }}-\frac{h_{4}^{\ast }h_{3}^{\ast }}{4(h_{3})^{\ast }}-\frac{%
h_{4}^{\ast }}{4h_{3}}\right) .  \label{vhcdric4}
\end{equation}

For another group of N-adapted coefficients, we have the formulas 
\begin{equation*}
\widehat{R}_{\ jka}^{i}=\frac{\partial \widehat{L}_{jk}^{i}}{\partial y^{k}}%
-(\frac{\partial \widehat{C}_{ja}^{i}}{\partial x^{k}}+\widehat{L}_{lk}^{i}%
\widehat{C}_{ja}^{l}-\widehat{L}_{jk}^{l}\widehat{C}_{la}^{i}-\widehat{L}%
_{ak}^{c}\widehat{C}_{jc}^{i})+\widehat{C}_{jb}^{i}\widehat{T}_{ka}^{b}.
\end{equation*}
Such coefficients are zero because $\widehat{C}_{jb}^{i}=0$ and $\widehat{L}%
_{jk}^{i}$ do not depend on $y^{k}.$ Corresponingly, we obtain $\widehat{R}%
_{ja}=\widehat{R}_{\ jia}^{i}=0$.

Contracting the indices in $\widehat{R}_{\ bcd}^{a},$ the Ricci
v-coefficients are computed 
\begin{equation*}
\widehat{R}_{bc}=\frac{\partial \widehat{C}_{bc}^{d}}{\partial y^{d}}-\frac{%
\partial \widehat{C}_{bd}^{d}}{\partial y^{c}}+\widehat{C}_{bc}^{e}\widehat{C%
}_{e}-\widehat{C}_{bd}^{e}\widehat{C}_{ec}^{d}.
\end{equation*}%
We get (summarizing indices) 
\begin{equation*}
\widehat{R}_{bc}=(\widehat{C}_{bc}^{3})^{\ast }+(\widehat{C}%
_{bc}^{4})^{\diamond }-\partial _{c}\widehat{C}_{b}+\widehat{C}_{bc}^{3}%
\widehat{C}_{3}+\widehat{C}_{bc}^{4}\widehat{C}_{4}-\widehat{C}_{b3}^{3}%
\widehat{C}_{3c}^{3}-\widehat{C}_{b4}^{3}\widehat{C}_{3c}^{4}-\widehat{C}%
_{b3}^{4}\widehat{C}_{4c}^{3}-\widehat{C}_{b4}^{4}\widehat{C}_{4c}^{4}.
\end{equation*}%
These formulas result in nontrivial 
\begin{eqnarray*}
\widehat{R}_{33} &=&\left( \widehat{C}_{33}^{3}\right) ^{\ast }+\left( 
\widehat{C}_{33}^{4}\right) ^{\diamond }-\widehat{C}_{3}^{\ast }+\widehat{C}%
_{33}^{3}\widehat{C}_{3}+\widehat{C}_{33}^{4}\widehat{C}_{4}-\widehat{C}%
_{33}^{3}\widehat{C}_{33}^{3}-2\widehat{C}_{34}^{3}\widehat{C}_{33}^{4}-%
\widehat{C}_{34}^{4}\widehat{C}_{43}^{4} \\
&=&-\frac{1}{2}\frac{h_{4}^{\ast \ast }}{h_{4}}+\frac{1}{4}\frac{%
(h_{4}^{\ast })^{2}}{(h_{4})^{2}}+\frac{1}{4}\frac{h_{3}^{\ast }}{h_{3}}%
\frac{h_{4}^{\ast }}{h_{4}}, \\
\widehat{R}_{44} &=&\left( \widehat{C}_{44}^{3}\right) ^{\ast }+\left( 
\widehat{C}_{44}^{4}\right) ^{\diamond }-\partial _{4}\widehat{C}_{4}+%
\widehat{C}_{44}^{3}\widehat{C}_{3}+\widehat{C}_{44}^{4}\widehat{C}_{4}-%
\widehat{C}_{43}^{3}\widehat{C}_{34}^{3}-2\widehat{C}_{44}^{3}\widehat{C}%
_{34}^{4}-\widehat{C}_{44}^{4}\widehat{C}_{44}^{4} \\
&=&-\frac{1}{2}\frac{h_{4}^{\ast \ast }}{h_{3}}+\frac{1}{4}\frac{h_{3}^{\ast
}h_{4}^{\ast }}{(h_{3})^{2}}+\frac{1}{4}\frac{h_{4}^{\ast }}{h_{3}}\frac{%
h_{4}^{\ast }}{h_{4}}.
\end{eqnarray*}%
For certain applications, these formulas can be rewritten in the form%
\begin{equation}
\widehat{R}_{~3}^{3}=\frac{1}{h_{3}}\widehat{R}_{33}=\frac{1}{2h_{3}h_{4}}%
(-h_{4}^{\ast \ast }+\frac{(h_{4}^{\ast })^{2}}{2h_{4}}+\frac{h_{3}^{\ast
}h_{4}^{\ast }}{2h_{3}}),\widehat{R}_{~4}^{4}=\frac{1}{h_{4}}\widehat{R}%
_{44}=\frac{1}{2h_{3}h_{4}}(-h_{4}^{\ast \ast }+\frac{(h_{4}^{\ast })^{2}}{%
2h_{4}}+\frac{h_{3}^{\ast }h_{4}^{\ast }}{2h_{3}}).  \label{vcdric}
\end{equation}%
Here we note that originally such computations were provided in \cite{sv11}%
); more details are provided in \cite{nonassocFinslrev25}.

So, a quasi-stationary d-metric ansatz (\ref{dmq}) is characterized by
nontrivial N-adapted coefficients of the canonical d-connection $\widehat{R}%
_{1}^{1}=\widehat{R}_{2}^{2}$ (\ref{hcdric}), $\ \widehat{R}_{3k}$ (\ref%
{vhcdric3}), $\widehat{R}_{4k}$ (\ref{vhcdric4}) and $\widehat{R}_{~3}^{3}=%
\widehat{R}_{~4}^{4}$ (\ref{vcdric}). For such ansatz, other classes of
coefficients are trivial with respect to N-adapted frames: $\ \widehat{R}%
_{ka}\equiv 0$ for any $k=1,2$ and $a=3,4.$ Such values may be not zero in
other systems of reference or coordinates.

We compute the canonical Ricci d-scalar using above N-adapted nontrivial
coefficients of the canonical Ricci d-tensor, 
\begin{equation*}
\widehat{R}sc:=\widehat{\mathbf{g}}^{\alpha \beta }\widehat{\mathbf{R}}_{\
\alpha \beta }=\widehat{g}^{ij}\widehat{R}_{ij}+\widehat{g}^{ab}\widehat{R}%
_{ab}=\widehat{R}_{~i}^{i}+\widehat{R}_{~a}^{a}=2(\widehat{R}_{2}^{2}+%
\widehat{R}_{~4}^{4}).
\end{equation*}%
In this formula, we consider nontrivial (\ref{hcdric}) and (\ref{vcdric}).
We can compute also the nontrivial components of the canonical Einstein
d-tensor, 
\begin{equation*}
\widehat{\mathbf{E}}n:=\{\widehat{\mathbf{R}}_{\ \gamma }^{\beta }-\frac{1}{2%
}\delta _{\gamma }^{\beta }\ \widehat{R}sc\}=\{-\widehat{R}_{~4}^{4},-%
\widehat{R}_{~4}^{4};\ \widehat{R}_{ak};\widehat{R}_{ka}\equiv 0;-\widehat{R}%
_{2}^{2},-\widehat{R}_{2}^{2}\}.
\end{equation*}%
Such symmetries are important for a general decoupling and integration of
the Einstein equations written in canonical dyadic variables in a form (\ref%
{eq1})-(\ref{e2c}).

\subsection{ Off-diagonal integration of decoupled Einstein equations}

\label{appendixab}

We note that $g_{i}=e^{\psi (x^{k})},$ i.e the h-components of d-metric (\ref%
{dmq}) \ are defined in general form as solutions a 2-d Poisson equation (%
\ref{eq1}) with a generating source $\ ^{h}\Upsilon (x^{k}).$ We omit
further details but show how to find general formulas for v-coefficient and
the N-connection coefficients.

Introducing $h_{3}$ and $h_{4}$ in explicit form in the coefficients (\ref%
{coeff}) from (\ref{e2a})-(\ref{e2c}), we obtain such a nonlinear system: 
\begin{eqnarray}
\Psi ^{\ast }h_{4}^{\ast } &=&2h_{3}h_{4}\ ^{v}\Upsilon \Psi ,  \label{auxa1}
\\
\sqrt{|h_{3}h_{4}|}\Psi &=&h_{4}^{\ast },  \label{auxa2} \\
\ \Psi ^{\ast }w_{i}-\partial _{i}\Psi &=&\ 0,  \label{aux1ab} \\
\ n_{i}^{\ast \ast }+\left( \ln \frac{|h_{4}|^{3/2}}{|h_{3}|}\right) ^{\ast
}n_{i}^{\ast } &=&0.\   \label{aux1ac}
\end{eqnarray}%
Prescribing a generating function, $\Psi (x^{i},y^{3}),$ and a generating
source, $\ ^{v}\Upsilon (x^{i},y^{3}),$ we can integrate recurrently these
equations if the conditions $h_{4}^{\ast }\neq 0$ and $^{v}\Upsilon \neq 0.$
If such conditions are not satisfied, more special analytic methods have to
be applied. We do not consider such case because we can always choose
certain N-adapted frames of reference when "good" conditions allows us to
find necessary smooth class solutions. Let us define 
\begin{equation}
\rho ^{2}:=-h_{3}h_{4}  \label{rho}
\end{equation}%
and re-write (\ref{auxa1}) and (\ref{auxa2}), respectively, as a system 
\begin{equation}
\Psi ^{\ast }h_{4}^{\ast }=-2\rho ^{2}\ ^{v}\Upsilon \ \Psi \mbox{
and }h_{4}^{\ast }=\rho \ \Psi .  \label{auxa3a}
\end{equation}%
So, we can substitute the value of $h_{4}^{\ast }$ from the second equation
into the first equation and express 
\begin{equation}
\rho =-\Psi ^{\ast }/2\ ^{v}\Upsilon .  \label{rho1}
\end{equation}%
This $\rho $ can be used for the second equation in (\ref{auxa3a}). We can
compute (integrating on $y^{3})$ 
\begin{equation}
\ h_{4}(x^{k},y^{3})=h_{4}^{[0]}(x^{k})-\int dy^{3}[\Psi ^{2}]^{\ast }/4(\
^{v}\Upsilon ).  \label{g4}
\end{equation}%
Then, we introduce this coefficient and formulas in (\ref{rho}) and (\ref%
{rho1}), which allows us to compute%
\begin{equation}
h_{3}(x^{k},y^{3})=-\frac{1}{4h_{4}}\left( \frac{\Psi ^{\ast }}{\
^{v}\Upsilon }\right) ^{2}=-\left( \frac{\Psi ^{\ast }}{2\ \ ^{v}\Upsilon }%
\right) ^{2}\left( h_{4}^{[0]}(x^{k})-\int dy^{3}\frac{[\Psi ^{2}]^{\ast }}{%
4\ \ ^{v}\Upsilon }\right) ^{-1}.  \label{g3}
\end{equation}

Using $h_{3}$ (\ref{g3}) and $h_{4}$ (\ref{g4}), we can integrate two times
on $y^{3}$ and generate solutions of (\ref{aux1ac}): 
\begin{eqnarray}
n_{k}(x^{k},y^{3}) &=&\ _{1}n_{k}+\ _{2}n_{k}\int dy^{3}\ \frac{h_{3}}{|\
h_{4}|^{3/2}}=\ _{1}n_{k}+\ _{2}n_{k}\int dy^{3}\left( \frac{\Psi ^{\ast }}{%
2\ \ ^{v}\Upsilon }\right) ^{2}|\ h_{4}|^{-5/2}  \notag \\
&=&\ _{1}n_{k}+\ _{2}n_{k}\int dy^{3}\left( \frac{\Psi ^{\ast }}{2\ \
^{v}\Upsilon }\right) ^{2}\left\vert h_{4}^{[0]}(x^{k})-\int dy^{3}[\Psi
^{2}]^{\ast }/4\ \ ^{v}\Upsilon \right\vert ^{-5/2}.  \label{gn}
\end{eqnarray}%
In (\ref{gn}), we consider two integration functions $\ _{1}n_{k}=\
_{1}n_{k}(x^{i})$ and (re-defining introducing certain coefficients) $\
_{2}n_{k}=\ _{2}n_{k}(x^{i}).$

The linear on $w_{i}$ algebraic system (\ref{aux1ab}) allows us to compute%
\begin{equation}
w_{i}=\partial _{i}\ \Psi /(\Psi )^{\ast }.  \label{gw}
\end{equation}

Putting together above values for the coefficients of the d-metric and
N-connection (as defined by formulas (\ref{g3}),(\ref{g4}) and (\ref{gw}), (%
\ref{gn}) and together with a solution of 2-d Poisson equations for $\psi
(x^{k})$), we can generate quasi-stationary off-diagonal solutions (\ref{dmq}%
) of the Einstein equations written in canonical nonholonomic variables.

In a similar form, for underlined v-coefficients depending generically on $%
y^{4}=t,$ above formulated integration procedure can be performed for
generating locally anisotropic cosmological solutions with d-metrics (\ref%
{dmc}). We omit such incrimental computations in this work.

\subsection{Off-diagonal quasi-stationary solutions with small parameters}

\label{appendixac}For various applications in modern cosmology and DE and DM
physics, we can considering $\kappa $-linear nonlinear transforms (\ref%
{nonlintrsmalp}) with generating functions ivolving $\chi $-polarizations in
(\ref{offdiagpolfr}). This way, we can define small nonholonomic
deformations of a prime d-metric $\mathbf{\mathring{g}}$ into so-called $%
\kappa $-parametric solutions with $\zeta $- and $\chi $-coefficients
derived from approximations: 
\begin{eqnarray}
\psi &\simeq &\psi (x^{k})\simeq \psi _{0}(x^{k})(1+\kappa \ _{\psi }\chi
(x^{k})),\mbox{ for }\   \label{epsilongenfdecomp} \\
\ \eta _{2} &\simeq &\eta _{2}(x^{k_{1}})\simeq \zeta _{2}(x^{k})(1+\kappa
\chi _{2}(x^{k})),\mbox{ we can consider }\ \eta _{2}=\ \eta _{1};  \notag \\
\eta _{4} &\simeq &\eta _{4}(x^{k},y^{3})\simeq \zeta
_{4}(x^{k},y^{3})(1+\kappa \chi _{4}(x^{k},y^{3})).  \notag
\end{eqnarray}%
In these formulas, $\psi $ and $\eta _{2}=\ \eta _{1}$ are such way chosen
to be related to the solutions of the 2-d Poisson equation $\partial
_{11}^{2}\psi +\partial _{22}^{2}\psi =2\ ^{v}\Upsilon (x^{k}),$ see (\ref%
{eq1}).

We compute $\kappa $-parametric deformations to quasi-stationary d-metrics
with $\chi $-generating functions by introducing formulas (\ref%
{epsilongenfdecomp}) for respective coefficients of d-metrics: 
\begin{equation*}
d\ \widehat{s}^{2}=\widehat{g}_{\alpha \beta }(x^{k},y^{3};\psi
,g_{4};^{v}\Upsilon )du^{\alpha }du^{\beta }=e^{\psi _{0}}(1+\kappa \ ^{\psi
}\chi )[(dx^{1})^{2}+(dx^{2})^{2}]
\end{equation*}%
\begin{eqnarray*}
&&-\{\frac{4[\partial _{3}(|\zeta _{4}\mathring{g}_{4}|^{1/2})]^{2}}{%
\mathring{g}_{3}|\int dy^{3}\{\ \ ^{v}\Upsilon \partial _{3}(\zeta _{4}%
\mathring{g}_{4})\}|}-\kappa \lbrack \frac{\partial _{3}(\chi _{4}|\zeta _{4}%
\mathring{g}_{4}|^{1/2})}{4\partial _{3}(|\zeta _{4}\mathring{g}_{4}|^{1/2})}%
-\frac{\int dy^{3}\{\ ^{v}\Upsilon \partial _{3}[(\zeta _{4}\mathring{g}%
_{4})\chi _{4}]\}}{\int dy^{3}\{\ ^{v}\Upsilon \partial _{3}(\zeta _{4}%
\mathring{g}_{4})\}}]\}\mathring{g}_{3} \\
&&\{dy^{3}+[\frac{\partial _{i}\ \int dy^{3}\ ^{v}\Upsilon \ \partial
_{3}\zeta _{4}}{(\mathring{N}_{i}^{3})\ ^{v}\Upsilon \partial _{3}\zeta _{4}}%
+\kappa (\frac{\partial _{i}[\int dy^{3}\ ^{v}\Upsilon \ \partial _{3}(\zeta
_{4}\chi _{4})]}{\partial _{i}\ [\int dy^{3}\ ^{v}\Upsilon \partial
_{3}\zeta _{4}]}-\frac{\partial _{3}(\zeta _{4}\chi _{4})}{\partial
_{3}\zeta _{4}})]\mathring{N}_{i}^{3}dx^{i}\}^{2}
\end{eqnarray*}%
\begin{eqnarray}
&&+\zeta _{4}(1+\kappa \ \chi _{4})\ \mathring{g}_{4}\{dt+[(\mathring{N}%
_{k}^{4})^{-1}[\ _{1}n_{k}+16\ _{2}n_{k}[\int dy^{3}\frac{\left( \partial
_{3}[(\zeta _{4}\mathring{g}_{4})^{-1/4}]\right) ^{2}}{|\int dy^{3}\partial
_{3}[\ ^{v}\Upsilon (\zeta _{4}\mathring{g}_{4})]|}]  \label{offdncelepsilon}
\\
&&+\kappa \frac{16\ _{2}n_{k}\int dy^{3}\frac{\left( \partial _{3}[(\zeta
_{4}\mathring{g}_{4})^{-1/4}]\right) ^{2}}{|\int dy^{3}\partial _{3}[\
^{v}\Upsilon (\zeta _{4}\mathring{g}_{4})]|}(\frac{\partial _{3}[(\zeta _{4}%
\mathring{g}_{4})^{-1/4}\chi _{4})]}{2\partial _{3}[(\zeta _{4}\mathring{g}%
_{4})^{-1/4}]}+\frac{\int dy^{3}\partial _{3}[\ ^{v}\Upsilon (\zeta _{4}\chi
_{4}\mathring{g}_{4})]}{\int dy^{3}\partial _{3}[\ ^{v}\Upsilon (\zeta _{4}%
\mathring{g}_{4})]})}{\ _{1}n_{k}+16\ _{2}n_{k}[\int dy^{3}\frac{\left(
\partial _{3}[(\zeta _{4}\mathring{g}_{4})^{-1/4}]\right) ^{2}}{|\int
dy^{3}\partial _{3}[\ ^{v}\Upsilon (\zeta _{4}\mathring{g}_{4})]|}]}]%
\mathring{N}_{k}^{4}dx^{k}\}^{2}.  \notag
\end{eqnarray}%
Such parametric solutions allow to define, for instance, ellipsoidal
deformations of BH metrics into BE ones. For other classes of solutions with
small parameters, we can compute quasi-classical off-diagonal deformations
of some solutions in GR. Various corrections from MGTs can be also defined
and computed as certain off-dagonal $\kappa $-parametric solutions of type (%
\ref{offdncelepsilon}).

\setcounter{equation}{0} \renewcommand{\theequation}
{B.\arabic{equation}} \setcounter{subsection}{0} 
\renewcommand{\thesubsection}
{B.\arabic{subsection}}

\section{Tables 1-3 for generating off-diagonal solutions}

\label{appendixb}

In this appendix, we summarize the main steps on general decoupling and
integrating of (modified) Einstein equations with generic off-diagonal
quasi-stationary and locally anisotropic cosmological metrics in GR. Details
of geometric constructions and proofs with various generalizations for MGTs
were reviewed in \cite{sv11,vacaru18,vbubuianu17,nonassocFinslrev25}. The
AFCDM for GR in canonical nonholonomic variabes introduced in section \ref%
{sec02} is outlined below in Tables 1-3. \ We show how to use 2+2
nonholonomic variables and corresponding ansatz for metrics which allow us
to construct quasi-stationary and, for respective $t$-dual symmetries,
locally anisotropic cosmological solutions.

\subsection{Off-diagonal ansatz and nonlinear PDEs}

In Table 1, we provide necessary two types of parameterizations of frames
and coordinates on Lorentz manifolds enabled with N-connection h- and
v-splitting.



{\scriptsize 
\begin{eqnarray*}
&&%
\begin{tabular}{l}
\hline\hline
\begin{tabular}{lll}
& {\ \textsf{Table 1:\ Diagonal and off-diagonal ansatz for systems of
nonlinear ODEs and PDEs} } &  \\ 
& the Anholonomic Frame and Connection Deformation Method, \textbf{AFCDM}, & 
\\ 
& \textit{for constructing generic off-diagonal exact and parametric
solutions} & 
\end{tabular}%
\end{tabular}
\\
&&{%
\begin{tabular}{lll}
\hline
diagonal ansatz: PDEs $\rightarrow $ \textbf{ODE}s &  & AFCDM: \textbf{PDE}s 
\textbf{with decoupling; \ generating functions} \\ 
radial coordinates $u^{\alpha }=(r,\theta ,\varphi ,t)$ & $u=(x,y):$ & 
\mbox{ nonholonomic 2+2
splitting, } $u^{\alpha }=(x^{1},x^{2},y^{3},y^{4}=t)$ \\ 
LC-connection $\mathring{\nabla}$ & [connections] & $%
\begin{array}{c}
\mathbf{N}:T\mathbf{V}=hT\mathbf{V}\oplus vT\mathbf{V,}\mbox{ locally }%
\mathbf{N}=\{N_{i}^{a}(x,y)\} \\ 
\mbox{ canonical connection distortion }\widehat{\mathbf{D}}=\nabla +%
\widehat{\mathbf{Z}};\widehat{\mathbf{D}}\mathbf{g=0,} \\ 
\widehat{\mathcal{T}}[\mathbf{g,N,}\widehat{\mathbf{D}}]%
\mbox{ canonical
d-torsion}%
\end{array}%
$ \\ 
$%
\begin{array}{c}
\mbox{ diagonal ansatz  }g_{\alpha \beta }(u) \\ 
=\left( 
\begin{array}{cccc}
\mathring{g}_{1} &  &  &  \\ 
& \mathring{g}_{2} &  &  \\ 
&  & \mathring{g}_{3} &  \\ 
&  &  & \mathring{g}_{4}%
\end{array}%
\right)%
\end{array}%
$ & $\mathbf{g}\Leftrightarrow $ & $%
\begin{array}{c}
g_{\alpha \beta }=%
\begin{array}{c}
g_{\alpha \beta }(x^{i},y^{a})\mbox{ general frames / coordinates} \\ 
\left[ 
\begin{array}{cc}
g_{ij}+N_{i}^{a}N_{j}^{b}h_{ab} & N_{i}^{b}h_{cb} \\ 
N_{j}^{a}h_{ab} & h_{ac}%
\end{array}%
\right] ,\mbox{ 2 x 2 blocks }%
\end{array}
\\ 
\mathbf{g}_{\alpha \beta }=[g_{ij},h_{ab}],\mathbf{g}=\mathbf{g}%
_{i}(x^{k})dx^{i}\otimes dx^{i}+\mathbf{g}_{a}(x^{k},y^{b})\mathbf{e}%
^{a}\otimes \mathbf{e}^{b}%
\end{array}%
$ \\ 
$\mathring{g}_{\alpha \beta }=\left\{ 
\begin{array}{cc}
\mathring{g}_{\alpha }(r) & \mbox{ for BHs} \\ 
\mathring{g}_{\alpha }(t) & \mbox{ for FLRW }%
\end{array}%
\right. $ & [coord.frames] & $g_{\alpha \beta }=\left\{ 
\begin{array}{cc}
g_{\alpha \beta }(x^{i},y^{3}) & \mbox{ quasi-stationary configurations} \\ 
\underline{g}_{\alpha \beta }(x^{i},y^{4}=t) & 
\mbox{locally anisotropic
cosmology}%
\end{array}%
\right. $ \\ 
&  &  \\ 
$%
\begin{array}{c}
\mbox{coord. transf. }e_{\alpha }=e_{\ \alpha }^{\alpha ^{\prime }}\partial
_{\alpha ^{\prime }}, \\ 
e^{\beta }=e_{\beta ^{\prime }}^{\ \beta }du^{\beta ^{\prime }},\mathring{g}%
_{\alpha \beta }=\mathring{g}_{\alpha ^{\prime }\beta ^{\prime }}e_{\ \alpha
}^{\alpha ^{\prime }}e_{\ \beta }^{\beta ^{\prime }} \\ 
\begin{array}{c}
\mathbf{\mathring{g}}_{\alpha }(x^{k},y^{a})\rightarrow \mathring{g}_{\alpha
}(r),\mbox{ or }\mathring{g}_{\alpha }(t), \\ 
\mathring{N}_{i}^{a}(x^{k},y^{a})\rightarrow 0.%
\end{array}%
\end{array}%
$ & [N-adapt. fr.] & $\left\{ 
\begin{array}{cc}
\begin{array}{c}
\mathbf{g}_{i}(x^{k}),\mathbf{g}_{a}(x^{k},y^{3}), \\ 
\mbox{ or }\mathbf{g}_{i}(x^{k}),\underline{\mathbf{g}}_{a}(x^{k},t),%
\end{array}
& \mbox{ d-metrics } \\ 
\begin{array}{c}
N_{i}^{3}=w_{i}(x^{k},y^{3}),N_{i}^{4}=n_{i}(x^{k},y^{3}), \\ 
\mbox{ or }\underline{N}_{i}^{3}=\underline{n}_{i}(x^{k},t),\underline{N}%
_{i}^{4}=\underline{w}_{i}(x^{k},t),%
\end{array}
& 
\end{array}%
\right. $ \\ 
$\mathring{\nabla},$ $Ric=\{\mathring{R}_{\ \beta \gamma }\}$ & Ricci tensors
& $\widehat{\mathbf{D}},\ \widehat{\mathcal{R}}ic=\{\widehat{\mathbf{R}}_{\
\beta \gamma }\}$ \\ 
$~^{m}\mathcal{L[\mathbf{\phi }]\rightarrow }\ ^{m}\mathbf{T}_{\alpha \beta }%
\mathcal{[\mathbf{\phi }]}$ & 
\begin{tabular}{l}
generating \\ 
sources%
\end{tabular}
& $%
\begin{array}{cc}
\widehat{\mathbf{\Upsilon }}_{\ \nu }^{\mu }=\mathbf{e}_{\ \mu ^{\prime
}}^{\mu }\mathbf{e}_{\nu }^{\ \nu ^{\prime }}\mathbf{\Upsilon }_{\ \nu
^{\prime }}^{\mu ^{\prime }}[\ ^{m}\mathcal{L}(\mathbf{\varphi ),}T_{\mu \nu
},\Lambda ] &  \\ 
=diag[\ ^{h}\Upsilon (x^{i})\delta _{j}^{i},\ ^{v}\Upsilon
(x^{i},y^{3})\delta _{b}^{a}], & \mbox{ quasi-stationary configurations} \\ 
=diag[\ ^{h}\Upsilon (x^{i})\delta _{j}^{i},\ ^{v}\underline{\Upsilon }%
(x^{i},t)\delta _{b}^{a}], & \mbox{ locally anisotropic cosmology}%
\end{array}%
$ \\ 
trivial equations for $\mathring{\nabla}$-torsion & LC-conditions & $%
\widehat{\mathbf{D}}_{\mid \widehat{\mathcal{T}}\rightarrow 0}=\mathbf{%
\nabla }\mbox{ extracting new classes of solutions in GR}$ \\ \hline\hline
\end{tabular}%
}
\end{eqnarray*}%
}This table can be extended for higher dimension Lorentz manifolds and (co)
tangent Lorentz bundles as considered in{\scriptsize \ }\cite%
{vacaru18,vbubuianu17,nonassocFinslrev25}.

\subsection{Quasi-stationary configurations}

The key steps for applying the AFCDM for generating stationary off-diagonal
exact solutions of (modified) Einstein equations are outlined in Table 2.

{\scriptsize 
\begin{eqnarray*}
&&%
\begin{tabular}{l}
\hline\hline
\begin{tabular}{lll}
& {\large \textsf{Table 2:\ Off-diagonal quasi-stationary configurations}} & 
\\ 
& Exact solutions of $\widehat{\mathbf{R}}_{\mu \nu }=\mathbf{\Upsilon }%
_{\mu \nu }$ (\ref{cdeq1}) transformed into a system of nonlinear PDEs (\ref%
{eq1})-(\ref{e2c}) & 
\end{tabular}
\\ 
\end{tabular}
\\
&&%
\begin{tabular}{lll}
\hline\hline
&  &  \\ 
$%
\begin{array}{c}
\\ 
\end{array}%
\begin{array}{c}
\mbox{d-metric ansatz with} \\ 
\mbox{Killing symmetry }\partial _{4}=\partial _{t} \\ 
\mbox{general or spherical coordinates}%
\end{array}%
$ &  & $%
\begin{array}{c}
ds^{2}=g_{i}(x^{k})(dx^{i})^{2}+g_{a}(x^{k},y^{3})(dy^{a}+N_{i}^{a}(x^{k},y^{3})dx^{i})^{2},%
\mbox{ for } \\ 
g_{i}=e^{\psi {(x}^{k}{)}%
},g_{a}=h_{a}(x^{k},y^{3}),N_{i}^{3}=w_{i}(x^{k},y^{3}),N_{i}^{4}=n_{i}(x^{k},y^{3});
\\ 
g_{i}=e^{\psi {(r,\theta )}},\,\,\,\,g_{a}=h_{a}({r,\theta },\varphi ),\
N_{i}^{3}=w_{i}({r,\theta },\varphi ),\,\,\,\,N_{i}^{4}=n_{i}({r,\theta }%
,\varphi ),%
\end{array}%
$ \\ 
&  &  \\ 
Effective matter sources &  & $\mathbf{\Upsilon }_{\ \nu }^{\mu }=[\
^{h}\Upsilon ({r,\theta })\delta _{j}^{i},\ ^{v}\Upsilon ({r,\theta }%
,\varphi )\delta _{b}^{a}],\mbox{ if }x^{1}=r,x^{2}=\theta ,y^{3}=\varphi
,y^{4}=t$ \\ \hline
Nonlinear PDEs (\ref{eq1})-(\ref{e2c}) &  & $%
\begin{array}{c}
\psi ^{\bullet \bullet }+\psi ^{\prime \prime }=2\ \ ^{h}\Upsilon ; \\ 
\varpi ^{\ast }\ h_{4}^{\ast }=2h_{3}h_{4}\ \ ^{v}\Upsilon ; \\ 
\beta w_{i}-\alpha _{i}=0; \\ 
n_{k}^{\ast \ast }+\gamma n_{k}^{\ast }=0;%
\end{array}%
$ for $%
\begin{array}{c}
\varpi {=\ln |\partial _{3}h_{4}/\sqrt{|h_{3}h_{4}|}|,} \\ 
\alpha _{i}=(\partial _{3}h_{4})\ (\partial _{i}\varpi ),\beta =(\partial
_{3}h_{4})\ (\partial _{3}\varpi ),\  \\ 
\ \gamma =\partial _{3}\left( \ln |h_{4}|^{3/2}/|h_{3}|\right) , \\ 
\partial _{1}q=q^{\bullet },\partial _{2}q=q^{\prime },\partial
_{3}q=\partial q/\partial \varphi =q^{\ast }%
\end{array}%
$ \\ \hline
$%
\begin{array}{c}
\mbox{ Generating functions:}\ h_{3}(x^{k},y^{3}), \\ 
\Psi (x^{k},y^{3})=e^{\varpi },\Phi ((x^{k},y^{3})); \\ 
\mbox{integration functions:}\ h_{4}^{[0]}(x^{k}),\  \\ 
_{1}n_{k}(x^{i}),\ _{2}n_{k}(x^{i}); \\ 
\mbox{\& nonlinear symmetries}%
\end{array}%
$ &  & $%
\begin{array}{c}
\ (\Psi ^{2})^{\ast }=-\int dy^{3}\ ^{v}\Upsilon h_{4}^{\ \ast }, \\ 
\Phi ^{2}=-4\ \Lambda h_{4},\mbox{ see }(\ref{nonlinsymrex}); \\ 
h_{4}=h_{4}^{[0]}-\Phi ^{2}/4\ \Lambda ,h_{4}^{\ast }\neq 0,\ \Lambda \neq
0=const%
\end{array}%
$ \\ \hline
Off-diag. solutions, $%
\begin{array}{c}
\mbox{d--metric} \\ 
\mbox{N-connec.}%
\end{array}%
$ &  & $%
\begin{array}{c}
\ g_{i}=e^{\ \psi (x^{k})}\mbox{ as a solution of 2-d Poisson eqs. }\psi
^{\bullet \bullet }+\psi ^{\prime \prime }=2~\ ^{h}\Upsilon ; \\ 
h_{3}=-(\Psi ^{\ast })^{2}/4\ ^{v}\Upsilon ^{2}h_{4},\mbox{ see }(\ref{g3}),(%
\ref{g4}); \\ 
h_{4}=h_{4}^{[0]}-\int dy^{3}(\Psi ^{2})^{\ast }/4\ ^{v}\Upsilon
=h_{4}^{[0]}-\Phi ^{2}/4\ \Lambda ; \\ 
\\ 
w_{i}=\partial _{i}\ \Psi /\ \partial _{3}\Psi =\partial _{i}\ \Psi ^{2}/\
\partial _{3}\Psi ^{2}|; \\ 
n_{k}=\ _{1}n_{k}+\ _{2}n_{k}\int dy^{3}(\Psi ^{\ast })^{2}/\ ^{v}\Upsilon
^{2}|h_{4}^{[0]}-\int dy^{3}(\Psi ^{2})^{\ast }/4\ ^{v}\Upsilon ^{2}|^{5/2}.
\\ 
\\ 
\end{array}%
$ \\ \hline
LC-configurations (\ref{zerot1}) &  & $%
\begin{array}{c}
\partial _{\varphi }w_{i}=(\partial _{i}-w_{i}\partial _{3})\ln \sqrt{|h_{3}|%
},(\partial _{i}-w_{i}\partial _{3})\ln \sqrt{|h_{4}|}=0, \\ 
\partial _{k}w_{i}=\partial _{i}w_{k},\partial _{3}n_{i}=0,\partial
_{i}n_{k}=\partial _{k}n_{i}; \\ 
\mbox{ see d-metric }(\ref{qellc})\mbox{ for } \\ 
\Psi =\check{\Psi}(x^{i},y^{3}),(\partial _{i}\check{\Psi})^{\ast }=\partial
_{i}(\check{\Psi}^{\ast })\mbox{ and } \\ 
\ ^{v}\Upsilon (x^{i},\varphi )=\Upsilon \lbrack \check{\Psi}]=\check{%
\Upsilon},\mbox{ or }\Upsilon =const. \\ 
\end{array}%
$ \\ \hline
N-connections, zero torsion &  & $%
\begin{array}{c}
w_{i}=\partial _{i}\check{A}=\left\{ 
\begin{array}{c}
\partial _{i}(\int dy^{3}\ \check{\Upsilon}\ \check{h}_{4}{}^{\ast }])/%
\check{\Upsilon}\ \check{h}_{4}{}^{\ast }; \\ 
\partial _{i}\check{\Psi}/\check{\Psi}^{\ast }; \\ 
\partial _{i}(\int dy^{3}\ \check{\Upsilon}(\check{\Phi}^{2})^{\ast })/(%
\check{\Phi})^{\ast }\ \check{\Upsilon};%
\end{array}%
\right. \\ 
\mbox{ and }n_{k}=\check{n}_{k}=\partial _{k}n(x^{i}).%
\end{array}%
$ \\ \hline
$%
\begin{array}{c}
\mbox{polarization functions} \\ 
\mathbf{\mathring{g}}\rightarrow \widehat{\mathbf{g}}\mathbf{=}[g_{\alpha
}=\eta _{\alpha }\mathring{g}_{\alpha },\ \eta _{i}^{a}\mathring{N}_{i}^{a}]%
\end{array}%
$ &  & $%
\begin{array}{c}
\\ 
ds^{2}=\eta _{1}(r,\theta )\mathring{g}_{1}(r,\theta )[dx^{1}(r,\theta
)]^{2}+\eta _{2}(r,\theta )\mathring{g}_{2}(r,\theta )[dx^{2}(r,\theta
)]^{2}+ \\ 
\eta _{3}(r,\theta ,\varphi )\mathring{g}_{3}(r,\theta )[d\varphi +\eta
_{i}^{3}(r,\theta ,\varphi )\mathring{N}_{i}^{3}(r,\theta )dx^{i}(r,\theta
)]^{2}+ \\ 
\eta _{4}(r,\theta ,\varphi )\mathring{g}_{4}(r,\theta )[dt+\eta
_{i}^{4}(r,\theta ,\varphi )\mathring{N}_{i}^{4}(r,\theta )dx^{i}(r,\theta
)]^{2}, \\ 
\end{array}%
$ \\ \hline
Prime metric defines a BH &  & $%
\begin{array}{c}
\\ 
\lbrack \mathring{g}_{i}(r,\theta ),\mathring{g}_{a}=\mathring{h}%
_{a}(r,\theta );\mathring{N}_{k}^{3}=\mathring{w}_{k}(r,\theta ),\mathring{N}%
_{k}^{4}=\mathring{n}_{k}(r,\theta )] \\ 
\mbox{diagonalizable by frame/ coordinate transforms.} \\ 
\end{array}%
$ \\ 
Example of a prime metric &  & $%
\begin{array}{c}
\\ 
\mathring{g}_{1}=(1-r_{g}/r)^{-1},\mathring{g}_{2}=r^{2},\mathring{h}%
_{3}=r^{2}\sin ^{2}\theta ,\mathring{h}_{4}=(1-r_{g}/r),r_{g}=const \\ 
\mbox{the Schwarzschild solution, or any BH solution.} \\ 
\mbox{ for new KdS solutions }\mbox{ with }\mathbf{\mathring{g}\simeq \breve{%
g}}(x^{i},y^{3})\mathbf{=}(\breve{g}_{\alpha };\breve{N}_{i}^{a}); \\ 
\end{array}%
$ \\ \hline
Solutions for polarization funct. &  & $%
\begin{array}{c}
\eta _{i}=e^{\ \psi (x^{k})}/\mathring{g}_{i};\eta _{3}\mathring{h}_{3}=-%
\frac{4[(|\eta _{4}\mathring{h}_{4}|^{1/2})^{\ast }]^{2}}{|\int dy^{3}\
^{v}\Upsilon \lbrack (\eta _{4}\mathring{h}_{4})]^{\ast }|\ }; \\ 
\eta _{4}=\eta _{4}(x^{k},y^{3})\mbox{ as a generating
function}; \\ 
\ \eta _{i}^{3}\ \mathring{N}_{i}^{3}=\frac{\partial _{i}\ \int dy^{3}\
^{v}\Upsilon (\eta _{4}\ \mathring{h}_{4})^{\ast }}{^{3}\ ^{v}\Upsilon \
(\eta _{4}\ \mathring{h}_{4})^{\ast }}; \\ 
\eta _{k}^{4}\ \mathring{N}_{k}^{4}=\ _{1}n_{k}+16\ \ _{2}n_{k}\int dy^{3}%
\frac{\left( [(\eta _{4}\mathring{h}_{4})^{-1/4}]^{\ast }\right) ^{2}}{|\int
dy^{3}\ ^{v}\Upsilon \lbrack (\eta _{4}\ \mathring{h}_{4})]^{\ast }|\ };%
\end{array}%
$ \\ \hline
Polariz. funct. with zero torsion &  & $%
\begin{array}{c}
\eta _{i}=e^{\ \psi (x^{k})}/\mathring{g}_{i};\eta _{4}=\check{\eta}%
_{4}(x^{k},y^{3})\mbox{ as a generating function}; \\ 
\eta _{3}=-\frac{4[(|\eta _{4}\mathring{h}_{4}|^{1/2})^{\ast }]^{2}}{%
\mathring{g}_{3}|\int dy^{3}\ ^{v}\Upsilon \lbrack (\check{\eta}_{4}%
\mathring{h}_{4})]^{\ast }|\ };\eta _{i}^{3}=\partial _{i}\check{A}/%
\mathring{w}_{k},\eta _{k}^{4}=\frac{\ \partial _{k}n}{\mathring{n}_{k}}. \\ 
\end{array}%
$ \\ \hline\hline
\end{tabular}%
\end{eqnarray*}%
}The formulas for generating quasi-stationary solutions stated in above
table can be dualized on a time like variable (see explanations for formula (%
\ref{dualcosm})) and used for generating locally anisotropic cosmological
solutions.

\subsection{Locally anisotropic cosmological solutions}

We summarize the main steps for constructing off-diagonal locally
anisotropic solutions of (modified) Einstein equations using the AFCDM in
this form:

{\scriptsize 
\begin{eqnarray*}
&&%
\begin{tabular}{l}
\hline\hline
\begin{tabular}{lll}
& {\large \textsf{Table 3:\ Off-diagonal locally anisotropic cosmological
models}} &  \\ 
& Exact solutions of $\widehat{\mathbf{R}}_{\mu \nu }=\underline{\mathbf{%
\Upsilon }}_{\mu \nu }$ (\ref{cdeq1}) transformed into a system of nonlinear
PDEs (\ref{dualcosm}) & 
\end{tabular}
\\ 
\end{tabular}
\\
&&%
\begin{tabular}{lll}
\hline\hline
$%
\begin{array}{c}
\mbox{d-metric ansatz with} \\ 
\mbox{Killing symmetry }\partial _{3}=\partial _{\varphi }%
\end{array}%
$ &  & $%
\begin{array}{c}
d\underline{s}^{2}=g_{i}(x^{k})(dx^{i})^{2}+\underline{g}%
_{a}(x^{k},y^{4})(dy^{a}+\underline{N}_{i}^{a}(x^{k},y^{4})dx^{i})^{2},%
\mbox{ for } \\ 
g_{i}=e^{\psi {(x}^{k}{)}},\,\,\,\,\underline{g}_{a}=\underline{h}_{a}({x}%
^{k},t),\ \underline{N}_{i}^{3}=\underline{n}_{i}({x}^{k},t),\,\,\,%
\underline{\,N}_{i}^{4}=\underline{w}_{i}({x}^{k},t),%
\end{array}%
$ \\ 
&  &  \\ 
Effective matter sources &  & $\underline{\mathbf{\Upsilon }}_{\ \nu }^{\mu
}=[~\ _{h}\underline{\Upsilon }({x}^{k})\delta _{j}^{i},~\ _{v}\underline{%
\Upsilon }({x}^{k},t)\delta _{b}^{a}];x^{1},x^{2},y^{3},y^{4}=t$ \\ \hline
Nonlinear PDEs &  & $%
\begin{array}{c}
\psi ^{\bullet \bullet }+\psi ^{\prime \prime }=2\ ^{h}\underline{\Upsilon };
\\ 
\underline{\varpi }^{\diamond }\ \underline{h}_{3}^{\diamond }=2\underline{h}%
_{3}\underline{h}_{4}\ ^{v}\underline{\Upsilon }; \\ 
\underline{n}_{k}^{\diamond \diamond }+\underline{\gamma }\underline{n}%
_{k}^{\diamond }=0; \\ 
\underline{\beta }\underline{w}_{i}-\underline{\alpha }_{i}=0;%
\end{array}%
$ for $%
\begin{array}{c}
\underline{\varpi }{=\ln |\partial _{t}\underline{{h}}_{3}/\sqrt{|\underline{%
h}_{3}\underline{h}_{4}|}|,} \\ 
\underline{\alpha }_{i}=(\partial _{t}\underline{h}_{3})\ (\partial _{i}%
\underline{\varpi }),\ \underline{\beta }=(\partial _{t}\underline{h}_{3})\
(\partial _{t}\underline{\varpi }), \\ 
\ \underline{\gamma }=\partial _{t}\left( \ln |\underline{h}_{3}|^{3/2}/|%
\underline{h}_{4}|\right) , \\ 
\partial _{1}q=q^{\bullet },\partial _{2}q=q^{\prime },\partial
_{4}q=\partial q/\partial t=q^{\diamond }%
\end{array}%
$ \\ \hline
$%
\begin{array}{c}
\mbox{ Generating functions:}\ \underline{h}_{4}({x}^{k},t), \\ 
\underline{\Psi }(x^{k},t)=e^{\underline{\varpi }},\underline{\Phi }({x}%
^{k},t); \\ 
\mbox{integr. functions:}\ h_{4}^{[0]}(x^{k}),\ _{1}n_{k}(x^{i}),\  \\ 
_{2}n_{k}(x^{i});\mbox{\& nonlinear symmetries}%
\end{array}%
$ &  & $%
\begin{array}{c}
\ (\underline{\Psi }^{2})^{\diamond }=-\int dt\ ^{v}\underline{\Upsilon }%
\underline{h}_{3}^{\diamond }, \\ 
\underline{\Phi }^{2}=-4\ \underline{\Lambda }\underline{h}_{3}; \\ 
\underline{h}_{3}=\underline{h}_{3}^{[0]}-\underline{\Phi }^{2}/4\ 
\underline{\Lambda },\underline{h}_{3}^{\diamond }\neq 0,\ \underline{%
\Lambda }\neq 0=const%
\end{array}%
$ \\ \hline
Off-diag. solutions, $%
\begin{array}{c}
\mbox{d--metric} \\ 
\mbox{N-connec.}%
\end{array}%
$ &  & $%
\begin{array}{c}
\ g_{i}=e^{\ \psi (x^{k})}\mbox{ as a solution of 2-d Poisson eqs. }\psi
^{\bullet \bullet }+\psi ^{\prime \prime }=2\ ^{h}\underline{\Upsilon }; \\ 
\overline{h}_{4}=-(\overline{\Psi }^{2})^{\diamond }/4\ ^{v}\underline{%
\Upsilon }^{2}\underline{h}_{3}; \\ 
\underline{h}_{3}=h_{3}^{[0]}-\int dt(\underline{\Psi }^{2})^{\diamond }/4\
^{v}\underline{\Upsilon }=h_{3}^{[0]}-\underline{\Phi }^{2}/4\ \underline{%
\Lambda }; \\ 
\underline{n}_{k}=\ _{1}n_{k}+\ _{2}n_{k}\int dt(\underline{\Psi }^{\diamond
})^{2}/\ ^{v}\underline{\Upsilon }^{2}\ |h_{3}^{[0]}-\int dt(\underline{\Psi 
}^{2})^{\diamond }/4\ ^{v}\underline{\Upsilon }|^{5/2}; \\ 
\underline{w}_{i}=\partial _{i}\ \underline{\Psi }/\ \partial _{t}\underline{%
\Psi }=\partial _{i}\underline{\Psi }^{2}/\ \partial _{t}\underline{\Psi }%
^{2}. \\ 
\end{array}%
$ \\ \hline
LC-configurations &  & $%
\begin{array}{c}
\partial _{t}\underline{w}_{i}=(\partial _{i}-\underline{w}_{i}\partial
_{t})\ln \sqrt{|\underline{h}_{4}|},(\partial _{i}-\underline{w}_{i}\partial
_{4})\ln \sqrt{|\underline{h}_{3}|}=0, \\ 
\partial _{k}\underline{w}_{i}=\partial _{i}\underline{w}_{k},\partial _{t}%
\underline{n}_{i}=0,\partial _{i}\underline{n}_{k}=\partial _{k}\underline{n}%
_{i}; \\ 
\underline{\Psi }=\underline{\check{\Psi}}(x^{i},t),(\partial _{i}\underline{%
\check{\Psi}})^{\diamond }=\partial _{i}(\underline{\check{\Psi}}^{\diamond
})\mbox{ and } \\ 
\ ^{v}\underline{\Upsilon }(x^{i},t)=\underline{\Upsilon }[\underline{\check{%
\Psi}}]=\underline{\check{\Upsilon}},\mbox{ or }\underline{\Upsilon }=const.
\\ 
\end{array}%
$ \\ \hline
N-connections, zero torsion &  & $%
\begin{array}{c}
\underline{n}_{k}=\underline{\check{n}}_{k}=\partial _{k}\underline{n}(x^{i})
\\ 
\mbox{ and }\underline{w}_{i}=\partial _{i}\underline{\check{A}}=\left\{ 
\begin{array}{c}
\partial _{i}(\int dt\ \underline{\check{\Upsilon}}\ \underline{\check{h}}%
_{3}^{\diamond }])/\underline{\check{\Upsilon}}\ \underline{\check{h}}%
_{3}^{\diamond }{}; \\ 
\partial _{i}\underline{\check{\Psi}}/\underline{\check{\Psi}}^{\diamond };
\\ 
\partial _{i}(\int dt\ \underline{\check{\Upsilon}}(\underline{\check{\Phi}}%
^{2})^{\diamond })/\underline{\check{\Phi}}^{\diamond }\underline{\check{%
\Upsilon}};%
\end{array}%
\right. .%
\end{array}%
$ \\ \hline
$%
\begin{array}{c}
\mbox{polarization functions} \\ 
\mathbf{\mathring{g}}\rightarrow \underline{\widehat{\mathbf{g}}}\mathbf{=}[%
\underline{g}_{\alpha }=\underline{\eta }_{\alpha }\underline{\mathring{g}}%
_{\alpha },\underline{\eta }_{i}^{a}\underline{\mathring{N}}_{i}^{a}]%
\end{array}%
$ &  & $%
\begin{array}{c}
ds^{2}=\underline{\eta }_{i}(x^{k},t)\underline{\mathring{g}}%
_{i}(x^{k},t)[dx^{i}]^{2}+\underline{\eta }_{3}(x^{k},t)\underline{\mathring{%
h}}_{3}(x^{k},t)[dy^{3}+\underline{\eta }_{i}^{3}(x^{k},t)\underline{%
\mathring{N}}_{i}^{3}(x^{k},t)dx^{i}]^{2} \\ 
+\underline{\eta }_{4}(x^{k},t)\underline{\mathring{h}}_{4}(x^{k},t)[dt+%
\underline{\eta }_{i}^{4}(x^{k},t)\underline{\mathring{N}}%
_{i}^{4}(x^{k},t)dx^{i}]^{2}, \\ 
\end{array}%
$ \\ \hline
$%
\begin{array}{c}
\mbox{ Prime metric defines } \\ 
\mbox{ a cosmological solution}%
\end{array}%
$ &  & $%
\begin{array}{c}
\lbrack \underline{\mathring{g}}_{i}(x^{k},t),\underline{\mathring{g}}_{a}=%
\underline{\mathring{h}}_{a}(x^{k},t);\underline{\mathring{N}}_{k}^{3}=%
\underline{\mathring{w}}_{k}(x^{k},t),\underline{\mathring{N}}_{k}^{4}=%
\underline{\mathring{n}}_{k}(x^{k},t)] \\ 
\mbox{diagonalizable by frame/ coordinate transforms.} \\ 
\end{array}%
$ \\ 
$%
\begin{array}{c}
\mbox{Example of a prime } \\ 
\mbox{ cosmological metric }%
\end{array}%
$ &  & $%
\begin{array}{c}
\mathring{g}_{1}=a^{2}(t)/(1-kr^{2}),\mathring{g}_{2}=a^{2}(t)r^{2}, \\ 
\underline{\mathring{h}}_{3}=a^{2}(t)r^{2}\sin ^{2}\theta ,\underline{%
\mathring{h}}_{4}=c^{2}=const,k=\pm 1,0; \\ 
\mbox{ any frame transform of a FLRW or a Bianchi metrics} \\ 
\end{array}%
$ \\ \hline
Solutions for polarization funct. &  & $%
\begin{array}{c}
\eta _{i}=e^{\ \psi (x^{k})}/\mathring{g}_{i};\underline{\eta }_{4}%
\underline{\mathring{h}}_{4}=-\frac{4[(|\underline{\eta }_{3}\underline{%
\mathring{h}}_{3}|^{1/2})^{\diamond }]^{2}}{|\int dt\ ^{v}\underline{%
\Upsilon }[(\underline{\eta }_{3}\underline{\mathring{h}}_{3})]^{\diamond
}|\ };\mbox{ gener. funct. }\underline{\eta }_{3}=\underline{\eta }%
_{3}(x^{i},t); \\ 
\underline{\eta }_{k}^{3}\ \underline{\mathring{N}}_{k}^{3}=\ _{1}n_{k}+16\
\ _{2}n_{k}\int dt\frac{\left( [(\underline{\eta }_{3}\underline{\mathring{h}%
}_{3})^{-1/4}]^{\diamond }\right) ^{2}}{|\int dt\ ^{v}\underline{\Upsilon }[(%
\underline{\eta }_{3}\underline{\mathring{h}}_{3})]^{\diamond }|\ };\ 
\underline{\eta }_{i}^{4}\ \underline{\mathring{N}}_{i}^{4}=\frac{\partial
_{i}\ \int dt\ ^{v}\underline{\Upsilon }(\underline{\eta }_{3}\underline{%
\mathring{h}}_{3})^{\diamond }}{\ ^{v}\underline{\Upsilon }(\underline{\eta }%
_{3}\underline{\mathring{h}}_{3})^{\diamond }},%
\end{array}%
$ \\ \hline
Polariz. funct. with zero torsion &  & $%
\begin{array}{c}
\eta _{i}=e^{\ \psi }/\mathring{g}_{i};\underline{\eta }_{4}=-\frac{4[(|%
\underline{\eta }_{3}\underline{\mathring{h}}_{3}|^{1/2})^{\diamond }]^{2}}{%
\underline{\mathring{g}}_{4}|\int dt\ ^{v}\underline{\Upsilon }[(\underline{%
\eta }_{3}\underline{\mathring{h}}_{3})]^{\diamond }|\ };%
\mbox{ gener.
funct. }\underline{\eta }_{3}=\underline{\check{\eta}}_{3}({x}^{i},t); \\ 
\underline{\eta }_{k}^{4}=\partial _{k}\underline{\check{A}}/\mathring{w}%
_{k};\underline{\eta }_{k}^{3}=(\partial _{k}\underline{n})/\mathring{n}_{k}.
\\ 
\end{array}%
$ \\ \hline\hline
\end{tabular}%
\end{eqnarray*}%
}

Tables 1- 3 can be used for generating off-diagonal exact and parametric
solutions in GR for various types of prescribed generating functions and
(effective) generating sources. Typically, such solutions involve 6
independent components (from ten) of a Lorentzian metric depending at least
on 3 space-time coordinates. They descibe a generic nonlinear off-diagonal
dynamics subjected to various types of nonholonomic constraints and
nonlinear symmetries which allow to introduce effective cosmological
constants. The physical properties of such off-diagonal solutions in GR are
very different from those described with diagonalizable ansatz. This way, we
describe a new nonlinear physics with various applications, for instance, in
QG and MGTs, accelerating cosmology and DE and DM physics \cite%
{vacaru25,vacaru18,partner06,gheorghiuap16,vbubuianu17,vacaruplb16,nonassocFinslrev25}%
. In this work, some examples of physically important solutions are studied
in section \ref{sec03}.

\end{document}